\pdfoutput=1

\documentclass[a4paper,12pt,oneside,openright]{book}
\usepackage{import}
\usepackage{graphicx}

\usepackage{comment,soul}
\usepackage[dvipsnames]{xcolor} 
\usepackage[hidelinks]{hyperref}
\usepackage{enumitem}
\usepackage{amsmath,amssymb,epsf,cite,graphicx,subfigure}
\usepackage{amsmath,braket,tikzsymbols}
\usepackage[bbgreekl]{mathbbol}
\usepackage{comment}
\usepackage{cancel}
\usepackage{multicol}
\usepackage{pdfpages}

\usepackage{blindtext}
\usepackage{mdframed}

\usepackage{geometry}

\usepackage{titlesec}

\usepackage{setspace}

\hypersetup{
    linktoc=all,     
}

\usepackage{fancyhdr}
\newcommand{\diff}{\text{d}}

\newcommand{\rhon}{\rho_\text{n}}

\newcommand{\nn}{n_\text{n}}
\newcommand{\ns}{n_\text{s}}

\newcommand{\pdert}[3]{\frac{\partial #1}{\partial #2}{\bigg|_{#3}}}
\newcommand{\pdertmedium}[3]{\frac{\partial #1}{\partial #2}{\Big|_{#3}}}
\newcommand{\pdertsmall}[3]{\frac{\partial #1}{\partial #2}{\big|_{#3}}}

\newcommand{\dmu}{\partial_\mu}
\newcommand{\dnu}{\partial_\nu}

\newcommand{\intd}{\int \diff^d x \,}

\newcommand{\be}{\begin{equation}}
\newcommand{\ee}{\end{equation}}

\newcommand{\xio}{{\xi_0}}
\newcommand{\mud}{\mu_{\text{diss}}}
\newcommand{\mus}{\mu_{\text{s}}}

\newcommand{\na}{\langle n_a(\omega,\mathbf{k})\rangle }

\usepackage[mathscr]{euscript}

\selectfont

\allowdisplaybreaks

\title{
Revisiting the Landau criterion: a hydrodynamic and holographic approach to superfluid instabilities
}

\author{Filippo Sottovia}
\date{04/01/2024}

\makeatletter
\def\thickhrulefill{\leavevmode \leaders \hrule height 1ex \hfill \kern \z@}
\def\@makechapterhead#1{%
  \vspace*{10\p@}%
  {\parindent \z@ 
        \raggedleft
        \reset@font\huge\bfseries
        \begin{tabular}{c|p{15cm}}
          {\thechapter{}\  }
          &\quad
          \Huge #1
        \end{tabular}
        \par\nobreak
    \vskip 70\p@
  }}
\def\@makeschapterhead#1{%
  \vspace*{10\p@}%
  {\parindent \z@ 
        \raggedleft
        \reset@font\huge\bfseries
        \begin{tabular}{cp{15cm}}
          {\hphantom{\thechapter{}}\  }
          &\quad
          \Huge #1
        \end{tabular}
        \par\nobreak
    \vskip 70\p@
  }}

\begin{document}
\author{Filippo Sottovia}

\includepdf[pages=-]{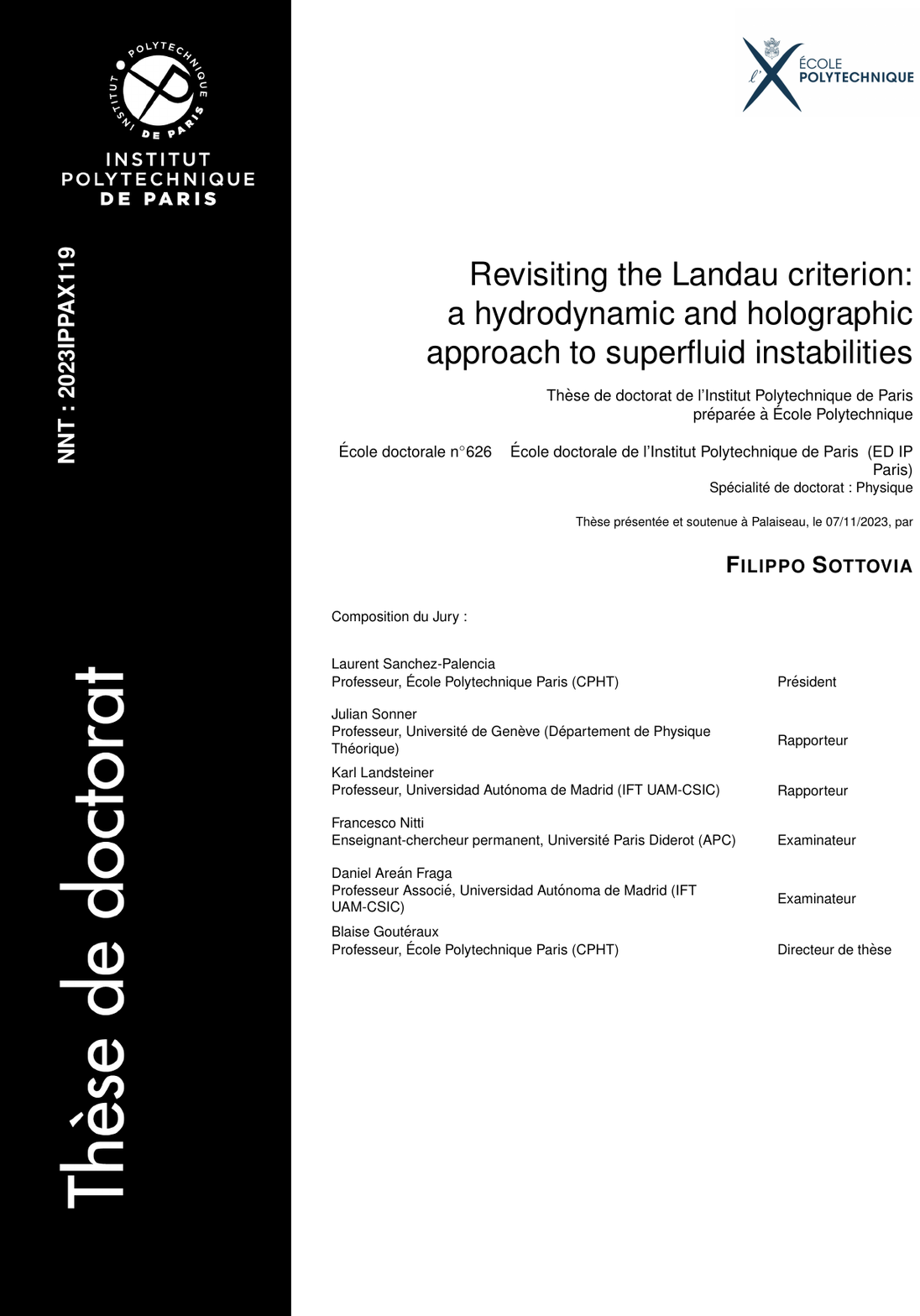}

\newpage
\
\thispagestyle{empty}
\newpage

\frontmatter
\maketitle

\chapter*{Acknowledgements}
First of all I would like to express my gratitude to Prof. Karl Landsteiner and Prof. Julian Sonner for accepting to read the manuscript and to be part of the committee. For the same reason I also want to thank Prof. Francesco Nitti, Prof. Laurent Sanchez-Palencia and Prof. Daniel Are\'an Fraga.
\\My supervisor Blaise Gout\'eraux deserves special credit for guiding me throughout these three years. I am especially grateful to him for his patience and all the opportunities he gave me, including the workshops and conferences that I had the chance to attend. I also learnt a lot from Eric Mefford and Daniel Are\'an Fraga. I would really like to thank them for bearing with me through my numerous attempts at tackling long calculations: their help was extremely valuable. 
\\I would also like to thank Balt van Rees and, again, Laurent Sanchez-Palencia, for being on my mini-defence committee.
\bigskip
\medskip
\\This journey would not have been possible without those with whom I shared the road. In the first place, the other PhD students and the postdoctoral researchers. Yorgo, Erik, Gabriele, Matthieu, Adrien, Majdouline, David, Vaios, Mikel, David, Aditya, Danny, Ashish: it has been a pleasure to share ideas (and laughs) with you.
\bigskip
\\My days would have been much duller without my wonderful friends. First, those who have always been there: gli amici del liceo, tra cui Luisa, Federico ed Eva, Chiara (i pranzi domenicali non saranno dimenticati) e tutti gli amici del Posta, und meine Schweizer Freunde, Stefania, Lidia, Nicolas, Robin und die Spitzenk\"oche. Thank you Dimitrios for our weekly dinners. Grazie Mendun per le lunghe chiacchierate.
\\Merci aussi \`a mes colocataires, Bianca, Lucie, Manon et Lucas pour tous les moments, quotidiens ou non, que nous avons partag\'es (mention sp\'eciale \`a Robin). 
\\Un grand merci aux scouts de Fresnes, chefs et jeunes, de m'avoir accueilli comme l'un des leurs. Vous \^etes trop nombreux pour que j'\'ecrive tous vos pr\'enoms, mais sachez que ce fut un plaisir et un grand honneur. 
\bigskip
\\Last, but not least, I shall thank my family for their constant support. A mamma Ilaria e pap\`a Lucio, a zia Cristina e a tutti gli altri zii e zie, ai miei nonni e cugini un grande grazie per l'affetto che mi avete dato in questi anni.

\newpage 
\ 
\newpage
\chapter*{Abstract}
Superfluidity is a state of matter whose main feature is the absence of viscous flow and heat transport. It originates from the spontaneous breaking of a $U(1)$ continuous global symmetry below a critical temperature $T_c$. The gradient of the Goldstone field associated to the symmetry breaking determines the superfluid velocity. At finite temperature a superfluid and a normal fluid component coexist; both of them contribute to charge flow. Landau predicted that the superfluid becomes unstable at large superfluid velocities. He determined the critical velocity at low temperature by assuming boost invariance and that the normal fluid is composed of free bosonic quasiparticles with Bose-Einstein statistics. 
\\Hydrodynamics, on the other hand, gives a valid description of interacting systems at length- and timescales longer than the typical scales at which local thermal equilibrium is reached. In this thesis we employ a hydrodynamical approach in order to investigate the instabilities of superfluids at finite superflow. In particular, we find that a hydrodynamic collective mode crosses to the upper half complex frequency plane, which signals a dynamical instability. At the same time, however, this instability is also thermodynamic, as its onset is controlled by one of the second derivatives of the free energy changing sign. 
\\We study the onset of this instability in two main setups: the ``probe limit'', where the fluctuations of the temperature and the normal fluid's velocity are frozen, and a complete approach, which includes them. 
\\In both cases we test our findings with the help of gauge-gravity duality (holography). Up to our numerical precision, we find agreement between the hydrodynamic modes of the boundary theory and the quasinormal modes of the gravity theory. We are therewith able to verify our criterion in a holographic setup.
\\Most importantly, our criterion, which is formulated in a model-independent way, applies to interacting systems irrespective of the strength of interactions, does not rely on boost invariance and does not assume any specific quantum statistics. As a final check, we also show that it yields the Landau critical velocity for Galilean superfluids with Bose-Einstein quasiparticles.

\newpage
\chapter*{R\'esum\'e en fran\c{c}ais}
La suprafluidit\'e est un \'etat de la mati\`ere caract\'eris\'e par l'absence de flux visqueux et de transfert de chaleur. Elle survient lorsqu'une sym\'etrie globale continue de type $U(1)$ est bris\'ee spontan\'ement. Le gradient du champ de Goldstone associ\'e d\'etermine la v\'elocit\'e suprafluide. \`A temperature non-nulle, le fluide poss\`ede une composante normale, qui contribue \'egalement au flot de la charge. Une pr\'ediction c\'el\`ebre de Landau est que le syst\`eme souffre d'une instabilit\'e lorsque la vitesse suprafluide d\'epasse une certaine valeur critique. Landau d\'etermina cette valeur dans un r\'egime de basse temp\'erature en supposant le syst\`eme invariant sous les transformations de Galil\'ee et que le fluide normal est constitu\'e de particules libres avec une statistique de Bose-Einstein. Cependant, afin d'\'evaluer la vitesse critique selon son crit\`ere, il est n\'ecessaire de conna\^itre le spectre des excitations microscopiques. 
\bigskip
\\L'approximation hydrodynamique permet de d\'ecrire les syst\`emes avec interactions aux \'echelles d'espace et de temps longues compar\'ees \`a celles o\`u l'\'equilibre thermodynamique est \'etabli localement. Dans cette th\`ese nous introduisons d'abord le formalisme hydrodynamique des syst\`emes suprafluides et donnons une classification compl\`ete des contributions dissipatives. En suite, nous appliquons cette approche hydrodynamique afin de determiner la vitesse critique des suprafluides. 
\bigskip
\\En particulier, aux grandes valeurs de la vitesse suprafluide, nous trouvons qu'un mode collectif hydrodynamique passe dans le demi-plan sup\'erieur des fr\'equences complexes, ce qui signale une instabilit\'e dynamique. Cette instabilit\'e est en fait d'origine thermodynamique, car elle est provoqu\'ee par un changement de signe d'une d\'eriv\'ee seconde de l'\'energie libre. Elle a lieu lorsque:
\begin{equation}\label{critf}
\xi + \partial_\xi n_\text{s}=0 \, \text{,}
\end{equation}
o\`u $\xi$ indique la partie spatiale du gradient du champ de Goldstone et $n_\text{s}$ la densit\'e de charge suprafluide.
\\Nous \'etudions l'occurrence de cette instabilit\'e dans deux configurations diff\'erentes : la ``limite-test'', o\`u la temperature et la vitesse du fluide normal sont fig\'ees, et une approche compl\`ete, qui les inclut. 
\bigskip
\\Dans les deux cas nous v\'erifions nos r\'esultats \`a l'aide de la dualit\'e entre th\'eorie de jauge et de gravit\'e. Formul\'ee en 1997 par Maldacena, cette conjecture postule l'\'equivalence entre une th\'eorie de gravit\'e en $d+1$ dimensions dans un espace asymptotiquement anti-de Sitter (c'est \`a dire, \`a courbature n\'egative constante) et une th\'eorie des champs invariante sous les transformations conformes (dite ``th\'eorie de bord''). En particulier, on peut montrer qu'une th\'eorie des champs fortement coupl\'ee correspond \`a une th\'eorie de (supra)gravit\'e classique, d\'ecrite par les \'equations d'Einstein. En outre, on peut d\'ecrire une th\'eorie de bord \`a temperature finie en \'etudiant une solution de trou noir des \'equations d'Einstein. La temperature et l'entropie du syst\'eme sont donn\'ees respectivement par la ``temperature de Hawking'' et ``l'entropie de Bekenstein-Hawking'' du trou noir.
\bigskip
\\D'un point de vue pratique, nous suivons l'approche introduite en 2008 par Hartnoll, Herzog et Horowitz et consid\'erons une g\'eometrie coupl\'ee \`a un champ de jauge et un champ scalaire. En dessous d'une temperature critique le champ scalaire forme un condensat en acqu\'erant une solution non nulle. Il y a donc brisure spontan\'ee d'une sym\'etrie : le gradient du champ de Goldstone qui en provient correspond \'a la velocit\'e suprafluide dans la th\'eorie de bord. Dans la ``limite-test'' nous neglig\'eons les interactions entre le tenseur de la m\'etrique et les champs de mati\`ere.
\\Nous trouvons une tr\`es bonne concordance entre les fr\'equences caract\'eristiques du syst\`eme hydrodynamique et les fr\'equences quasinormales de la th\'eorie de gravitation. Notamment, nous pouvons constater qu'une fr\'equence quasinormale va \`a z\'ero et acquiert une partie imaginaire positive quand la condition (\ref{critf}) est satisfaite. Cela nous permet donc de v\'erifier notre crit\`ere dans un mod\`ele holographique. 
\\Notamment, notre crit\'ere (formul\'e sans r\'ef\'erence \`a un mod\`ele en particulier) est valide ind\'ependamment du couplage et ne requiert pas d'invariance sous les transformations de Galil\'ee. En dernier lieu, nous montrons aussi qu'il reproduit la limite de Landau une fois appliqu\'e \`a des suprafluides galil\'eens avec statistiques de Bose-Einstein.

\newpage
\tableofcontents

\mainmatter

\chapter{Introduction}
Superfluidity is a state of matter characterized by the absence of dissipative flow and of heat transport at zero temperature \cite{leggett, annett, Schmitt:2014eka}. It arises from the spontaneous breaking of a $U(1)$ global symmetry of a scalar field, caused by an order parameter acquiring a vacuum expectation value. The associated gapless Goldstone mode $\phi$ is responsible for the onset of an associated charge current: the corresponding velocity, i.e. the \textit{superfluid velocity}, is proportional to the gradient of the Goldstone: $\mathbf{v}_\text{s}\sim \nabla \phi$. In fact, under $U(1)$ transformations the Goldstone transforms non-linearly $\phi \to \phi + \lambda$ and as a consequence only the superfluid velocity can be measured, not the phase itself.
\bigskip
\\The conventional phenomenological framework adopted in the description of such systems, the so-called Landau-Tisza model, describes the coexistence of a superfluid component with a normal fluid one. At a critical temperature $T_c$ the superfluid component vanishes and the system transitions to a normal fluid. For $^4$He, the best-known superfluid, $T_c \approx 2.2$K. In such a composite system the superfluid component relies on a Bose-Einstein condensate (BEC). The normal fluid component, on the other hand, consists of the bath of thermal excitations above the ground state forming the condensate.
\\On top of the critical temperature superfluids also feature a critical velocity $v_c$. Landau determined it in terms of the microscopic excitation spectrum of the thermal theory \cite{landaubook, khal, pines,Amado:2013aea}:
\begin{equation} \label{landaucrit}
v_c = \min \frac{\epsilon_p}{p} \, .
\end{equation}
This criterion has a simple formulation; however, in order to apply it, one needs to know the microscopic details of the system. It is therefore best suited for systems with weakly interacting degrees of freedom, where the calculation of the energy spectrum $\epsilon(p)$ can be successfully carried out with field theory methods. 
\bigskip
\\Strongly coupled theories, on the other hand, entail a different approach. Rather than directly diagonalizing the Hamiltonian in perturbation theory, one usually studies them by means of effective field theories (EFT). A classical example of an effective field theory, and the one that we are going to employ in this thesis, is hydrodynamics \cite{landaubook, chlub, lamb}. Assuming small perturbations out of thermal equilibrium, it describes the collective excitations of thermal systems in the low energy limit (or, alternatively, at times and length scales much larger than the typical relaxation scales of the underlying microscopic theory). 
\\Another common strategy for tackling strongly coupled systems is the use of dualities. A prime example thereof is the AdS-CFT correspondence \cite{nastase, natsuume, erdmenger, zaanen}, or gauge-gravity duality, which states the equivalence between a $d+1$-dimensional gravity theory in an asymptotically anti-de Sitter (AdS) geometry and a conformal field theory (CFT) living on its $d$-dimensional boundary. To wit, one can send the string coupling to zero, thus obtaining classical supergravity on the gravity side, while also keeping a large effective coupling in the boundary theory. This is the so-called `t Hooft limit: it allows us to translate the AdS-CFT correspondence into a duality between a classical gravity theory, governed by Einstein's equations of motion, and a hydrodynamic effective field theory. Hence the jargon ``fluid-gravity correspondence''.
\bigskip
\\In this thesis we make use of both these approaches in order to determine a criterion for the critical velocity of superfluids that only depends on collective quantities and not on microscopical degrees of freedom. 
\\In short, we are going to show that the system undergoes an instability when:
\begin{equation} \label{ourcrit}
\partial_\xi \left( \xi n_\text{s}\right) = 0 \, ,
\end{equation}
where $\xi$ is the norm of the spatial gradient of the Goldstone $\phi$ and $n_\text{s}$ is the superfluid charge density. We verify this condition both in hydrodynamics and using gauge-gravity duality. In the latter approach, we use a Lagrangian involving (in its most complete form) fluctuations of the background metric, of a gauge field and of a complex scalar field, which will be dual to the fluctuations of the stress-energy tensor, current and condensate respectively.
\bigskip
\\This criterion, being formulated in terms of collective variables, manifestly holds for strongly coupled system described by effective theories. It is nevertheless also valid for weakly coupled systems: we show that it predicts the same critical velocity as the Landau criterion (\ref{landaucrit}) in the case of $^4$He with roton excitations.
\\Another relevant point concerns the nature of the instability signalled by (\ref{ourcrit}). We demonstrate that this is a dynamical and a thermodynamical instability at the same time. In fact, when the criterion is fulfilled the imaginary part of the eigenmodes of the collective excitations of the system becomes positive, giving rise to perturbations which grow exponentially at late times. Concurrently, however, the diagonal susceptibility associated to the superfluid velocity diverges and changes sign, indicating the onset of a thermodynamical instability. 
\\This, together with the agreement with the Landau criterion (\ref{landaucrit}) for $^4$He, is a non-trivial check that highlights the power of hydrodynamics as an effective field theory.
\bigskip
\\The ``holographic check'' (i.e. the verification by means of gauge-gravity duality) of our criterion gives a further assessment of a correspondence that has had an appreciable success since 2008, when Herzog, Hartnoll and Horowitz published their pioneering papers \cite{HHHprobe, Hartnoll:2008kx}. In particular, here we are adding a non-zero background superfluid velocity to the picture, which translates into a finite boundary value of the spatial part of the gauge field. In spite of its rather technical character, then, such a test of the AdS-CFT correspondence is significant in and of itself.
\bigskip
\\In this thesis, after introducing the necessary tools from hydrodynamics and the gauge-gravity correspondence, we are going to show criterion (\ref{ourcrit}) in a handful of frameworks. 
\\As a first example we are going to take a look at the so-called ``probe limit'', where we only consider perturbations to the Goldstone and the chemical potential. We compute the eigenmodes of the system and derive our criterion (\ref{ourcrit}) by determining the value of $\xi$ for which the imaginary part of one mode becomes positive. On the gravity side of the AdS-CFT correspondence this amounts to neglecting the backreaction between the metric and the matter fields. In the probe limit we also present an analytical solution to the Einstein equations of motion, valid in the limit of small condensate and background superfluid velocity. We are able to explicitly show the correspondence of such a model with the hydrodynamical description of probe superfluids, and to check the consistency of our criterion. 
\\We then proceed with the hydrodynamical description of the full Landau-Tisza model, including fluctuations of the temperature and the normal fluid velocity. Correspondingly, on the gravity side we are going to include backreaction. The eigenmodes of the system are much too involved to be written down, but we are still able to check the validity of our criterion. 
\bigskip
\\Throughout this thesis we are using natural units; in particular, we set $c=\hbar=1$. Some of the results are given in \cite{shortpaper}, while a more detailed analysis is presented in \cite{longpaper}.

\chapter{Hydrodynamics: an overview}
\normalsize{}In the introduction to their book \cite{landaubook} Landau and Lifschitz  define fluid mechanics as ``the theory of fluids and gases''. Although this definition works for most practical applications, modern hydrodynamics deserves a more precise framing. 
\\In general,  complex systems such as fluids and gases can behave quite differently depending on the scales at which they are probed. In particular, typical microscopic operators relax on length- and timescales $\ell_{th}$, $\tau_{th}$.  Thus fast variables, which do no overlap with conserved operators, decay on short scales, becoming irrelevant for the effective description of the state at long length- and timescales. This implies that the system can be described by studying the operators $n_a$ overlapping with quantities protected by conservation laws. These operators form the \textit{ensemble} of the state. \footnote{Note that the choice of ensemble is not unequivocal, different ensembles can be used to describe the same state.}
\section{Constitutive relations and frame choice}
Assuming thermal equilibrium and the existence of a thermal density matrix $\rho$ one can define the expectation values $\langle n_a\rangle$ of the conserved operators $\hat{n}_a$:
\be
\langle n_a\rangle \equiv \frac{1}{Z}\, \text{Tr} \, \left[ \hat{n}_a \rho \right] \, \text{,}
\ee
where $Z\equiv \text{Tr} \rho$ and the trace is taken over a complete set of energy eigenstates of the system (for instance in the canonical ensemble $\rho = e^{-\beta H}$ and in the grand-canonical ensemble $\rho =e^{-\beta H + N\mu}$).
Then one can write the conservation laws as:
\begin{equation} \label{hydroeoms}
\partial_t \, \langle n_a(t,\mathbf{x}) \rangle + \nabla \cdot \langle\mathbf{j}_a(t,x)\rangle = 0 
\end{equation}
The currents $\mathbf{j}_a$ are fast variables; as such, their expectation values in the thermal ensemble can be locally expanded in gradients of the conserved densities: 
\begin{equation} \label{hydroexp}
\langle \mathbf{j}_a \rangle = \alpha_{ab}\, \langle n_b\rangle - D_{ab} \,\nabla\langle n_b\rangle + \mathcal{O}(\nabla^2)
\end{equation}
Such relations are called constitutive relations. They are expressed up to first order in gradients (or, in Fourier space, momenta). The zeroth order $\alpha_{ab}$ terms define an \textit{ideal} hydrodynamic system. The first-order terms $D_{ab}$, on the other hand, are the so-called \textit{transport coefficients}.  They can be hydrostatic (that is, originating from higher derivative corrections to the hydrostatic pressure and signalling deviations out of homogeneous equilibrium), or non-hydrostatic, describing slight deviations out of equilibrium. Furthermore, they may or may not contribute to entropy production (which has to stay non-negative, as dictated by the second law of thermodynamics). Those which do are called \textit{dissipative} terms. By imposing parity and time-reversal invariance \cite{Bhattacharyya:2012xi} one can show that relativistic (and Galilean) superfluids do not admit any non-dissipative transport coefficients at first order in gradients. 
\\Note that the convergence of the hydrodynamic series with contributions at arbitrarily high orders in gradients is not guaranteed a priori. For the purpose of this thesis we will assume that a truncation of the series to linear terms in gradients gives a valid low-energy description of the system at hand.
\bigskip  
\\The constitutive relations are constrained by the symmetries of the system. 
\bigskip
\\First of all, notice that there is usually some ambiguity in the constitutive relations beyond ideal order:  global thermodynamic variables such as temperature, chemical potential, or fluid velocity, are not well-defined out of equilibrium. Rather, they are to be understood as parameters occurring in the definition of the conserved densities and the related currents, which do have a microscopic definition.  This leaves room for different, equivalent definitions of first order hydrodynamics, which come under the name of \textit{frame choices}. In relativistic hydrodynamics one can for instance require \cite{landaubook} that the expectation values of the current and the stress-energy tensor be orthogonal to the fluid velocity. This frame choice is called the Landau frame \cite{Kovtun:2012rj, kovtun19}.
\bigskip
\\In addition, according to the second law of thermodynamics entropy production must be positive definite. This leads to constraints on the components of the matrix $D_{ab}$.
\bigskip
\\Finally, time reversal invariance gives rise to the so-called Onsager relations, which are essentially symmetry relations between the Green's functions (which we are going to introduce in the next sections). As such, they translate into symmetry conditions for thermodynamic derivatives.
\section{Susceptibilities and Green's functions}
Hydrodynamics describes deviations from thermodynamic equilibrium. Therefore together with writing down the constitutive relations the first step in describing a hydrodynamic system is the definition of the pressure, i.e. the first law. It will be written as 
\begin{equation} \label{firstlawsources}
\diff p = \sum_{a} n_a\, \diff s_a 
\end{equation}
$s_a$ is called the \textit{source} of $n_a$. One can express the fluctuations of the ensemble in terms of those of the sources. The matrix $\chi_{ab}$ responsible for the change of basis is the \textit{susceptibility matrix}:
\begin{equation} \label{susceptibilities}
\delta  n_a = \chi_{ab}\,\delta s_b
\end{equation}
Owing to its definition and recalling that in the grand-canonical ensemble the free energy density $f$ equals the pressure up to a sign, we see that the susceptibility matrix is the Hessian matrix of $f$. As such, a necessary condition for thermodynamic equilibrium is its positive definiteness. In particular, all of its diagonal components have to be nonnegative. 
\bigskip  
\\We can now take a look back at the equations of motion (\ref{hydroeoms}) and introduce the retarded Green's functions $G^R_{ab}(t,\mathbf{x}) \equiv G^R_{n_a,n_b}(t,\mathbf{x})$. 
Formally, the retarded functions are given by the commutator of the two operators involved \cite{hartnolllectures, chlub, altland_simons_2010, sonnersolvay}: \be \label{defwithcomm}
G_{ab}^R(t-t', \mathbf{x}-\mathbf{x}') \,\equiv i\theta(t-t')\,\langle[n_a(t,\mathbf{x}),n_b(t',\mathbf{x}')] \rangle
\ee
On perturbing the sources at $t\geq0$ by  $\delta s_i(t, \mathbf{x})$ the solution of the equations of motion can be written as: 
\begin{equation} \label{greendef}
\delta \langle n_a(t, \mathbf{x})\rangle =  \int\diff t' \diff^d \mathbf{x}' \,G^R_{ab}(t-t', \mathbf{x}-\mathbf{x}')\, \delta s_b(t', \mathbf{x}')
\end{equation}
From their definition (\ref{defwithcomm}) one can see that $G^R(t,\mathbf{k})$ vanishes at negative times. Hence, upon Fourier transforming $(t,x)\to (\omega, k)$, we see that $G^R(\omega, \mathbf{k})$ is analytic in the upper half-plane of complex $\omega$.
%
%
%
\subsection{Onsager relations}
Let $\Theta$ denote the time-reversal operator, under which any hermitian operator $n_a$ behaves as
\begin{equation}
\Theta n_a(t,\mathbf{x})\Theta^{-1} = \eta_a n_a(-t, \mathbf{x}) \quad \quad \quad \quad \text{with the time-reversal eigenvalue } \eta_a = \pm 1 
\end{equation}
The expectation value of an operator is taken over a complete set of states (labeled as $|\phi\rangle$) with the density operator $\hat{\rho}$. Imposing that the system be invariant under time-reversal one gets $[H, \Theta]=0$ and so $\Theta H \Theta^{-1}=H$. Similarly, $\Theta \hat{\rho} \Theta^{-1}=\hat{\rho}$. 
\\We can now compute the correlation function $\langle n_a(t, \mathbf{x}) n_b(0, \mathbf{x})\rangle$:
\begin{align}
\begin{split}
\langle n_a(t, \mathbf{x}) n_b(0, \mathbf{0})\rangle &\equiv \sum_\phi \langle \phi| \hat{\rho} \, n_a(t, \mathbf{x}) n_b(0, \mathbf{0})|\phi\rangle\\
&=\sum_\phi \langle \phi| \Theta^{-1} \Theta\, \hat{\rho}\, \Theta^{-1} \Theta n_a(t, \mathbf{x}) \Theta^{-1} \Theta n_b(0, \mathbf{0})\Theta^{-1} \Theta |\phi\rangle\\
&=\eta_a \eta_b \sum_\phi \langle \phi| \Theta^{-1}\, \hat{\rho}\, n_a(-t, \mathbf{x})  n_b(0, \mathbf{0})\Theta |\phi\rangle\\
&=\eta_a \eta_b \sum_\phi \langle \Theta \phi|\hat{\rho}\, n_b(0, \mathbf{0})   n_a(-t, \mathbf{x}) \,|\Theta \phi\rangle\\
\end{split}
\end{align}
where we exploited the invariance of $\hat{\rho}$ under time-reversal. The time-reversed states $| \Theta \phi\rangle$ form a complete set, so we have
\begin{equation}
\langle n_a(t, \mathbf{x}) n_b(0, \mathbf{0})\rangle = \eta_a \eta_b \, \langle n_b(0, \mathbf{0}) n_a(-t, \mathbf{x})\rangle 
\end{equation}
We can rewrite $\langle  n_b(0, \mathbf{0}) n_a(t, \mathbf{x})\rangle$ in a similar fashion.  We then get:
\begin{equation}
G^R_{ab}(t,\mathbf{x}) = -i\theta(t)\langle[ n_a|(t,\mathbf{x}),n_b(0,\mathbf{0})]\rangle = -i\theta(t)\, \eta_a \eta_b\, \langle [ n_b(t,-\mathbf{x}, n_a(0, \mathbf{0})]\rangle = \eta_a \eta_b \, G^R_{ba}(t,-\mathbf{x})
\end{equation}
\\If the constitutive relations depend on a quantity $\xi$ (such as a magnetic field or, in our case, the background superfluid velocity) that flips under time-reversal we have
\begin{equation}
\Theta H(\xi) \Theta^{-1} = H(-\xi)
\end{equation}
giving
\begin{equation}
G^R_{ab}(t,\mathbf{x},\xi) =  \eta_a \eta_b \, G^R_{ba}(t,-\mathbf{x},-\xi)
\end{equation}
and in Fourier space:
\begin{equation}
G^R_{ab}(\omega,\mathbf{k},\xi) =  \eta_a \eta_b \, G^R_{ba}(\omega,-\mathbf{k},-\xi)
\end{equation}
%
%
%
%
%
%
%
\subsection{Computation of the Green's functions}
To compute the retarded Green's functions one first defines a matrix $M_{ab}$ so that the equations of motion can be written as: \footnote{$M_{ab}$ should not depend on time. In order to be able to write the equations of motion as in (\ref{eqwithM}), one should choose a frame where non-ideal contributions to the constitutive relations do not carry time derivatives. In the rest of this chapter we are going to assume that this is the case.}
\begin{equation}\label{eqwithM}
\partial_t \, \langle n_a(t,\mathbf{x})\rangle + M_{ab}\,\delta \langle  n_b(t,\mathbf{x}) \rangle = 0
\end{equation}
In Fourier space:
\begin{equation}\label{defGab}
G^R_{ab}(\omega, \mathbf{k}) = (\mathbf{1} + i\omega K^{-1})_{ac}\, \chi_{cb} \, \text{,}
\end{equation}
with 
\be \label{Kdef}
K \equiv -i\omega\, \mathbf{1} + M. 
\ee
\\Note that $\lim_{\omega, \mathbf{k}\to 0}G^R_{ab}=\chi_{ab}$: hence the Onsager relations for the susceptibilities 
\be
\chi_{ab}(\xi) = \eta_a \eta_b \, \chi_{ba}(-\xi)
\ee
\\To show (\ref{defGab}) we follow Kadanoff and Martin's approach \cite{KadanoffandMartin, Kovtun:2012rj}:
\begin{itemize}
\item First, we Fourier transform (\ref{eqwithM}). $\langle \delta n_b(t,\mathbf{x}) \rangle$ is only defined for $t>0$, so we have to employ a Laplace transformation in time. In frequency-momentum space we obtain
\be
  \int_0^\infty \diff t\, e^{i\omega t}  \partial_t \delta\langle  n_a(t,\mathbf{k})\rangle = -\delta \langle n_a(t=0,\mathbf{k})\rangle -i\omega \, \delta \langle n_a(\omega,\mathbf{k})\rangle 
\ee
And so (\ref{eqwithM}) becomes:
\be
-i\omega \,\delta\na + M_{ab}\, \delta\langle n_b(\omega,\mathbf{k})\rangle =\delta\langle n_a(t=0,\mathbf{k})\rangle 
\ee
Recalling the definition of $K$ and $\chi$ we can invert this relation and find:
\be \label{naevo}
\delta\na = (K^{-1})_{ab}\chi_{bc}\,\delta  s_b(t=0,\mathbf{k})
\ee
\item We want to compare this expression with its equivalent formulation in terms of the retarded Green's functions $G^R_{ab}$ in frequency-momentum space. 
\\Suppose that at $t=0$ we turn off the sources $ \delta \langle s_{b}(t, \mathbf{k}) \rangle$.\footnote{We could as well turn them on at $t=0$, and get the same final result. This particular choice comes from the fact that  $G^R(\omega, \mathbf{k})$ is analytic in the upper half-plane of complex $\omega$; otherwise we would have to define an analytical continuation in the lower-half plane. } Then, by definition of the retarded Green's functions, we obtain:
\be \label{deltana1}
\delta \langle n_a(t, \mathbf{k}) \rangle = -\int_{-\infty}^{0}\diff t' \, e^{\epsilon t'} G^R_{ab}(t-t',\mathbf{k}) \,\delta  s_b(t'=0,\mathbf{k}) \quad \text{,}
\ee
where we imposed that the time dependence of the source is given by an adiabatic damping $e^{\epsilon t'}$. This choice ensures convergence of the Fourier integrals.
\item The Fourier transformation of the retarded function satisfies:
\be
G^R_{ab}(t-t', \mathbf{k}) = \frac{1}{2\pi} \int_{-\infty}^{\infty} \diff \omega' G^R_{ab}(\omega', \mathbf{k})\, e^{-i\omega'(t-t')}
\ee
Together with (\ref{deltana1}), after integrating over $\diff t'$, this yields:
\be
\delta \langle n_a(t, \mathbf{k}) \rangle = -\frac1{2\pi}\langle s_b(t'=0,\mathbf{k})\rangle \int_{-\infty}^{\infty} \diff \omega' G^{R}_{ab}(\omega', \mathbf{k}) \, \frac{e^{-i\omega' t}}{i\omega' + \epsilon}
\ee
\item By taking another Laplace transformation we get (again adding a factor $e^{-\tilde{\epsilon}t}$ to ensure convergence):
\be
\begin{split}
\delta \langle n_a(\omega, \mathbf{k}) \rangle &= -\frac1{2\pi}\langle s_b(t'=0,\mathbf{k})\rangle \int_{-\infty}^{\infty} \diff \omega' \int_{0}^{\infty} \diff t \, G^{R}_{ab}(\omega', \mathbf{k}) \, \frac{e^{it(\omega-\omega' +i\tilde{\epsilon})}}{i\omega' + \epsilon} \\
&= -\frac1{2\pi}\langle s_b(t'=0,\mathbf{k})\rangle \int_{-\infty}^{\infty} \diff \omega' G^{R}_{ab}(\omega', \mathbf{k}) \, \frac{1}{(\omega' -\epsilon)(\omega' - \omega - i \tilde{\epsilon})} 
\end{split}
\ee
We evaluate this integral by means of Cauchy's theorem in the upper half-plane, where $G^{R}_{ab}(\omega, \mathbf{k})$ is analytic. In the $\epsilon, \tilde{\epsilon} \to 0$ limit  we finally find:
\be \label{relforG}
\delta \langle n_a(\omega, \mathbf{k}) \rangle =  -\frac1{i\omega} s_b(t'=0,\mathbf{k}) \left(G^{R}_{ab}(\omega=0, \mathbf{k})-G^{R}_{ab}(\omega, \mathbf{k})  \right)
\ee
\item By comparing (\ref{relforG}) with (\ref{naevo}) we finally obtain (\ref{defGab}).
\end{itemize}
\bigskip  
From (\ref{greendef}) and (\ref{relforG}) one can see that the poles of the Green's functions determine the eigenfrequencies of the system. In momentum space the poles $\omega(\mathbf{k})$ of $G^R_{ab}(\omega,\mathbf{k})$ are called the \textit{modes} of the system. They are usually written in powers of $k \equiv |\mathbf{k}|$:
\begin{equation} \label{generalmode}
\omega(\mathbf{k}) = v\, k +\frac{i\Gamma}{2}\, k^2 + \mathcal{O}(k^3)
\end{equation}
where $v$ is called the speed of sound and $\Gamma$ the attenuation constant. In ideal hydrodynamics $\omega(\mathbf{k})$ is real for real $\mathbf{k}$, giving $\Gamma = 0$.
\\Notably, the modes $\omega_j$ ($1\leq j \leq \text{dimension of the ensemble}$) of the system determine the propagation of (small) initial perturbations out of equilibrium, which is given by the wave functions $e^{-i \omega_j t +ikx }$. One sees that whenever $\text{Im}(\omega)$ crosses the real axis and becomes positive a dynamical instability is produced, signalled by the exponential growth in time of the corresponding perturbation \cite{hartnolllectures}.
\\Note: in complex Fourier space the collision of the hydrodynamic modes with the first non-gapped non-hydrodynamic mode, as a function of complex wavevector, determines the radius of convergence of the hydrodynamic series \cite{withers18, groz1, groz2}. In fact, one can write the propagators as a Laurent series around the modes, which are their poles. It is a known property of Laurent series that their radius of convergence is set by the next pole. If the latter does not correspond to a hydrodynamic mode, then hydrodynamics will not be able to describe the physics beyond it. 
%
%
%
%
\bigskip  
\\A final note: practically, one can compute the retarded Green's functions by means of the so-called \textit{variational method} \cite{Kovtun:2012rj}. 
\\To do so one couples the currents to external fields $s_\mathcal{O}$, so that the variation of the grand potential \footnote{We work in the grandcanonical ensemble, where the system is assumed to be in thermodynamic equilibrium with a reservoir with which it can exchange heat and particles. The grand potential $\Omega = -\ln Z$ satisfies $\Omega = U - TS - \mu N$, $U$ being the internal energy.} $\Omega\equiv-\ln Z$ ($Z$ being the static partition function) receives contributions of the form: 
\be
\Omega \to \Omega + \int \diff^d x\sqrt{|g|}\,  \langle \mathcal{O}\rangle \phi_\mathcal{O}
\ee
Let us take relativistic superfluid hydrodynamics as an example. As we are going to see, it includes a charge current $j^\mu$ coupling to a gauge field $A_\mu$, a stress-energy tensor $T^{\mu \nu}$ coupling to a perturbation to the metric $g_{\mu \nu}$ and Goldstone field coupling to a scalar source $s_\phi$. 
\\The external fields will appear on the right-hand side of the equations of motion. For instance, in our case: 
\be \label{eomsvar}
\dmu j^\mu = s_\phi  \qquad \qquad \nabla_\mu T^{\mu \nu} = j_\rho F^{\rho \sigma} + s_\phi \left(\partial^\nu \phi - A^\nu\right)
\ee
with the antisymmetric field strength tensor $F_{\mu \nu} \equiv \dmu A_\nu - \dnu A_\mu$. These relations can be derived by imposing gauge and diffeomorphism invariance:
\begin{itemize}
\item Under a gauge transformation the Goldstone field and the gauge field are shifted by $\phi \to \phi + \lambda$ and $A_\mu \to A_\mu + \dmu \lambda$. Then the action varies by:
\be
S \to S + \int\diff^{d}x  \sqrt{-g} \, s_\phi \lambda + \int\diff^{d}x  \sqrt{-g} \, j^\mu \partial_\mu \lambda \sim \int\diff^{d}x  \sqrt{-g} \,  \lambda   (s_\phi - \dmu j^\mu) 
\ee
Imposing invariance we get $\dmu j^\mu = s_\phi$.
\item Similarly, under a diffeomorphism parametrized by a vector $\theta_\mu$ the fields are shifted by their Lie derivative with respect to $\theta_\mu$: $\phi \to \phi + \theta^\mu \dmu \phi$, $A_\mu \to A_\mu + \theta^\rho \partial_\rho A_\mu + A_\rho \dmu \theta^\rho$ and $g_{\mu \nu} \to g_{\mu \nu} + \nabla_\mu \theta_\nu + \nabla_\nu \theta_\mu$. Then upon integration per parts the action varies by:
\be
S \to S + \int\diff^{d}x  \sqrt{-g} \, \theta_\mu \left(s_\phi \partial^\mu \phi + j_\nu F^{\mu\nu} -  \nabla_\nu T^{\mu \nu}  \right)
\ee
After restoring gauge invariance by shifting $\partial^\nu  \phi \to  \partial^\nu  \phi  -A^\nu$ (the $s_\phi A^\nu$ term is quadratic in the external fields, and thus does not contribute at linear order in perturbation theory) we then obtain the second equation in (\ref{eomsvar}).
\end{itemize}
\bigskip
One then solves the equations of motion and computes the expectation values of the currents on shell, whose variation with respect to the external fields yields the Green's functions. 
\bigskip
\\Starting from the expectation values in thermal equilibrium we define:
\begin{equation}\label{varmetdef}
\mathcal{J}^\mu \equiv \sqrt{|g|}\langle j^\mu \rangle \quad  \quad \quad \mathcal{T}^{\mu \nu} \equiv \sqrt{|g|}\langle T^{\mu \nu} \rangle  \quad  \quad \quad \Phi \equiv \sqrt{|g|}\langle \phi \rangle
\end{equation}
The factor of $\sqrt{-g}$ arises due to the functional form of the action in theories with non-trivial metric: $S=\int \diff^d x\, \sqrt{-g}\mathcal{L}$.
\\The retarded functions can then be computed as follows:
\be\label{defvar}
G^R_{T^{\mu \nu}T^{\rho \sigma}} = 2\, \pdert{\mathcal{T}^{\mu \nu}}{h_{\rho \sigma}}{\substack{h_{\alpha \beta \neq \rho \sigma}=\\A_\alpha=s_\phi=0}}   \quad \, \, G^R_{T^{\mu \nu}J^{ \sigma}}=  \pdert{\mathcal{T}^{\mu \nu}}{{A_{ \sigma}}}{\substack{h_{\alpha \beta}=A_{\alpha\neq \nu}\\=s_\phi=0}} \quad \, \, G^R_{T^{\mu \nu}\phi} =\pdert{\mathcal{T}^{\mu \nu}}{s_\phi}{h_{\alpha \beta}=A_\alpha=0} 
\ee
The definitions of $G^R_{j^\mu \star}$ and $G^R_{\phi \star}$ are analogous. 
\\The retarded Green's functions computed with this procedure are found to agree with (\ref{defGab}) up to so-called contact terms, which are independent of $\omega$. 
\section{Instabilities: dynamical and thermodynamic}
As we anticipated above, a necessary condition for the system to be in a stable state is that the imaginary part of the modes be negative. This can be seen by considering the evolution of the vacuum expectation value of a quantity $n_a$ in the ensemble, after turning on a source $s_b$ at $t=0$. 
\bigskip
\\Recall equation (\ref{greendef}):
\begin{equation}
\begin{split}
\delta \langle n_a(t, \mathbf{x})\rangle &=  \int\diff t' \diff^d \mathbf{x}' \,G^R_{ab}(t-t', \mathbf{x}-\mathbf{x}')\, \delta s_b(t', \mathbf{x}') \\
&= -i \int\diff t' \diff^d \mathbf{x}' \,\theta(t-t')\rho_{ab}(t-t', \mathbf{x}-\mathbf{x}')\, \delta s_b(t', \mathbf{x}')
\end{split}
\end{equation}
where in the second line we have employed the spectral function $\rho_{ab}$ \cite{Kovtun:2012rj}: 
\begin{equation*}
G_{ab}^R(t-t',\mathbf{x}-\mathbf{x'}) \equiv -i \theta(t-t')\rho_{ab}(t-t',\mathbf{x}-\mathbf{x'})
\end{equation*}
The above integral is the convolution of $G_{ab}^R$ with $\delta s_b$, so upon Fourier transforming this becomes:
\be
\delta \langle n_a(t, \mathbf{x})\rangle = -i \int\diff t' \diff^d \mathbf{x}' \,\int\diff \omega \diff^d \mathbf{k} \,e^{-i\omega(t-t')+i\mathbf{k}(\mathbf{x}-\mathbf{x'})}\theta(t-t')\rho_{ab}(\omega, \mathbf{k})\, \delta s_b(\omega, \mathbf{k})
\ee
We can evaluate the integral over $\diff \omega$ by means of Cauchy's integral theorem. Let us choose a contour in the upper half plane consisting of the real axis, together with the half circle $\omega = Re^{i\theta}$, with $0\leq\theta\leq \pi$ and $R\to \infty$. Assuming $\rho_{ab}(\omega, \mathbf{k})\, \delta s_b(\omega, \mathbf{k})$ to fall off sufficiently quickly at $|\omega| \to \infty$, the integral over the whole contour reduces to the integral over the real axis, which is the one that we want to compute. With Cauchy's theorem we find then:
\be
\delta \langle n_a(t, \mathbf{x})\rangle = -\frac{1}{2\pi} \int\diff t' \diff^d \mathbf{x}' \,\int \diff^d \mathbf{k} \,e^{i\mathbf{k}(\mathbf{x}-\mathbf{x'})}\theta(t-t') \sum_{\substack{\omega_i \text{ pole of }  \rho_{ab}\\ \text{Im}(\omega_i)>0}} e^{-i\omega_i(t-t')}\delta s_{b}(\omega_i) \text{Res}\,\rho_{ab}(\omega_i)
\ee
The Heaviside function $\theta(t-t')$ implies that we have to check whether $\delta \langle n_a(t, \mathbf{x})\rangle $ is well-behaved for $t\to +\infty$. But in this limit whenever $\text{Im}(\omega)>0$ we get an exponential growth in time. Therefore if we assume the perturbation $\delta \langle n_a(t, \mathbf{x})\rangle$ to be well-behaved at large times we see that the retarded propagator $G_{ab}^R$ shall not have any poles in the upper half plane.
\bigskip
\\Whenever the imaginary part of a hydrodynamic mode crosses over to the upper half plane one obtains a \textit{dynamic instability}.
\bigskip
\\In thermodynamics, on the other hand, instabilities are signalled by the free energy not being at a minimum anymore. 
\\This condition can be detected by looking at the susceptibility matrix (\ref{susceptibilities}). Since it is the Hessian matrix of the free energy, it has to be positive definite for the system to sit in a (local) minimum of the free energy. This also implies that all its diagonal entries must be positive. 
\bigskip
\\In this thesis we are going to show that at a critical superfluid velocity $\xi_c$ one encounters a dynamical instability signalled by the imaginary part of a hydrodynamic mode crossing over to the upper half-plane. Together with the imaginary part, also the real part of the mode is found to undergo a sign change.
Furthermore, we demonstrate that this dynamic instability is also a thermodynamic instability by showing that the susceptibility associated to the superfluid velocity also changes sign. 
\\Finally we show that our criterion reduces to the well-known Landau criterion for superfluids with roton excitations.

\section{Superfluid hydrodynamics}
Superfluidity is a phase of matter whose main feature is frictionless charge transport (vanishing shear viscosity). 
It was first observed in $^4$He at temperatures below $T_c \approx 2.2$K \cite{landaubook, kapitza, allen} and more recently in ultracold atomic gases \cite{pethsmith}. It is also expected to take place inside neutron stars \cite{pagereddy}, in quark plasma above $10^{11}$K and in metals at low temperature \cite{leggett}. 
\bigskip
\\A relativistic superfluid is a charged fluid with a spontaneously broken global symmetry (in its simplest form a $U(1)$ symmetry) and a related Goldstone boson. The superfluid velocity is proportional to the gradient of the Goldstone:
\be
 {v_\text{s}}_\nu \equiv  \frac{ \partial_\nu \phi}{\mu}
 \ee
 This makes its spacelike component irrotational: 
 $\nabla \times \mu \mathbf{v}_\text{s} = 0$,
  hence the absence of unbound vortices in the system \footnote{The formation of vortices after applying external angular momentum is a quantum effect, stemming from the fact that the condensate vanishes in the core of the vortex. This effectively change the topology of the fluid - see \cite{simon}, section 2.3.1 for more details. The vortices counterbalance the externally applied angular momentum. }. 
\bigskip
\\\begin{minipage}{0.58\textwidth}
The conventional phenomenological framework adopted in the description of such systems, the so-called Landau-Tisza model \cite{landau41,Tisza}, considers the coexistence of a superfluid component with a normal fluid one. At a critical temperature $T_c$ the superfluid component vanishes and the system transitions to a normal fluid. For $^4$He, the best-known superfluid, $T_c \approx 2.2$K. On the other hand, at $T=0$ the charge fraction carried by the normal component vanishes and one is left with a pure superfluid.
\\In such a composite system the charge density associated with the superfluid component is given by a Bose-Einstein condensate (BEC). The normal fluid, on the other hand, consists of the bath of thermal excitations above the ground state forming the condensate. 
\end{minipage}
\hspace*{0.6cm}
\begin{minipage}{0.36\textwidth}
\hspace*{-0.4cm}
\includegraphics[scale=0.25]{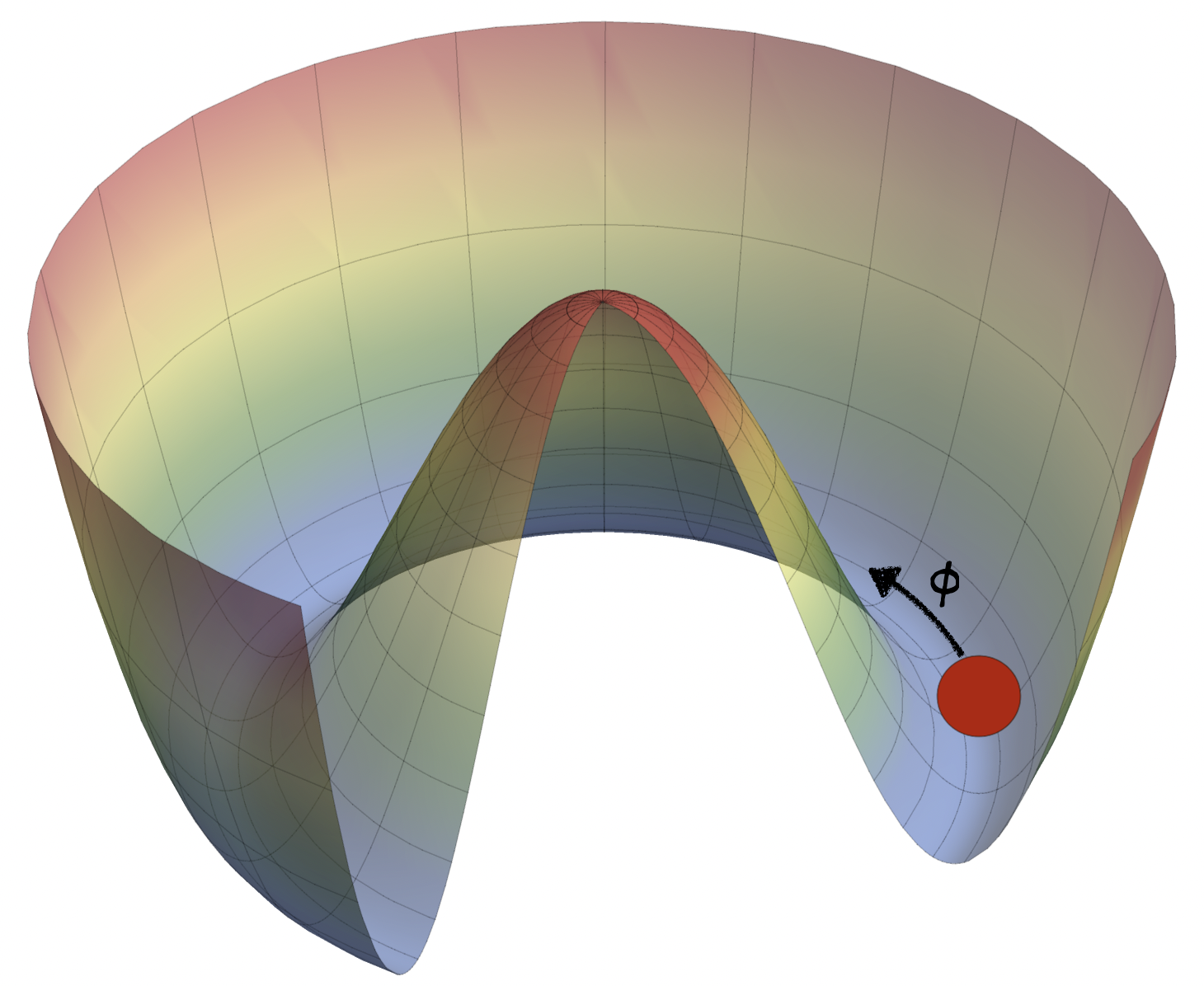}
\\Graphical representation of the $U(1)$ spontaneous symmetry breaking mechanism.
\end{minipage}

\subsection{The Landau criterion}
Superfluids become unstable when the relative velocity between the normal fluid and the superfluid component exceeds a critical velocity $v_c$. A first criterion to compute this bound was first derived by Landau and is called the \textit{Landau criterion} \cite{Schmitt:2014eka, simon} .
\bigskip
\\Let us assume that the superfluid moves through a capillary with velocity $v_\text{s}$;  $\epsilon_p$ is its excitation spectrum in its rest frame. Imposing Galilean boost invariance, its spectrum in the capillary's reference frame is given by $\epsilon_p'=\epsilon_p + \mathbf{p}\cdot \mathbf{v}_\text{s}$. At zero temperature, in order for it not to be energetically convenient to thermally excite the condensate, the excitation spectrum in the capillary's rest frame must be nonnegative. Its minimum is attained when $\mathbf{p}$ and $\mathbf{v}_s$ are antiparallel \cite{Schmitt:2014eka}:
\begin{equation}
\min \epsilon_p' =\epsilon_p -p\,v_\text{s}
\end{equation}
This quantity has to be nonnegative:
\begin{equation} \label{landaucrit}
v_\text{s} \leq v_c \equiv\min \frac{\epsilon_p}{p}
\end{equation}
This criterion, albeit generally effective in foreseeing the critical velocity of common superfluids such as Helium 4, has some limitations. First of all, it assumes Galilei invariance, which is necessary to compute the excitation spectrum in the capillary's rest frame. Also, the derivation given above is only valid at low temperatures, and does not take dissipation into account. From a practical point of view, moreover, in order to actually compute the bound one needs to know the microscopic theory responsible for the excitation spectrum. 
\bigskip
\\ \textbf{Hence the main purpose of this thesis: to formulate a criterion for the critical velocity of superfluids which does not rely on the microscopic details of the system or invariance under boosts.}
\bigskip

\subsection{Relativistic superfluid hydrodynamics}
The hydrodynamics of a relativistic superfluid is built upon two background scalar quantities, the chemical potential $\mu$ and the temperature $T$, together with the Goldstone boson $\phi$ and $u^\mu$, the normal fluid velocity. It is normed: $u^\mu u_\mu = -1$.
\bigskip
\\The superfluid velocity is proportional to the gradient of the Goldstone, denoted as $\xi^\mu$:
\begin{equation}
{v_\text{s}}_\nu = \frac{\xi_\nu}{\mu} \equiv \frac{\partial_\nu \phi}{\mu}
\end{equation}
The Josephson relation links the gradient of the Goldstone boson and the chemical potential: \footnote{Following \cite{Delacretaz:2019brr}, one can express the Josephson relation as the constitutive relation for a higher-form current, putting all hydrodynamic equations on equal footing. In this thesis, however, we are going to stick to the conventional formulation of superfluid hydrodynamics.}
\begin{equation}\label{jos}
u^\mu \xi_\mu + \mu = 0
\end{equation}
Slightly moving out of equilibrium one can modify this relation by adding dissipative contributions packaged in $\mud$:
\begin{equation}\label{josdiss}
u^\mu \xi_\mu + \mu + \mu_\text{diss} = 0
\end{equation}
We define the projector onto the subspace perpendicular to the normal fluid velocity $u^\mu$:
\begin{equation}
\Delta^{\mu \nu} \equiv \eta^{\mu \nu} + u^\mu u^\nu
\end{equation}
This allows us to write down the component of the superfluid velocity orthogonal to the normal fluid velocity:
\begin{equation} \label{zeta}
\zeta^\mu \equiv \Delta^\mu_\nu \xi^\nu 
\end{equation}
We can now write down the constitutive relations up to linear order in gradients (labeled by the $_\text{diss}$ subscript). We will justify the form of the constitutive relations at ideal order in the box at the end of this section.
\\The constitutive relation for the current can then be written as:
 \begin{equation} \label{jmu}
j^\mu = n u^\mu + \frac{n_\text{s}}{\mu}\zeta^\mu  + j_\text{diss}^\mu
\end{equation}
Note that at ideal order, given the Josephson relation (\ref{jos}), this is equivalent to writing 
\begin{equation} 
j^\mu = n_n u^\mu + n_\text{s}\frac{\xi^\mu}{\mu} + \mathcal{O(\partial)}
\end{equation}
with the normal charge density $n_n \equiv n - n_\text{s}$.
\bigskip
\\The stress-energy tensor takes the following form: 
\begin{equation}\label{Tmunu}
T^{\mu\nu} = (\epsilon + p)\, u^\mu u^\nu + p\,\eta^{\mu \nu} + n_\text{s} \left( \zeta^\mu u^\nu + \zeta^\nu u^\mu + \frac{1}{\mu}\zeta^\mu \zeta^\nu\right)+\pi_\text{diss}^{\mu \nu}
\end{equation}
For example, with zero background superfluid velocity (that is, when the spatial part of $\xi^\mu$ appears at first order in fluctuations) we consider the following dissipative terms: 
\be \label{dissnobkg}
j_\text{diss}^\mu = -\frac{\kappa}{T}\,\Delta^{\mu \nu}\dnu \frac{\mu}{T} \qquad \quad \mu_{\text{diss}} = -\zeta_3 \, \dmu \left( n_\text{s} \zeta^\mu\right) \qquad \quad \pi_\text{diss}^{\mu \nu} = -\eta \, \sigma^{\mu \nu} \, ,
\ee
where $\sigma_{\mu \nu}\equiv \Delta^\alpha_\mu \Delta^\beta_\nu \left(\partial_{(\alpha} u_{\beta)} -\eta_{\alpha \beta}\frac{\partial_\rho u^\rho}{d} \right)$is the traceless symmetric part of $\partial_\mu u_\nu$, projected perpendicular to $u_\mu$ \cite{Bhattacharya:2011eea} \, \footnote{We define $A_{(\mu}B_{\nu)} \equiv( A_\mu B_\nu + A_\nu B_\mu)/2$} . $\eta$ is the shear viscosity, $\kappa$ the thermal diffusivity and $\zeta_3$ the superfluid diffusivity \cite{Arean:2021tks}. 
\bigskip
\\The Smarr relation relates energy density, pressure, charge density and entropy density:
\be \label{smarr1}
\epsilon + p = n\mu + sT
\ee
Following \cite{callen}, it can be proven by recalling that the entropy $S$, the internal energy $E$, the volume $V$ and the charge $N$ are extensive quantities, meaning that $E(\lambda z) = \lambda E(z)$ (and similarly for $S,V$, and $N$) for a scaling parameter $\lambda$ which rescales the ``size'' $z$ of the system. But then the energy is given by $E(z)= \frac{\partial E(z, \lambda)}{\partial \lambda}\big|_{\lambda=1}$, and therefore $E = \pdertsmall{E}{\lambda}{\lambda=1} =  (\pdertsmall{E}{S}{V, N}S +  \pdertsmall{E}{V}{S, N}V+  \pdertsmall{E}{N}{V, S}N)\big|_{\lambda=1}$ (since just as we do for $E$, we have $S(z)= \frac{\partial S(z, \lambda)}{\partial \lambda}\big|_{\lambda=1}$ etc.).
\\The first law of thermodynamics states $\diff E \supset T\diff S +  \mu \diff N - p\diff V$: hence we find $E = TS - pV + \mu N$. Dividing both sides of this equation by $V$ we finally get (\ref{smarr1}) since the densities are defined as $\epsilon \equiv E/V$, $s\equiv S/V$ and $\rho \equiv N/V$.
\bigskip
\\We impose conservation of the current and the stress-energy tensor:
\begin{align}
\begin{split}
\partial_\mu j^\mu &= 0 \\
\partial_\mu T^{\mu \nu} &= 0
\end{split}
\end{align}
\begin{mdframed}
\textbf{Ideal superfluid hydrodynamics: fixing the constitutive relations and the Josephson equation}
\medskip
\\In this box we motivate the constitutive relation (\ref{Tmunu}) and derive the Josephson equation (\ref{jos}) by imposing that entropy production vanish at ideal level (following \cite{Herzog:2011ec}).
\bigskip
\\First, let us give the most general parametrization for the stress-energy tensor. For a normal fluid it is just given by $T^{\mu \nu} =  (\epsilon + p)\, u^\mu u^\nu + p\,\eta^{\mu \nu} $.
Adding the superfluid component to the picture we can build two symmetric tensors, $\zeta^\mu u^\nu + \zeta^\nu u^\mu$ and $\zeta^\mu \zeta^\nu$ and one more scalar, $\partial_\mu \phi \partial^\mu \phi$ or, equivalently, $\zeta^2$ (from the definition of $\zeta^\mu$ it follows that $\zeta^\mu u_\mu=0$). We can then write: 
\begin{equation}
T_{\mu \nu} =(\epsilon + p)\, u^\mu u^\nu + p\,\eta^{\mu \nu}  + \alpha_1 \zeta^2 \, \Delta_{\mu \nu} + 2\alpha_2 \, \zeta_{(\mu}u_\nu) + \alpha_3  \, \zeta_\mu \zeta_\nu
\end{equation}
for some coefficients $\alpha_i(\mu, T, \zeta^2)$. 
\\We define \cite{Herzog:2011ec}:
\be
\mu_\text{s}\equiv - u^\mu \xi_\mu
\ee
Then (\ref{zeta}) becomes $\zeta^\mu = \xi^\mu - \mus u^\mu$.
\\The constitutive relation for the current is simpler - we assume the normal and superfluid components to carry an amount of charge given by their charge densities:
\begin{equation}
j^\mu = n_n u^\mu + n_\text{s} \frac{\partial^\mu \phi}{\mu_s}=n u^\mu + n_\text{s} \zeta^\mu \,
\end{equation}
with the total charge $n\equiv n_n + n_\text{s}$.
Finally, we parametrize the first law as:
\begin{equation}\label{dPsuperfluid}
\diff p = s \diff T+ n\diff \mu + \frac{X}{2} \diff (\partial_\mu \phi \partial ^\mu \phi+ \mu^2) \,
\end{equation}
where $X = X(\mu, T, \zeta^2)$.
\bigskip
\\We now make a crucial assumption: entropy is only carried by the normal component. In fact, the charge carriers composing the superfluid sit in the Bose-Einstein condensate, thereby only occupying the ground state: at sufficiently small temperature there will be no disorder in the superfluid component (the total number of possible microstates being 1) and thus zero entropy.  With the entropy density $s$ the entropy current will then be proportional to the normal fluid's four-velocity $u^\mu$:
 \begin{equation}
 j_s^\mu = su^\mu
 \end{equation}
We make use of the equations of motion in order to determine the coefficients $\alpha_i$ and $X$. Consider the conservation of the current, multiplied by $\mu$:
 \begin{equation}\label{mudmuJ}
 \mu \partial_\mu j^\mu =0
 \end{equation}
With the Smarr relation $\epsilon + p = n \mu + sT$ it becomes:
 \begin{align} \label{ugo}
 \begin{split}
 \mu \partial_\mu j^\mu &= \mu \partial_\mu \big( n_\text{s} \zeta^\mu \big) + \partial_\mu \big( (\epsilon + p) u^\mu \big) - T\partial_\mu (su^\mu)-s\partial_\mu (Tu^\mu)-n u^\mu \partial_\mu \mu
 \\&=\mu_\text{s}\mu \partial_\nu \frac{n_\text{s}}{\mu_\text{s}} + \mu \frac{n_\text{s}}{\mu_\text{s}}\partial_\mu \big( \mu_\text{s}\zeta^\mu \big)+ \partial_\mu \big( (\epsilon + p) u^\mu \big) - T\partial_\mu (su^\mu)-s\partial_\mu (Tu^\mu)-n u^\mu \partial_\mu \mu
 \end{split}
 \end{align}
 Now, we assume the entropy current to be conserved at ideal order, so $\partial_\mu(su^\mu)=0$. Extracting $T\partial_\mu(su^\mu)$ from (\ref{ugo}) this allows us to write:
  \begin{equation}\label{dmuJnoent}
T\partial_\mu(su^\mu)= \mu_\text{s}u^\nu \partial_\nu \frac{n_\text{s}}{\mu_\text{s}} + \mu \frac{n_\text{s}}{\mu_\text{s}}\partial_\mu \big( \mu_\text{s}\zeta^\mu \big)+ \partial_\mu \big( (\epsilon + p) u^\mu \big) -s\partial_\mu (Tu^\mu)-n u^\mu \partial_\mu \mu = 0
 \end{equation}
 Note that this equation is not automatically equivalent with the equation of motion $\mu \partial_\mu J^\mu=0$ because of the additional assumption that $T\partial_\mu (su^\mu)=0$. Still, it could be a trivial equation if the other equation of motion, $\partial_\nu T^{\mu \nu}=0$, implied it. To investigate this, we contract the derivative of the stress-energy tensor with $u_\mu$. We find:
 \begin{align}
\begin{split}
u_\mu \partial_\nu T^{\mu \nu} = &u^\nu \partial_\nu p - \partial_\mu \big( (\epsilon + p)u^\mu\big) \\
&- \alpha_1\zeta^2\partial_\mu u^\mu - \frac{\alpha_2}{\mus} \partial_\nu(\mu_\text{s}n^\nu)-\mu_\text{s} n^\nu \partial_\nu \frac{\alpha_2}{\mus} \\
&+ \frac{\alpha_3}{\mus^2} \left(\frac{1}{2}u_\nu \partial^\nu(\mu_\text{s}^2 \zeta^2) - \mu_\text{s} u_\mu u_\nu \partial^\nu \big(\mu_\text{s}\zeta^\mu\big) + \mu_\text{s}\zeta^\mu \partial_\nu \mu_\text{s}\right)
\end{split}
\end{align}
where we used that $u^\mu u_\mu = -1$ and so $u^\mu \partial_\nu u_\mu=0$. 
\\Now, imposing $u_\mu \partial_\nu T^{\mu \nu} =0$ we can substitute for $ \partial_\mu \left( (\epsilon + p)u^\mu\right)$ in (\ref{dmuJnoent}). Furthermore, from the first law (\ref{dPsuperfluid}) we can compute the derivatives of $p$. We finally get:
 \begin{align}
 \begin{split}
T\partial_\mu(su^\mu) \,= 0 =\,& \frac{\mus^2 X+\alpha_3}{2\mus^2}\,u_\nu \partial^\nu(\mu_\text{s}^2 n^2)\\
 & + \frac{\alpha_2-\alpha_3}{2\mus}\,u_\mu u_\nu \partial^\nu (\mu_\text{s} n^\mu) \\
 &+ \mu_\text{s}n^\nu\bigg(-\partial_\nu \frac{\alpha_2}{\mus} - \alpha_3\,\partial_\nu \frac{1}{\mu_\text{s}} + \mu\partial_\nu \frac{\rho_\text{s}}{\mu_\text{s}}\bigg) \\
 &  + \frac{\mu n_\text{s}-\alpha_2}{\mu_\text{s}}  \partial_\nu(\mu_\text{s} n^\nu)\\
 & - \alpha_1 \, \zeta^2\partial_\rho u^\rho\\
 &+ \frac{X}{2}u_\nu \partial^\nu(\mu^2 - \mu_\text{s}^2)
 \end{split}
 \end{align}
The important point to stress here is that in order to obtain this equation we already employed the equations of motion: therefore this equation, which describes the conservation of the entropy and is an additional requirement that we impose, constrains $\alpha_i$ and $X$.
\\We see that in order to let the third and the last line vanish we need $\mu = \mu_\text{s}$. This is the Josephson relation (\ref{jos}).
\\We also find $\alpha_1 =0$ and $\alpha_3 = \alpha_2 = \mu n_\text{s}$ and so $X = -\frac{\alpha_3}{\mu} = -\frac{n_\text{s}}{\mu}$. We thus recover the constitutive relation (\ref{Tmunu}) for the stress-energy tensor.
\bigskip

\bigskip

\end{mdframed}
\subsubsection{Modified phase frame and entropy production}
We work in the so-called modified phase frame. Assuming a dissipative contribution to the chemical potential as in (\ref{josdiss}) we define:
\begin{align}
\begin{split}
\xi_0^\mu &\equiv \xi^\mu - \mu_{\text{diss}}u^\mu \\
\xio &\equiv \sqrt{-\xio_\mu \xio^\mu}
\end{split}
\end{align}
The Josephson relation gives $u^\mu \xi_{0,\mu} = -\mu$. Furthermore, using $\xio_\mu$ instead of $\xi_\mu$ the definition of $\zeta^\mu$ is unchanged: $\zeta^\mu \equiv \Delta^{\mu \nu}\xi_\nu = \Delta^{\mu \nu}\xio_\nu$.
\\Notice that in the modified phase frame we can rewrite the constitutive relations (\ref{jmu}) and (\ref{Tmunu}) as (with $f\equiv n_\text{s}\mu$):
\begin{align} \label{current}
\begin{split}
j^\mu &= \nn u^\mu + f\xio^\mu + j_\text{diss}^\mu \\
T^{\mu \nu}&= (\rhon + p )u^\mu u^\nu + p\, \eta^{\mu \nu} + f \xio^\mu \xio^\nu + \pi^{\mu \nu}_\text{diss}
\end{split}
\end{align}
Note that, due to the change of frame, the dissipative corrections are in general going to be different from those defined previously.
\\In order to write down dissipative coefficients that ensure positivity of entropy production one first has to compute the entropy current. Starting from the first law (\ref{diffp}) one can show that in the modified phase frame its divergence is given by:
\begin{equation} \label{entropyproduction}
\dmu j_s^\mu = - \bigg( \dmu \frac{\mu}{T} \bigg)j_\text{diss}^\mu -\mu_\text{diss} \, \Delta^{\mu \nu} \dnu \bigg( \frac fT \zeta_\mu \bigg)- \bigg( \dmu \frac{u_\nu}{T} \bigg) \pi_{\text{diss}}^{\mu \nu} \, \text{,}
\end{equation}
where $f\equiv n_\text{s}/\mu$. We are going to show this formula in a box at the end of this section. 
\bigskip
\\One can easily check that the formula for entropy production (\ref{entropyproduction}) does not carry any time derivatives. This is the main practical advantage of the modified phase frame. In fact, as we are going to see, it is a natural choice to define $j_\text{diss}^\mu$, $ \pi_{\text{diss}}^{\mu \nu}$ and $\mus$ as linear combinations of the factors they come with in $\dmu j_s^\mu$, since in this way it is immediate to impose positivity of entropy production. Thus we see that working in the modified phase frame we can cast the equations of motion as in (\ref{eqwithM}).
\\Moreover, from a formal point of view the ideal level constitutive relations are expressed in terms of the chemical potential $\mu$, not of $u^\mu \xi_\mu$. Hence the use of $\xio_\mu$, which satisfies $u^\mu \xio_\mu = -\mu$, provides for a cleaner formulation of the constitutive relations. 

%
%
%
%
\bigskip
\begin{mdframed}
\textbf{Entropy production in the modified phase frame: derivation}
\medskip
\\In this box we follow \cite{Bhattacharya:2011eea} (with a different sign convention for the Josephson relation).
\\As a first step we want to show that:
\be \label{myA17}
\xio_\mu (u^\rho \partial_\rho)\xio^\mu +  \xio u^\mu \partial_\mu \xio =-\mud \, \xi_\mu u^\rho \partial_\rho( u^\mu)-\zeta^\rho \partial_\rho \mud
\ee
We start by computing (up to linear order in $\mud$):
\begin{align}
\begin{split}
\xio_\mu (u^\rho \partial_\rho)\xio^\mu &= \,  \xi_\mu u^\rho \partial_\rho \xi^\mu -\mud \, u_\mu u^\rho \partial_\rho \xi^\mu - \xi_\mu u^\rho \partial_\rho( \mud u^\mu)  \\
&\stackrel{u^\mu \xi_\mu = -\mu -\mud}{=}  \,  \xi_\mu u^\rho \partial_\rho \xi^\mu - \mud \, \big(u_\mu u^\rho \partial_\rho \xi^\mu + \xi_\mu u^\rho \partial_\rho u^\mu \big)  -  \mu \, u^\rho \partial_\rho \mud  
\end{split}
\end{align}
Now consider: 
\begin{align}
\begin{split}
u_\mu \xio^\rho \partial_\rho \xio^\mu =& \,  u_\mu \xi^\rho \partial_\rho \xi^\mu -\mud \, u_\mu u^\rho \partial_\rho\xi^\mu - u_\mu \xi^\rho \partial_\rho( \mud u^\mu)  \\
&\stackrel{u_\mu \partial_\rho u^\mu = 0}{=} u_\mu \xi^\rho \partial_\rho \xi^\mu -\mud \,u_\mu u^\rho \partial_\rho\xi^\mu  + \xi^\rho \partial_\rho \mud   
\end{split}
\end{align}
Now subtract the two terms:
\begin{align}
\begin{split}
\xio_\mu u^\rho \partial_\rho\xio^\mu - u_\mu \xio^\rho \partial_\rho \xio^\mu =&\,  \xi_\mu u^\rho \partial_\rho \xi^\mu  - u_\mu \xi^\rho \partial_\rho \xi^\mu \\
&-\mud \, \xi_\mu u^\rho \partial_\rho u^\mu + \big( \mu u^\rho-\xi^\rho \big)\partial_\rho \mud 
\end{split}
\end{align}
$\xi^\mu$ is irrotational, so $\xi_\mu u^\rho \partial_\rho \xi^\mu  - u_\mu \xi^\rho \partial_\rho \xi^\mu =\xi_\mu u^\rho \partial_\rho \xi^\mu -\xi_\mu u^\rho \partial_\mu \xi^\rho = 0$. Then:
\be
\xio_\mu u^\rho \partial_\rho \xio^\mu - u_\mu \xio^\rho \partial_\rho \xio^\mu =-\mud \, \xi_\mu u^\rho \partial_\rho u^\mu  + \big( \mu u^\rho-\xi^\rho \big)\partial_\rho \mud \ee
So we get:
\be
\xio_\mu (u^\rho \partial_\rho)\xio^\mu - u_\mu \xio^\rho \partial_\rho \xio^\mu =-\mud \, \xi_\mu u^\rho \partial_\rho u^\mu -\zeta^\rho \partial_\rho \mud
\ee
But:
\be
\xio u^\mu \dmu \xio = \xio u^\mu \cdot \frac{1}{2\xio}\big( -2\xio^\nu \dmu \xio_\nu \big) = -u^\mu \xio^\nu \dmu \xio_\nu
\ee
And so we finally get (\ref{myA17}).
\bigskip
Our goal is to compute the divergence of the entropy current:
\be
\dmu (su^\mu) = s\,  \dmu u^\mu + u^\mu \, \dmu s
\ee
To this end we rewrite the Smarr relation as: 
\be \label{smarr}
sT + \mu \,\nn = \epsilon - \mu n_\text{s} + p = \rho_\text{n} + p \, ,
\ee
where $ \rho_\text{n} \equiv  \epsilon - \mu n_\text{s}$. We can then recast the first law as $\diff \rhon = \mu \,\diff \nn + T\, \diff s + f \xio \, \diff \xio
$.
Then:
\be
u^\mu \dmu s = \frac1T \big( u^\mu \dmu \rhon - \mu \, u^\mu \dmu \nn - f u^\mu \xio \dmu \xio  \big)
\ee
And thus:
\be \label{A14}
\dmu \big( s u^\mu\big) = \frac1T \big( u^\mu \dmu \rhon +(\rhon + p)\dmu u^\mu \big)   - \frac{\mu}{T} \big( u^\mu \dmu \nn + \nn \dmu u^\mu \big) - \frac{f \xio}{T} u^\mu \dmu \xio
\ee
Now we can make use of the equations of motion. From the conservation of the stress-energy tensor (\ref{current}) we obtain (using the Josephson relation $u_\mu \xio^\mu = - \mu$)
\be
\begin{split}
0 = u_\nu \dmu T^{\mu \nu} &= u_\nu u^\mu u^\nu \dmu (\rhon + p) +  \eta^{\mu \nu} u_\nu \dmu p + (\rhon + p)u_\nu u^\mu \dmu u^\nu + u_\nu \dmu \big( f \xio^\mu \xio^\nu\big) \\
& \quad \quad + u_\nu \dmu \pi^{\mu \nu}_\text{diss}\\
&= -u^\mu \dmu \rhon -(\rhon + p) \, \dmu u^\mu - \mu \, \dmu \big( f\xio^\mu \big) + f u_\nu\, \xio^\mu \dmu \xio^\nu + u_\nu \dmu \pi^{\mu \nu}_\text{diss} 
\end{split}
\ee
So we can write:
\be
u^\mu \dmu \rhon +\big( \rhon + p\big) \, \dmu u^\mu = - \mu \, \dmu \big( f \xio^\mu \big) + fu_\nu \, \xio^\mu \dmu \xio^\nu + u_\nu \dmu \pi^{\mu \nu}_\text{diss}
\ee
Furthermore current conservation $\dmu j^\mu = 0$ implies:
\be
\dmu j^\mu = 0 = \nn \dmu u^\mu + u^\mu \dmu \nn +\dmu\left( f\xio^\mu  \right)+ \dmu j_\text{diss}^\mu 
\ee
This yields:
\be
u^\mu \dmu \nn + \nn \dmu u^\mu = -\dmu \big( f\xio^\mu \big) - \dmu j_\text{diss}^\mu
\ee
Plugging everything together we find:
\be
\begin{split}
\dmu \big( su^\mu\big)& =  \frac1T \big(-\mu \dmu \big( f\xio^\mu\big) + f u_\nu  \xio^\mu \dmu \xio^\nu + u_\nu \dmu \pi^{\mu \nu}_\text{diss} \big) \\
&\quad \qquad + \frac{\mu}{T} \big( \dmu \big( f\xio^\mu\big) + \dmu j_\text{diss}^\mu \big) - \frac{\xio f}{T}u^\mu \dmu \xio \\
&=  \frac1T \big(f u_\nu  \xio^\mu \dmu \xio^\nu + u_\nu \dmu \pi^{\mu \nu}_\text{diss} \big) + \frac{\mu}{T}   \dmu j_\text{diss}^\mu  - \frac{\xio f}{T}u^\mu \dmu \xio
\end{split}
\ee
Using that $\partial_\nu \xio^\mu - \partial_\nu \xio^\mu = \mathcal{O}(\partial^2)$ this becomes:
\be
\dmu \big( su^\mu \big)= -\frac fT \big( u^\mu \xio^\nu \dmu \xio^\nu + \xio  u^\mu \dmu \xio \big) + \frac1T u_\nu \dmu \pi^{\mu \nu}_\text{diss} + \frac{\mu}{T} \dmu j_\text{diss}^\mu
\ee
Now we make use of (\ref{myA17}) and obtain:
\be \label{newdsmu}
\dmu \big( su^\mu \big)= \frac fT \big( \mud \, \xi_\mu u^\rho \partial_\rho( u^\mu)+\zeta^\rho \partial_\rho \mud
 \big) + \frac{u_\nu}{T} \dmu \pi^{\mu \nu}_\text{diss} + \frac{\mu}{T} \dmu j_\text{diss}^\mu
\ee
Notice that, as expected, this expression vanishes at linear order in fluctuations.
\bigskip
\\We can now determine the divergence of the entropy current.
\\First we compute
\be
\frac{f}{\mu} \zeta^\nu \dnu \mud = \dnu \bigg( \frac{f}{T} \zeta^\nu \mud\bigg)- \mud \, \dnu \bigg(\frac{f}{T}\zeta^\nu \bigg)
\ee
and (using that $\zeta^\mu = \xio^\mu - \mu u^\mu$ and $u^\mu \zeta_\mu = 0$ so that $u^\mu \dnu \zeta_\mu = -\zeta^\mu \dnu u_\mu$):
\be
\begin{split}
 \frac{f}{T} \mud \xio^\mu u^\nu \dnu u_\mu  = -  \frac fT \mud u^\mu u^\nu \dnu \zeta_\mu
\end{split}
\ee
And so:
\be
  \frac fT \big(  \mud \, \xio^\mu u^\nu \dnu u_\mu + \zeta^\nu \dnu \mud  \big)  =  \dnu \bigg( \frac fT \zeta^\nu \mud \bigg) - \mud \, \left( \dnu \left( \frac fT\zeta^\nu\right) + \frac fT u^\mu u^\nu \dnu \zeta_\mu \right)
\ee
The last term can be rewritten as: 
\be
\begin{split}
 \dnu \bigg( \frac fT\zeta^\nu\bigg) + \frac fT u^\mu u^\nu \dnu \zeta_\mu = \Delta ^{\mu \nu}\dnu \bigg( \frac fT\zeta_\mu \bigg) 
 \end{split}
\ee
So that (\ref{newdsmu}) finally becomes:
\be
\dmu \big( su^\mu \big) = - \dmu \bigg( \frac fT \zeta^\mu \mud \bigg) - \mud \, \Delta^{\mu \nu}\dnu \bigg( \frac fT \zeta_\mu \bigg) + \frac{u_\nu}{T}\dmu \pi^{\mu \nu}_\text{diss} + \frac \mu T \dmu j_\text{diss}^\mu
\ee
Now we define the entropy current as:
\be
j_s^\mu \equiv su^\mu - \frac{\mu}{T} j_\text{diss}^\mu - \frac{u_\nu}{T} \pi^{\mu \nu} - \frac fT  \zeta^\mu\,\mud \, ,
\ee
with which we finally recover (\ref{entropyproduction}):
\be \label{entropyproduction}
\dmu j_s^\mu = - \bigg( \dmu \frac{\mu}{T} \bigg)j_\text{diss}^\mu - \mud \, \Delta^{\mu \nu} \dnu \bigg( \frac fT \zeta_\mu \bigg)- \bigg( \dmu \frac{u_\nu}{T} \bigg) \pi_\text{diss}^{\mu \nu} \, .
\ee

\end{mdframed}
\subsubsection{Ensemble and sources}
A commonly used ensemble is the entropy ensemble, consisting of the timelike component of the charge current (which, with vanishing normal fluid velocity, corresponds to the charge density), entropy density, momentum density and the spatial gradient of the Goldstone boson $\boldsymbol{\xi}\equiv \nabla \phi$:
\begin{equation*}
(j^t, s, \boldsymbol{\pi}, \boldsymbol{\xi})
\end{equation*}

We saw at the beginning of this section that the first law takes the form: \cite{Herzog:2011ec} 
\begin{align} \label{diffp}
\begin{split}
\diff p &= s\, \diff T + n \, \diff \mu - \frac{n_\text{s}}{2\mu} \, \diff (\xi^2 + \mu^2) \\
&=s\, \diff T  +\left(  n + \frac{n_\text{s}}{\mu}\,\boldsymbol{\xi} \cdot \mathbf{u} \right)\,  \diff \mu   -\frac{n_\text{s}}{\mu}\boldsymbol{\xi} \cdot (\diff \boldsymbol{\xi}-\mu \, \diff \mathbf{u} )
\end{split}
\end{align}
Take the constitutive relation for the current (\ref{jmu}). At ideal order $j^t = n + (n_\text{s}/\mu)\zeta^t$. But the Josephson relation (\ref{jos}) yields $\xi^t = \boldsymbol{\xi}\cdot \mathbf{u}+\mu$ and so $\zeta^t = \xi^t-\mu$. Thus we can write:
\begin{equation}
j^t =n + \frac{n_\text{s}}{\mu}\,\boldsymbol{\xi} \cdot \mathbf{u} 
\end{equation}
We can then take a look back at the first law (\ref{diffp}) together with definition (\ref{firstlawsources}) and read off the sources (recall that the spacelike part of the normal fluid velocity $u^\mu$ comes at first order in fluctuations):
\begin{equation}
 s_{j^t} = \mu \qquad \qquad s_s= T
\end{equation}
\begin{equation}
s_\xi^i \equiv \mathbf{h}_\xi^i =  \frac{n_\text{s}\xi^i}{\mu}
 \ee
On top of this, the source of the momentum density is, as, usual, the normal fluid velocity:
\begin{equation}
s_{\boldsymbol{\pi}} =  \mathbf{u}
\end{equation}
\\We can now write down the susceptibility matrix $\chi_{ab}$, so that:
\begin{equation}
\begin{pmatrix} \delta j^t \\ \delta s \\ \delta \boldsymbol{\pi} \\ \delta \boldsymbol{\xi} \end{pmatrix} = \chi \cdot \begin{pmatrix} \delta \mu \\ \delta T \\ \delta \mathbf{u} \\ \delta \mathbf{h}_\xi\end{pmatrix}
\end{equation}
\\Onsager reciprocity fixes some of the susceptibilities. In particular, in the collinear limit $\boldsymbol{\zeta} // \mathbf{u} // \mathbf{k} // \hat{e}_x$ one finds ($\xi \equiv \boldsymbol{\xi}^x$):  \begin{equation}
\partial_\mu n_\text{s}|_{\xi,T,u_x}=\frac{n_\text{s}}{\mu} - \,\frac{\mu}{\xi}\chi_{nh_\xi} \end{equation}
\begin{equation}
 \partial_T n_\text{s} |_{\xi, \mu, u_x} = -\frac{\mu}{\xi}\chi_{sh_\xi}|_{\mu, T, u_x}
\end{equation}
\begin{equation}
 \partial_{u_x} n_\text{s} |_{\xi, \mu,T} =- \mu\,\partial_{\xi}n_\text{s}|_{\mu, T, u_x}
\end{equation}
\begin{equation}
\chi_{n\pi} = - \mu \, \chi_{nh_\xi}
\end{equation}
\begin{equation}
\chi_{s\pi} = - \mu \, \chi_{sh_\xi}
\end{equation}
\begin{equation}
\chi_{p h_\xi} = \partial_{\xi}n_\text{s}|_{\mu, T, u_x} + \frac{n_\text{s}}{\xi}
\end{equation}
These relations yield the following susceptibility matrix: 
\begin{equation} \label{chicomplete}
\begin{pmatrix}
\delta j^t \\
\delta s \\
\delta \pi^x \\
\delta \xi
\end{pmatrix} =
\begin{pmatrix}
\chi_{nn} + \chi_{nh_\xi}^2\chi_{\xi\xi} & \chi_{ns} + \chi_{nh_\xi}\chi_{sh_\xi}\chi_{\xi\xi} & \frac{\xi n_\text{s}}{\mu} & - \chi_{nh_\xi}\chi_{\xi \xi} \\
 \chi_{ns} + \chi_{nh_\xi}\chi_{sh_\xi}\chi_{\xi\xi} & \chi_{ss}+\chi_{sh_\xi}^2\chi_{\xi \xi} & 0 & \chi_{sh_\xi} \\
  \frac{\xi n_\text{s}}{\mu}  & 0 & \epsilon + p + \frac{\xi}{n_\text{s}} \xi^2 & \mu \\
 \chi_{nh_\xi}\chi_{\xi \xi}  &  \chi_{sh_\xi}\chi_{\xi \xi}  &\mu & \chi_{\xi \xi}
\end{pmatrix}
\begin{pmatrix}
\delta \mu \\
\delta T \\
\delta u_x \\
\delta h_\xi
\end{pmatrix}
\end{equation}
Where
\begin{equation}
\chi_{\xi \xi} = \frac{\mu}{n_\text{s} + \xi\, \partial_{\xi}n_\text{s}|_{T,\mu, u_x}}
\end{equation}
Note that here we are expressing the static susceptibilities at fixed $h_\xi$ in terms of those computed at fixed $\xi$; for instance $\chi_{nn}^{h_\xi}= \chi_{nn}+ \chi_{nh_\xi}^2\chi_{\xi \xi}$.

\subsubsection{Defining the dissipative coefficients}
The contribution of the stress-energy tensor to entropy production (\ref{entropyproduction}) can be rewritten as \cite{Bhattacharya:2011eea}
\begin{equation}
\begin{split}
-\frac1T\, \pi_{diss}^{\mu \nu} \dmu u_\nu 
&=  -\frac1T\,  \pi_{diss}^{\mu \nu} \sigma_{\mu \nu} - \frac1{d\,T} \pi_{diss, \, \mu}^\mu \, \partial_\rho u^\rho
\end{split}
\end{equation}
Note that $\sigma_{\mu \nu}$, by construction, only couples to the traceless part of $\pi_{diss}^{\mu \nu}$, which we call $\pi_{diss, \, TL}^{\mu \nu}$. This implies that the contribution of $\pi_{diss}^{\mu \nu}$ to entropy production consists of two independent terms (corresponding to two irreducible representations of $SO(1,d)$): 
\begin{equation} \label{entropyproduction2}
\dmu \tilde{J}_s^\mu = - \bigg( \dmu \frac{\mu}{T} \bigg)\tilde{J}_{diss}^\mu +\mud \, \Delta^{\mu \nu} \dnu \bigg( \frac fT \zeta_\mu \bigg)- \frac1T \pi_{diss, \, TL}^{\mu \nu}\sigma_{\mu\nu} - \frac1{d\,T} \partial_\rho u^\rho {\pi_{diss}}^\mu_\mu
\end{equation}
%
%
%
%
%
This motivates the following definition of the dissipative contributions: 
\begin{equation} \label{dissmat}
\begin{pmatrix}J_{diss}^\mu \\ \mu_{diss}\\ \pi_{diss, TL}^{\mu \nu} \\ \pi^\rho_{diss, \rho}\end{pmatrix} = 
-\begin{pmatrix} A^{ \mu \rho} & B^\mu & C^{\mu \rho \sigma} & D^\mu \\
E^\rho & F & G^{\rho \sigma} & H \\
I^{\mu \nu \rho} & K^{\mu \nu} & L^{\mu \nu \rho \sigma} & M^{\mu \nu}\\
N^{\rho}  & Q & R^{\rho \sigma} & S
\end{pmatrix} 
\begin{pmatrix}
\Delta_{\rho}^{\alpha}\partial_\alpha \frac \mu T 
\\ \Delta^{\mu \nu} \dnu \big( \frac {\ns}{\mu T} \zeta_\mu \big) 
\\ \frac1T \, \sigma_{\rho \sigma} \\ \frac{1}{d\,T}\,\partial_\alpha u^\alpha
\end{pmatrix}
\end{equation}
This matrix will have to be positive (semi)definite in order to achieve positivity of entropy production. In particular, its diagonal elements must be nonnegative. 
\\We work in the transverse frame: $u_\mu j_{diss}^\mu = u_\mu \pi_{diss}^{\mu \nu}=0$, so that its elements can be parametrized as follows:
\begin{align}\label{Aons}
A^{\mu \rho } &= \alpha_A\,\Delta^{\mu \rho} + \beta_A \, \tilde{\zeta}^\mu \tilde{\zeta}^\rho      \\                                                                      
B^\mu &=\alpha_B  \zeta^\mu                       \\                                               
C^{\mu \rho \sigma} &= \alpha_C \, \big( \tilde{P}^{\mu \sigma}\tilde{\zeta}^\rho+ \tilde{P}^{\mu \rho}\tilde{\zeta}^\sigma  \big) + \beta_C \, \zeta^\mu P^{\rho \sigma} \\
D^\mu &= \alpha_D \, \zeta^\mu \\
E^\sigma &= \alpha_E \, \zeta^\sigma        \\   
G^{\rho \sigma} &= \alpha_G\, P^{\rho \sigma} \\
I^{\mu \nu \rho}  &= \alpha_I \, \big( \tilde{P}^{\mu \rho}\tilde{\zeta}^\nu+ \tilde{P}^{\nu \rho}\tilde{\zeta}^\mu  \big) + \beta_I \, \zeta^\rho P^{\mu \nu} \\
K^{\mu \nu} &= \alpha_K \, P^{\mu \nu} \\
L^{\mu \nu  \rho \sigma} &= \alpha_L \, P^{\mu \nu}P^{\rho \sigma} + \beta_L \,\Delta^{\mu \rho}\Delta^{\nu \sigma} + \gamma_L \, \big( \zeta^\mu \tilde{\zeta}^\sigma \tilde{P}^{\nu \rho}+\zeta^\nu \tilde{\zeta}^\sigma \tilde{P}^{\mu \rho}\big) \\
M^{\mu \nu} &= \alpha_M \, P^{\mu \nu} \\
N^\rho &= \alpha_N \, \zeta^\rho \\
R^{\rho \sigma} &= \alpha_R \, P^{\rho \sigma} \\
F &= \alpha_F \qquad H = \alpha_H \qquad Q = \alpha_Q \qquad S = \alpha_S \, \text{,}
\end{align}
where $\tilde{\zeta}^\mu\equiv\zeta^{\mu}/\sqrt{\zeta^\nu \zeta_\nu}$, $P^{\mu \nu} \equiv  \tilde{\zeta}^\mu \tilde{\zeta}^\nu - \Delta^{\mu \nu}/d$ is a traceless, transverse tensor and $\tilde{P}^{\mu \nu}\equiv \Delta^{\mu \nu} - \tilde{\zeta}^\mu \tilde{\zeta}^\nu$ is the projector onto the subspace orthogonal to both $u^\mu$ and $\zeta^\mu$.
\\Onsager reciprocity requires the matrix appearing in (\ref{dissmat}) to be symmetric. We get:
\be
\alpha_B = \alpha_E \quad \alpha_C = \alpha_I \quad \beta_C = \beta_I \quad \alpha_D = \alpha_N \quad \alpha_G = \alpha_K \quad \alpha_H = \alpha_Q \quad \alpha_M = \alpha_R 
\ee
We thus obtain a total of 14 dissipative parameters, as stated in \cite{Bhattacharya:2011tra}.
\bigskip
\\Note that in the limit of vanishing background superfluid velocity $\zeta^\mu \to 0$ the normed vector $\tilde{\zeta}^\mu$ is not well-defined \cite{Bhattacharya:2011eea}, so that one must set $\alpha_G=\alpha_L=0$.  After imposing Onsager invariance this means that the only non-vanishing dissipative contributions to the constitutive relations will be parametrized by a total of five coefficients: $\alpha_A$, $\beta_L$, $\alpha_F$, $\alpha_H$ and $\alpha_S$.
\bigskip
\\Imposing Weyl invariance amounts to requiring that the stress-energy tensor be traceless; in this case this condition translates to $\alpha_N = \alpha_Q = \alpha_R = \alpha_S = 0$, leaving us with a total of 10 dissipative coefficients after imposing Onsager reciprocity, as in  \cite{Bhattacharya:2011tra}. In the limit of zero background superfluid velocity this set further reduced to three coefficients: $\alpha_A$, $\beta_L$ and $\alpha_F$.
%
%
%
%
%
\bigskip
\\In the collinear limit, where the spatial parts of $u^\mu$ and $\zeta^\mu$ are both parallel to the wavevector, the linearized equations of motion can be written as follows:
\begin{equation} \label{eomsMtilde}
-i\omega \, \begin{pmatrix} \delta J^t \\ \delta s \\ \delta \pi^x \\ \delta \xi  \end{pmatrix} +  \tilde{M} \cdot \begin{pmatrix} \delta \mu \\ \delta T \\ \delta u_x \\ \delta h_\xi \end{pmatrix} = 0
\end{equation}
with
\small{
\begin{equation}
\hspace*{-2.3cm}
 \tilde{M}=\begin{pmatrix} \frac{\alpha_A+\beta_A}{T}\,k^2 & \frac{-\mu^2(\alpha_A+\beta_A )+ \xi^2 n_\text{s} \,\alpha_B}{T^2 \mu}\, k^2 & ikn + \frac{(\alpha_D+2\beta_C)\xi}{3T}\, k^2 & ik +\frac{k^2\alpha_B \xi}{T} \\
\frac{-\mu^2(\alpha_A+\beta_A )+ \xi^2 n_\text{s} \,\alpha_B}{T^2 \mu}\, k^2 & \frac{\mu^4(\alpha_A + \beta_A) + n_\text{s} ( n_\text{s}\alpha_F -2 \mu^2\alpha_B )}{T^2 \mu^2}\, k^2 & iks - \frac{\left( n_\text{s}(2\alpha_G + \alpha_H) + (\alpha_D+2\beta_C)\mu^2\right)\xi}{3T^2\mu}\,k^2 & -\frac{k^2\xi\left( n_\text{s} \alpha_F  + \mu^2\alpha_B \right)}{T^2 \mu} \\
ikn + \frac{(\alpha_D+2\beta_C)\xi}{3T}\, k^2 &  iks - \frac{\left( n_\text{s}(2\alpha_G + \alpha_H) + (\alpha_D+2\beta_C)\mu^2\right)\xi}{3T^2\mu}\,k^2 &2n_\text{s} \xi \,ik  + \frac{6\beta_L +4(\alpha_L +\alpha_M)  + \alpha_S}{9T}\,k^2 & ik\xi +\frac{k^2(2\alpha_G + \alpha_H)}{3T} \\
ik +\frac{k^2\alpha_B \xi}{T} &\frac{k^2\xi\left(n_\text{s} \alpha_F  + \mu^2\alpha_B \right)}{T^2 \mu}  & ik \xi  + \frac{k^2(2\alpha_G + \alpha_H)}{3T} & \frac{\alpha_F}{T}k^2
\end{pmatrix} 
\end{equation}} 
\normalsize{}Note that this matrix is Onsager symmetric, as expected. 
\\We see that in the collinear limit the derivative coefficients appear in the equations of motion in only six independent linear combinations. We can thus set:
\begin{equation} \label{disscolllim}
\sigma_0=\frac{\alpha_A+\beta_A}{T}\,,\quad\zeta_1=\frac{\alpha_G+\alpha_H}{2T}\,,\quad\zeta_2=\frac{\alpha_B}T\,,\quad \zeta_3=\frac{\alpha_F}{T}\,,\quad \zeta_6=\frac{(\alpha_D+\beta_C)}{2T}\,,
\end{equation}
\begin{equation}
\eta=\frac{\beta_L+\xi\gamma_L}{2T}\,,\quad\zeta=\frac{\alpha_S+\alpha_L+\alpha_M-2\xi\gamma_L}{4T}
\end{equation}
Notice that since, as we saw above, $\alpha_G=0$ for vanishing background superfluid velocity, for a Weyl invariant theory, featuring $\alpha_H=0$, also $\zeta_1$ will vanish at $\zeta^\mu=0$.
\\Interestingly, at nonzero superfluid velocity, the bulk viscosity is non-vanishing when conformal invariance is imposed.
\bigskip
\\As we mentioned earlier, positivity of entropy production sets bounds on the dissipative coefficients. In particular, the diagonal components of (\ref{dissmat}) must be nonnegative: here we can immediately conclude that $\zeta_3, \sigma_0\geq 0$. Furthermore, all principal minors of (\ref{dissmat}) must be positive definite. For example, this yields $\sigma_0 \zeta_3 \geq \xi^2 \zeta_2^2$.
\bigskip
\\Finally, in the collinear case the matrix $\tilde{M}$ describing the equations of motion (\ref{eomsMtilde}) becomes:
\begin{equation} \label{eomscollinear}
\hspace*{-1cm}
 \tilde{M}=\begin{pmatrix} \sigma_0\,k^2 & \frac{ -\mu^2 \, \sigma_0 - n_\text{s} \xi^2 \zeta_2}{T \mu}\, k^2 & ikn + \xi\zeta_6\, k^2 & ik +k^2\zeta_2 \xi \\
\frac{ -\mu^2 \, \sigma_0 - n_\text{s} \xi^2 \zeta_2}{T \mu}\, k^2 & \frac{n_\text{s}\xi^2(n_\text{s} \zeta_3 + 2\mu^2 \zeta_2)+ \mu^4\sigma}{T^2 \mu^2}\, k^2 & iks + \frac{\xi\left(n_\text{s} \zeta_1 +\mu^2 \zeta_6 \right)}{T\mu}\,k^2 & -\frac{k^2\xi\left( n_\text{s}\zeta_3 + \mu^2\zeta_2 \right)}{T \mu} \\
ikn + \xi\zeta_6\, k^2  &  iks + \frac{\xi\left(n_\text{s} \zeta_1 +\mu^2 \zeta_6 \right)}{T\mu}\,k^2 & 2n_\text{s} \xi \,ik  + (\eta+\zeta_b) \, k^2  & ik\xi k^2\, \zeta_1 \\\
ik +k^2\zeta_2 \xi& -\frac{k^2\xi\left( n_\text{s}\zeta_3 + \mu^2\zeta_2 \right)}{T \mu} &ik \xi k^2\,\zeta_1 & \zeta_3\,k^2
\end{pmatrix} 
\end{equation}
\subsubsection{Kubo relations}
The dissipative coefficients can be extracted from the $\omega \to 0$, $k\to 0$ behaviour of the retarded Green's functions. As we are going to see these relations are especially useful  in the context of the applications of gauge-gravity duality. Simply put, they allow us to extract the dissipative coefficients appearing in the gradient expansion of the vacuum expectation values of the boundary theory.
\bigskip
\\In the collinear limit we have:
\begin{equation} \label{kubo}
\begin{split}
\sigma_0 &= -\lim_{\omega \to 0, k\to 0} \, \frac1{\omega}\, \text{Im} \,G^R_{J^xJ^x}(\omega, k) \\
 \zeta_1 &=-\lim_{\omega \to 0, k\to 0} \, \frac{1}{k\omega}\, \text{Im} \,G^R_{T^{xx}\xi^x}(\omega, k) \\
\xi \, \zeta_2 &= -\lim_{\omega \to 0, k\to 0} \, \frac{1}{k}\, \text{Im} \,G^R_{J^x\xi^x}(\omega, k) \\
\zeta_3 &= -\lim_{\omega \to 0, k\to 0} \, \frac{\omega}{k^2}\,  \text{Im} \,G^R_{\xi^x \xi^x}(\omega, k) \\
\xi\, \zeta_6 &=- \lim_{\omega \to 0, k\to 0} \, \frac{1}{\omega}\, \text{Im}\, G^R_{J^xT^{xx}}(\omega, k) \\
\eta+\zeta_b &=- \lim_{\omega \to 0, k\to 0} \, \frac1{\omega}\, \text{Im} \,G^R_{T^{xx}T^{xx}}(\omega, k)\\
\eta &=- \lim_{\omega \to 0, k\to 0} \, \frac1{\omega}\, \text{Im} \,G^R_{T^{xy}T^{xy}}(\omega, k)
\end{split}
\end{equation}
We can check that in the $\xi\to 0$ limit $\zeta_1$ vanishes. In fact, we can rewrite the above Kubo relation as $ \zeta_1 = \lim_{\omega \to 0} \, \text{Re} \,G^R_{T^{xx}\phi}(\omega, k=0)$. Now, $T^{xx}$ is even under time-reversal, while $\phi$ is odd (since the superfluid velocity is odd and is given by its spatial gradient). Thus their correlator must be odd. At vanishing wavevector the only background quantity that is odd under time reversal is $\xi$ itself; hence $\text{Re} G^R_{T^{xx}\phi}(\omega, k=0)$ must have an expansion in odd powers of $\xi$, implying that it vanishes at $\xi \to 0$. This confirms what we saw above, i.e. that in the limit of vanishing background superfluid velocity only three dissipative coefficients enter the equations of motion.
\subsubsection{Back to the start: the conductivity}
Having defined the propagators, given the constitutive relations and written down the dissipative coefficients of superfluid hydrodynamics it is now a good time to take a step back and reflect on the physical implications of the \textit{superfluid} component. 
\\In particular, we should take a look at the conductivity. It is obtained from the current-current propagator in the vanishing momentum limit. In superfluid hydrodynamics with zero background superfluid velocity $\xi$ and a momentum relaxation rate $\Gamma$ (which appears in the equation of motion for the momentum density $T^{tx}$ as $(\partial_t + \Gamma)T^{tx}+\partial_x T^{xx} = 0$ ) one can show that it is given by \footnote{Non-dissipative, non-hydrostatic terms contribute to $\sigma$ at the same order at $\sigma_0$, see \cite{blaiseashish} for the case of hydrodynamic metals. Here we are assuming the absence of such terms.}:
\cite{KadanoffandMartin, Gouteraux:2019kuy}
\be
\sigma(\omega) \equiv \frac{i}{\omega}G_{J^xJ^x}(\mathbf{k}=0,\omega) =\frac{1}{\Gamma-i\omega} \frac{n_\text{n}^2}{\mu n_\text{n}+sT} + \frac{i}{\omega}\frac{n_\text{s}}{\mu} + \sigma_0 + \mathcal{O}(\omega)
\ee
We can see that, in the presence of a finite $\Gamma$, the conductivity $\sigma(\omega)$ has a pole at $\omega = 0$ if $n_\text{s}\neq0$. This is indeed a token of superconductivity. 
\\On the other hand, if $\Gamma=0$, $\sigma(\omega)$ does have a pole at $\omega=0$ even in the absence of a superfluid component. This feature, however, arises from translation invariance, as an electric field applied onto a translationally invariant system produces uniform acceleration \cite{Hartnoll:2008kx}.
\subsubsection{Computing the modes}
The hydrodynamic modes $\omega(k)$ (\ref{generalmode}) can be computed by solving for:
\begin{equation}\label{modesformula}
\text{det}(-i\omega \mathbb{1} + \tilde{M}\chi^{-1})=0
\end{equation}
For small values of the wavevector $k$ the modes are sound modes.
\\Their real part, linear in $k$, depends on thermodynamic quantities such as the charge and densities, the temperature and the susceptibilities. The imaginary part, one the other hand, is quadratic in $k$ and depends on the dissipative coefficients, too.
\\One commonly identifies two ``first sound'' and two ``second sound'' modes, as we are going to see in the coming section. 
\subsection{Modes in conformal limit with no background superfluid velocity}
At this point it would come natural to write down the modes of superfluid hydrodynamics with the full set of dissipative coefficients described above. Such modes can be computed; their analytic form, however, is too long and complicated to write down.
\\Nevertheless, the modes do simplify in the limit where the background superfluid velocity $\xi$ vanishes. In this section we are going to list them in this particular case, for a conformal theory: from a qualitative point of view, we would encounter the same types of mode (two pairs of sound modes and a diffusive mode) with $\xi \neq 0$, too. Moreover, this is a demonstration of the power of conformal symmetry, which allows us to derive relations between the thermodynamic derivatives. 
\\We take (\ref{dissnobkg}) as the definition of the three dissipative terms which are relevant in this limit.
\bigskip
\\As a first step, we recall \cite{Herzog:2011ec} that in the absence of a background superfluid velocity conformal symmetry implies that the $\mu/T$ ratio is the only available scale invariant background scalar quantity. Then any scalar can be expressed as a function of $\mu/T$, multiplied with a power of $T$ (or, equivalently, $\mu$), determining its dimensionality. So, in particular,  
\begin{equation}\label{scalingofp}
p = T^{d+1}f\Big( \frac{\mu}{T}\Big)
\end{equation}
But from the first law (\ref{diffp}) one gets:
\begin{equation}
s = \pdert{p}{T}{\mu} \quad \quad \quad \quad n = \pdert{p}{\mu}{T} 
\end{equation}
By further differentiating with respect to $\mu$ and $T$ one can verify that: \footnote{These relations can also be derived by imposing that the trace of the stress-energy tensor vanish: $\langle T_\mu^\mu \rangle = 0$. In fact, under a conformal transformation the (infinitesimal) variation of the metric is proportional to the metric itself: $\delta g^{\mu \nu} = \lambda g^{\mu \nu}$. We can write $\langle T_{\mu \nu} \rangle= -\frac{2}{\sqrt{-g}}\frac{\delta \Omega}{\delta g^{\mu \nu}}$. But then under a variation of the metric one has that $\delta \Omega = -\frac{\sqrt{-g}}{2}\delta g^{\mu \nu}\langle T_{\mu \nu} \rangle \propto \lambda g^{\mu \nu}\langle T_{\mu \nu} \rangle \propto \langle T_{\mu}^\mu \rangle $. So invariance  under conformal transformations implies that the stress-energy tensor is traceless.}
 \begin{equation}\label{thsusc2}
\pdert{s}{T}{\mu} = d \, \frac{s}{T} - \frac{\mu}{T}\pdert{s}{\mu}{T} \qquad \qquad \qquad  \pdert{n}{\mu}{T} = d \, \frac{n}{\mu} - \frac{T}{\mu}\pdert{s}{\mu}{T}
\end{equation}
Furthermore, one can define the reduced entropy $\sigma$ as \cite{landaubook}:
\be
\sigma \equiv \frac{s}{n}
\ee
and express the susceptibilities in terms of its thermodynamic derivatives. For instance:
\be
\pdert{s}{\mu}{T} = \pdert{n}{T}{\mu} = \frac{n\left( d\,s - nT\,\pdertsmall{\sigma}{T}{\mu}\right)}{sT + n\mu}
\ee
\bigskip
\\Writing $\omega = ck - \frac{i}{2}\Gamma k^2+ \mathcal{O}(k^3)$ as in (\ref{generalmode}), one obtains the following modes \cite{landaulif, chlub}:
\begin{itemize}
\item One diffusion mode:
\begin{equation}\label{diffnobkg}
v = 0 \quad \quad \quad \quad \Gamma = \frac{2\eta}{sT+\mu n_n}
\end{equation}
\item Two first sound modes: 
\begin{equation}\label{firstsoundnobkg}
v = \pm \frac{1}{\sqrt{d}} \quad \quad \quad \quad \Gamma = \frac{2(d-1)\eta}{d(sT+n\mu)}
\end{equation}
\item Two second sound modes \cite{Herzog:2009md,Herzog:2011ec,Gouteraux:2019kuy, Arean:2021tks, donnelly} :
\begin{align}\label{secondsoundnobkg}
\begin{split}
v &= \pm\frac{\sigma^2 n_\text{s}}{(sT+\mu n_n) \pdertsmall{\sigma}{T}{\mu}  } \\[1ex]
 \Gamma &= \frac{\mu (sT+\mu n)}{T^2 \,n^2 \, \pdertsmall{\sigma}{T}{\mu} }\, \sigma_0 + \frac{2(d-1)\,\mu n_\text{s}}{d(sT+\mu  n)(sT+\mu  n_n)}\,\eta + \frac{n_\text{s}(sT+\mu  n)}{\mu (sT + \mu n_n)}\,\zeta_3
 \end{split}
\end{align}
\end{itemize}
Notice that the imaginary parts of the modes are nonpositive, provided the dissipative coefficients are nonnegative. This, in turn, directly follows from positivity of entropy production (\ref{entropyproduction}). 
%
%
%
%
%
%
%

%
%
%

\chapter{Gauge-gravity duality}
\maketitle

The AdS-CFT correspondence was first proposed by Maldacena in 1997 \cite{Maldacena:1997re}. It posits the dynamical equivalence between a $d+1$-dimensional gravity theory in an asymptotically anti-de-Sitter (AdS) space and a $d$-dimensional conformal field theory (CFT). AdS spaces are maximally symmetric spaces with constant curvature; a theory is called conformal if it is invariant under conformal maps, i.e. transformations that locally preserve angles. 
\bigskip
\\The original correspondence involves type IIB string theory on AdS$_5\times S^5$ and $\mathcal{N}=4$ Super Yang-Mills theory with gauge group $SU(N)$. Yang-Mills theories are gauge theories with a non-abelian symmetry group; super Yang-Mills theories make use of supersymmetry to associate fermionic partners to the bosonic fields. In this case the gauge bosons can be mapped onto $\mathcal{N}=4$ different fermionic partners.
\\Owing to the nature of the theories it links together, the AdS-CFT correspondence is thus also called ``gauge-gravity duality''.
\bigskip
\\On top of $N$ the main parameters governing the involved theories are the string coupling $g_s$, the field theory coupling $g_{\text{YM}}$ and the AdS curvature radius $L$.
\\Writing down the full Yang-Mills Lagrangian is beyond the scope of this introduction; it will be sufficient to know that the Yang-Mills coupling $g_{\text{YM}}$ appears in front of the kinetic term for the gauge field: $\mathcal{L}\supset-\text{Tr}F_{\mu \nu}F^{\mu \nu}/(2g_\text{YM}^2)$. On the other hand, the string coupling $g_s$ is given by the expectation value of the dilaton field in the gravity theory emerging from the string theory. It controls the genus expansion of string theory scattering (a small $g_s$ suppresses the contributions of topologies with large genus to the scattering amplitude) and is related to Newton's gravitational constant in $d$ dimensions $G_d$ by $G_d\sim l_s^{d-2}g_s^2$, with $l_s$ being the string length.
\bigskip
\\The coupling of the field theory is related to the string coupling by:
\begin{equation}\label{couplingmap}
g_{\text{YM}}^2 =2\pi g_s
\end{equation}
Furthermore, the string length $l_s$ satisfies the relation
\begin{equation} \label{gstringl}
g_{\text{YM}}^2 \sim \frac{L^4}{l_sg_sN}
\end{equation}
In particular, one can consider a matrix theory with fields in the adjoint representation of $SU(N)$. Owing to the two-index structure of the matrix theory the Feynman diagrams are going to be surfaces with an intrinsic topology.  One defines: 
\begin{equation}\label{thooft}
\lambda \equiv g_{\text{YM}}^2N = \text{const}
\end{equation}
One can show that the Feynman diagrams go like $N^{2-2G}\lambda^{E-V}$, with $G$ being the genus of the surface \cite{mcgreevy, erdmenger}. In this case one can take the `t Hooft limit: $ N \to \infty $ with $g_{YM}^2\to 0$, so that $\lambda$ stays finite. In this case planar diagrams with $G=0$ will dominate and $N$-point correlation functions factorize \cite{mcgreevy}:
\begin{equation}
<\mathcal{O}\cdots \mathcal{O}>\, = \,<\mathcal{O} >\cdots< \mathcal{O}>  +\, \mathcal{O}(1/N^2)
\end{equation}
We thus obtain a large-$N$ theory with nontrivial interactions, although we see that correlators behave classically up to subleading corrections in $1/N$.
\\Indeed, from (\ref{couplingmap}) and (\ref{thooft}) we see that the $1/N$ expansion of the field theory corresponds to the genus expansion of the string theory: in the $N\to \infty$ limit one recovers classical string theory.
\\One can also assume $\lambda$ to be large. (\ref{gstringl}) and (\ref{thooft}) imply that in this limit one sends the string length to zero, yielding classical supergravity on the gravity side of the correspondence. This ``strong/weak coupling duality'' \cite{zaanen, skenderis} comes in handy because it allows us to study strongly coupled conformal field theories in $d$ dimensions by solving the classical Einstein's equations of motion in $d+1$ dimensions. 
\bigskip
\\In this thesis we are going to make use of gauge-gravity duality to study the hydrodynamics of superfluids. Now, a hydrodynamic theory is the effective description at long distances and late times of the universal behaviour at finite temperature of some underlying interacting theory. So far, however, we have not yet mentioned how one can obtain a thermal theory from a gravitational one. As we are going to see, the main idea will be to consider a black hole solution of the Einstein equations of motion. One can build a thermodynamic theory of black holes; most notably, their entropy is proportional to their area, and one can associate to them a temperature depending on the horizon behaviour of the metric. This will allow us to construct a finite temperature theory on the asymptotically AdS boundary of such a space.
\\We will do so by establishing a dictionary relating the fields and the propagators on both sides of the correspondence. In particular, in this thesis we will consider a gravitational theory coupled to a gauge field and a complex scalar field undergoing spontaneous symmetry breaking. The metric fluctuations will be dual to those of the stress-energy tensor and the gauge field to the current. 

\section{Prerequisites}
\subsection{Conformal symmetry}
\textit{Note: this section closely follows some parts of the introduction to conformal symmetry in Erdmenger and Ammon's book \cite{erdmenger}}.
\bigskip
\\Conformal field theories (CFT's) are invariant under conformal symmetry. The conformal symmetry group contains the transformations that locally preserve angles: dilatations and special conformal transformations. Such transformations leave the metric invariant up to an arbitrary (positive) scale factor:
\be \label{conftraf}
g_{\mu \nu}(x) \mapsto \Omega(x)^{-2}g_{\mu \nu}(x) \equiv e^{2\sigma(x)}g_{\mu \nu}(x)
\ee
\\Infinitesimal metric transformations are generated by a vector $\xi^\mu$:
\be
g_{\mu \nu} \to g_{\mu \nu} + \nabla_{(\mu}\xi_{\nu)}
\ee
A \textit{Killing vector} leaves the metric unaltered. One can show \cite{erdmenger} that in $d$-dimensional flat spacetime $g_{\mu \nu} = \eta_{\mu \nu}$ any conformal vector $\xi_\mu$ satisfies the  \textit{conformal Killing equation}:
\be \label{confkill}
\left( \eta_{\mu \nu} \partial_\rho \partial^\rho + (d-2)\, \partial_\mu \partial_\nu \right) \partial_\rho \xi^\rho = 0
\ee
In $d>2$ dimensions (in this thesis we will mainly work with conformal field theories in 3 spacetime dimensions) the most general solution to (\ref{confkill}) may be parametrized as:
\be
\xi^\mu(x) = a^\mu + \omega^\mu_\nu x^\nu + \lambda x^\mu + b^\mu x^2 - 2(b\cdot x)x^\mu
\ee
From this expression we can extract the generators of the conformal algebra in $d>2$ dimensions and read off the generators:
\begin{itemize}
\item  $a^\mu \Rightarrow$ translation: $\sigma(x)=0$. Generator: $P_\mu=-i\dmu$
\item  $\omega^\mu_\nu x^\nu \Rightarrow$ Lorentz transformation: $\sigma(x)=0$. Note that $\omega_{\mu \nu}=-\omega_{\nu \mu}$. 
\\Generator: $J_{\mu\nu} = i(x_\mu \dnu -x_\nu \dmu)$
\item  $\lambda x^\mu \Rightarrow$ dilatation:. $\sigma(x)=\lambda$. Generator: $D=-ix^\mu \dmu$
\item  $ b^\mu x^2 - 2(b\cdot x)x^\mu \Rightarrow$ special conformal transformation: $\sigma(x)=-2(b\cdot x)$. \\Generator: $K_\mu = i(x^2\dmu -2x_\mu x^\nu \dnu)$
\end{itemize}
$a^\mu$ and $b^\mu$ have $d$ independent components, while $\omega_{\mu \nu}$, being antisymmetric, has $d(d-1)/2$. So, together with $\lambda$, we are left with $(d+1)(d+2)/2$ components. Notice that this is the dimension of $SO(2,d)$, which gives a hint at the AdS-CFT correspondence. 
\bigskip
\\The only non-vanishing commutation relations between the generators are the following:
\be
[J_{\mu \nu},K_\rho] = i\left(\eta_{\mu \rho}K_\nu -\eta_{\nu \rho}K_\mu  \right)   \qquad  [P_\nu, K_\mu]=2i\left(\eta_{\mu \nu}D-J_{\mu \nu}\right)
\ee
\be
 [D,P_\mu] = iP_\mu \qquad [D,K_\mu] = -iK_\mu 
\ee
With these commutators one can construct the representations of the conformal algebra. A field $\phi(x)$ has \textit{scaling dimension} $\Delta$ if:
\be
[D,\phi(0)]= -i\Delta \phi(0) \, ,
\ee
so that under a rescaling $x \mapsto \lambda x$: 
\be
\phi(x) \mapsto \phi'(x')= \lambda^{-\Delta}\phi(x)
\ee
But then, using the commutation relations, it follows that:
\be
[D,P_\mu \, \phi(0)]= -i(\Delta-1) \phi(0) \qquad \text{and}\qquad   [D,K_\mu\, \phi(0)]= -i(\Delta+1) \phi(0)
\ee
Thus one can define \textit{primary operators}, i.e. operators that are annihilated by $K_\mu$, and construct a tower of operators of higher dimension (called \textit{conformal descendants}) by using $P_\mu$ as a creation operator.
\\The conformal dimension $\Delta$ is subject to unitarity bounds. In particular, for a scalar operator one can show that $\Delta \geq (d-2)/2$ \cite{erdmenger}.
\bigskip
\\Conformal field theories have many interesting properties which we are not going to examine here. For instance, two- and three-point correlation functions of primary operators are determined by the scaling dimensions of the fields. In particular, for a primary operator of dimension $\Delta$, one has:
\be
\langle \mathcal{O}(x_1)\mathcal{O}(x_2) \rangle = (x_1-x_2)^{-2\Delta}
\ee
Last but not least, an important feature of conformally invariant theories is that the trace of the stress-energy tensor vanishes. This can be seen as follows: the stress-energy tensor is given by the variation of the action with respect to the metric. But then (\ref{conftraf}) gives:
\begin{equation}
\delta S = -\frac{\sqrt{-g}}{2}\delta g^{\mu \nu}T_{\mu \nu} \propto \Omega^{-2} g^{\mu \nu}T_{\mu \nu} \propto T_{\mu}^\mu
\end{equation}
So in order for the theory to be conformally invariant we need $T_{\mu}^{\mu} = 0$.
\subsection{AdS spaces}
\textit{In this section we are going to follow the standard introduction to AdS spaces given in textbooks such as \cite{zaanen, erdmenger, natsuume}.}
\bigskip
\\AdS spaces are maximally symmetric and feature a negative curvature. In particular, in $d+1$ dimensions they possess $(d+1)(d+2)/2$ Killing vectors (as many as the number of independent components of the metric). Notice that the the scalar curvature has to be constant owing to the space being maximally symmetric.
\bigskip
\\A $(d+1)$-dimensional AdS space can be embedded into $(d+2)$-dimensional Minkowski space with metric $\tilde{\eta}=\text{diag}(-,+,\dots,+,-)$ by imposing the hyperboloid equation ($L$ is called the \textit{AdS radius}):
\be \label{adseq}
-X_0^2 + \sum_{i=1}^d X_i^2 -X_{d+1}^2  =-L^2
\ee
Note that the hyperboloid defined by this equations possesses the isometry group $SO(d,2)$.
\\The solution to (\ref{adseq}) can be parametrized as \cite{erdmenger}:
\be
(X^0,X^{i},X^{d+1})= L\, (\cosh \rho  \cos \tau,\,\Omega_i\, \sin \rho  \cos\tau,\,\cosh \rho \sin\tau) \qquad i=1,\dots,d   \\
\ee
The $\Omega_i$ coordinates parametrize the $S^{d-1}$ sphere: $\sum_{i=1}^d \Omega_i^2=1$. Then the metric takes the following form:
\be
\diff s^2 = \frac{L^2}{\cos^2\theta}\left( -\diff \tau^2 + \diff \theta^2 + \sin^2\theta \diff \Omega_{d-1}\right) 
\ee
where we defined a new coordinate $\theta$ by $\tan \theta = \sinh \rho$. Note that $\theta \in [0,\pi/2]$, so in this way we only cover half of the space. In particular, the \textit{AdS boundary} is located at $\theta=\pi/2$.
\\Another possible parametrization reads ($t\in \mathbb{R}$, $r\in \mathbb{R}_+$, $\mathbf{x}\in \mathbb{R}^{d-1}$):
\be
X^0 = \frac{L^2}{2r} + \frac{r}{2L^2}(\mathbf{x}^2-t^2+L^2)\,\quad X^i= \frac{rx^i}{L}\,\quad X^d = \frac{L^2}{2r} + \frac{r}{2L^2}(\mathbf{x}^2-t^2-L^2) \,\quad X^{d+1} = \frac{rt}{L}
\ee
These coordinates are called the \textit{Poincar\'e patch}. Notice that, since $r$ is positive, they also only cover half of the space. In this case the metric becomes ($\eta_{\mu \nu}$ is the Minkowski metric):
\be \label{ppatch}
\diff s^2 = \frac{L^2}{r^2}\diff r^2 +\frac{r^2}{L^2} \eta_{\mu \nu}\, \diff x^\mu \diff x^\nu
\ee
The AdS boundary is located at $r\to \infty$: in this limit one can rewrite the defining equation (\ref{adseq}) as $-X_0^2 + \sum_{i=1}^d X_i^2 -X_{d+1}^2  =0$.
\\As we are going to do in the rest of this thesis, with a simple coordinate transformation $u \equiv 1/r$ we obtain:
\be \label{ppatchu}
\diff s^2 = \frac{L^2}{u^2}\left( \diff u^2+ \eta_{\mu \nu}\, \diff x^\mu \diff x^\nu \right)
\ee
In this case the AdS boundary is obviously located at $u \to 0$. 
\bigskip
\\Notice \cite{zaanen} that this metric is invariant under rescalings $(u,x^\mu)\to \lambda (u,x^\mu)$, and under the transformation $(u,x^\mu)\to (u,x^\mu)(u^2+\eta_{\mu \nu}x^\mu x^\nu)$. In the $u\to 0$ limit this last transformation becomes inversion symmetry. 
\\Now recall that the generating algebra of the conformal group consists of the generator of dilatations $D$, those of translations $P_i$ and those of special conformal transformations $K_i$. Notably, one can write $K_i = IP_iI$, where $I$ is an inversion: $I: x^\mu \to x^\mu/x^2$.
Therefore we see that the isometry group $SO(d,2)$ acts on the boundary as the conformal group in $d$ dimensions.
\\Furthermore, note that the continuation of the metric (\ref{ppatchu}) to the boundary $u=0$ is realized by multiplying it with a \textit{defining function} which has a second order zero at $u=0$ \cite{erdmenger}. The choice of the defining function is not univocal: different defining functions describe boundary metrics related to each other by conformal transformations. The equivalence class of such boundary metrics is called \textit{conformal structure}.
\bigskip
\\The AdS$_{d+1}$ scalar curvature amounts to
\be \label{ricciads}
R = -\frac{d(d+1)}{L^2}
\ee
It is then easy to show that the AdS metric satisfies the Einstein equations of motion with the following cosmological constant:
\be
\Lambda = -\frac{d(d-1)}{2L^2}
\ee
\\A final note: all theories we consider in this thesis feature gauge symmetries: they are invariant under diffeomorphisms and gauge transformations (when adding scalar and gauge fields to the picture). In particular, we are going to use $U(1)$ gauge transformations, parametrized by a scalar field, and diffeomorphisms whose infinitesimal version is obtained with a Lie derivative with respect to a vector field.

\section{The GKPW prescription}
The AdS/CFT conjecture can be expressed as an equality between the generating functional of the conformal field theory on one side and that of the dual string theory on the other side. This induces a mapping, the so-called \textit{field-operator correspondence}, between the fields in the gravity theory and the operators of the CFT.
\bigskip
\\Let us first consider the simple example of a massless scalar field $\phi$ in the bulk. Let $\phi_0$ its restriction on the AdS boundary and $\mathcal{O}$ the conformal field to which it couples. The Gubser-Klepanov-Polyakov-Witten (GKPW) prescription links the partition function of the CFT to the Euclidean gravitational action of the bulk theory: \cite{witten1998, mcgreevy, erdmenger, Son:2002sd} 
\begin{equation}\label{GKPWscalar}
\left \langle e^{-\int_{\text{Bdy}} \phi_0 \mathcal{O}} \right\rangle_{\text{CFT}_d} = \mathcal{Z}_{\text{strings in AdS}_{d+1}}[\phi_0]
\end{equation}
Note that for the right-hand side to be well-defined the boundary field $\phi_0$ has to uniquely determine the bulk field $\phi$. This is a known property of AdS$_{d+1}$: any square-integrable solution to the scalar equation of motions is automatically the trivial solution, so that the boundary condition at infinity $\phi_0$ fixes $\phi$ in the bulk \cite{witten1998}. In other words, the imposition of regularity in the bulk is the second boundary condition that fully determines the solution \cite{sonnersolvay}. (This is strictly the case in Euclidean signature; in Lorentzian signature there will be two independent solutions, as we are going to see at the end of this chapter).
\bigskip
\\The generalization for different types of fields is straightforward. For instance a gauge field $A^M$ with boundary restriction $A_{0}^\mu$ on the boundary coupling to a current operator will produce an action $\mathcal{S}=\int A_0^\mu J_\mu$, so the equivalent prescription to (\ref{GKPWscalar}) will be \cite{witten1998}: 
\begin{equation}\label{GKPWgauge}
\left \langle e^{-\int_{\text{Bdy}}A_0^\mu J_\mu} \right\rangle_{\text{CFT}_d} = \mathcal{Z}_{\text{strings in AdS}_{d+1}}[A_0]
\end{equation}
We see that changing the boundary conditions on the gravity side amounts to modifying the Lagrangian of the dual conformal field theory. Thus the choice of the boundary conditions has direct physical implications: tuning the leading behaviour of the bulk fields is equivalent to adjusting the sources of the boundary theory. Moreover, imposing the correct boundary conditions on the gravity side guarantees the realization of symmetries in the conformal field theory - for instance gauge invariant boundary conditions ensure that the CFT propagators respect the Ward identities. 
\bigskip
\\In the classical limit one can take a saddle point approximation to the gravitational path integral, yielding
\begin{equation}
e^{-W_{\text{CFT}_d}}= e^{-S_\text{bulk}}
\end{equation}
In practice this makes it possible to compute $n$-point Euclidean correlation functions \cite{skenderis} by computing the functional derivative of the action with respect to the sources of the fields. For example the $n$-point function of a scalar field is given by: 
\be \label{npointholo}
\left \langle \mathcal{O}(x_1) \cdots \mathcal{O}(x_n) \right \rangle = (-1)^{n+1} \frac{\delta \mathcal{S}}{\delta \phi_0(x_1)\cdots \delta \phi_0(x_n)}\bigg|_{\phi_0 = 0}
\ee
\section{Holographic thermodynamics}
Having presented the main ideas behind the AdS-CFT correspondence we can now introduce some notions on holographic thermodynamics, which are necessary in order to understand the emergence of a thermal theory on the boundary. Such fundamental concepts are the Hawking temperature, the entropy and the free energy of the black hole. We are going to take a look at the Schwarzschild and Reissner-Nordstr\"om black holes as specific examples. 
\bigskip
\\We denote our coordinates as $x^M =(u,x^\mu)$, where $\mu$ runs over $d$ indices.

\subsection{The Hawking temperature}
The temperature of the black hole is identified with the \textit{Hawking temperature}, which is computed by imposing periodicity of the Wick-rotated metric at the horizon. Here we are following \cite{zaanen, erdmenger, Ghosh:1995rv}
\\The metric of a static black hole with spherical symmetry takes the following form:
\begin{equation}\label{bhmetric}
\diff s^2 = -g_{tt}(r) \, \diff t^2 + \frac{1}{g^{rr}(r)} \, \diff r^2 + r^2\diff \Omega^2_{d-1}
\end{equation}
A Wick rotation to Euclidean space yields:
\begin{equation}
\diff s_E^2 = g_{tt}(r) \, \diff \tau^2 + \frac{1}{g^{rr}(r)}  \, \diff r^2 + r^2\diff \Omega^2_{d-1}
\end{equation}
At the horizon radius $r_h$ both and $g_{tt}$ and $g_{tt}$ vanish, so that for $r \approx r_h$ one can Taylor expand:
\begin{equation}
\diff s_E^2 \approx g_{tt}'(r)(r-r_h) \, \diff \tau^2 + \frac{1}{{g^{rr}}'(r_h)(r-r_h)}  \, \diff r^2 + r^2\diff \Omega^2_{d-1}
\end{equation}
We define the ``polar coordinates'' \cite{erdmenger}
\begin{equation}
\rho^2 \equiv \frac{4(r-r_h)}{{g^{rr}}'(r_h)} \qquad \qquad \phi \equiv \frac{1}{2}\sqrt{{g^{rr}}'(r_h){g_{tt}}'(r_h)} \, \tau
\end{equation}
Then
\begin{equation}
\diff \rho^2 = \frac{\diff r^2}{{g^{rr}}'(r_h)(r-r_h)} \qquad \qquad \diff \phi^2 =  \frac{1}{4}{g^{rr}}'(r_h){g_{tt}}'(r_h) \, \diff \tau^2
\end{equation}
And in the near-horizon region we can write the metric as
\begin{equation}
\diff s_E^2 \approx \rho^2 \, \diff \phi^2 +  \diff \rho^2 + \cdots
\end{equation}
We see that we can interpret $\phi$ as an angular coordinate and $\rho$ as a radial coordinate. In particular, in order to avoid conical singularities, $\phi$ has to be $2\pi$-periodic:
\begin{equation}
\phi \sim \phi + 2 \pi
\end{equation}
Converting back to the original coordinates this gives:
\begin{equation}
\tau \sim \tau + \frac{4\pi}{\sqrt{{g^{rr}}'(r_h){g_{tt}}'(r_h)}}
\end{equation}
Knowing that the periodicity of imaginary time corresponds to the inverse temperature, we finally identify the temperature as:
\begin{equation}\label{Tss}
T_H = \frac{\sqrt{{g^{rr}}'(r_h){g_{tt}}'(r_h)}}{4\pi}
\end{equation}
Note that the Hawking temperature of a (static) black hole was historically identified with its surface gravity \cite{wald, carroll}.
\bigskip
\\Let us consider an asymptotically AdS Schwarzschild black hole in $d+1$ dimensions with a planar horizon as an example. In this case the metric is given by (\ref{bhmetric}) with ($L$ being the AdS radius) 
\begin{equation}
\diff s^2 = -\frac{r^2}{L^2}f(r)\, \diff t^2 + \frac{L^2}{r^2f(r)}\, \diff r^2 + \diff \Omega^2
\end{equation}
with 
\begin{equation}
f(r) = 1- \frac{m}{r^d} 
\end{equation}
and  $\diff \Omega^2$ is the angular part of the metric. The horizon radius $r_h$ satisfies $f(r_h)=0$ and therefore:
\begin{equation}
m = r_h^d \, \text{,}
\end{equation}
so that one finds:
\begin{equation} \label{Tschwarzschild}
T = \frac{r_h^2}{4\pi L^2}\left( \frac{d\,m }{r_h^{d+1}}\right) = \frac{d\, r_h}{4\pi L^2}
\end{equation} 
\subsection{Black hole thermodynamics}
One of the motivations of the AdS-CFT correspondence is the realization that the Bekenstein-Hawking entropy of black holes is proportional to the area of their event horizon:
\begin{equation}
S = \frac{A}{4G} 
\end{equation}
As we are going to see this identity motivates the intuition that the thermal state of the field theory is dual to a black hole in the bulk. It is instructive to see how this formula comes about, starting from thermodynamic identities \cite{erdmenger, zaanen, natsuume}.
\bigskip
\\Let us start with the way thermodynamic entropy is computed, starting from the free energy:
\begin{equation}
S = - \frac{\partial F}{\partial T} = \beta^2 \,\frac{\partial F}{\partial \beta} = \left(\beta \frac{\partial}{\partial \beta} -1 \right)  \beta F
\end{equation}
We thus need an expression for the free energy. It is given in terms of the thermodynamic partition function $\mathcal{Z}$: 
\begin{equation}
F = - \frac{1}{\beta} \ln \mathcal{Z}(\beta)
\end{equation}
Recall the GKPW formula (\ref{GKPWscalar}), linking the gravity partition function to that of the field theory. In general, the canonical partition function for gravity at temperature $T=1/\beta$ reads
\begin{equation}
\mathcal{Z}(\beta) = \int \mathcal{D}g \, e^{-\mathcal{S}[g]} \simeq e^{-\mathcal{S}^*} \, ,
\end{equation}
where in the classical limit in Euclidean signature we substituted the path integral with its saddle point approximation ($\mathcal{S}^*$ being the on-shell gravitational action).
Then one can compute the free energy as:
\begin{equation}\label{freeen}
F \simeq \frac{\mathcal{S}^*}{\beta}
\end{equation}
Formally, one integrates over all metric with appropriate falloff; however, the partition function can be approximately computed with a saddle point approximation: $\mathcal{S}^*=\mathcal{S}(g^*)$, where $g^*$ is the metric solves the Einstein equations of motion.
\bigskip
\\For the purpose of this chapter the action $\mathcal{S}[g]$ will be the Einstein-Hilbert (EH) action for classical gravity in Euclidean signature:
\begin{equation}\label{EHaction}
\mathcal{S}_{\text{EH}}= - \frac{1}{16\pi G}  \int_{\mathcal{M}} \diff^{d+1} x \, \sqrt{|g|}\, (R-2\Lambda) 
\end{equation}
where $R$ is the scalar Ricci curvature, $\Lambda$ the cosmological constant and $\mathcal{M}$ the manifold over which the theory is defined. This action, however, does not yield a well-defined variational problem for the fluctuations of the bulk metric. In fact, after integration by parts it still carries terms proportional to the derivatives of the metric, which are incompatible with Dirichlet boundary conditions. These terms can by canceled by supplementing the Einstein-Hilbert action with the so-called Gibbons-Hawking-York boundary term \cite{gibbonshawking}: 
\begin{equation}
\mathcal{S}=\mathcal{S}_{\text{EH}} +  \mathcal{S}_{\text{GBY}} = - \frac{1}{16\pi G}  \int_{\mathcal{M}} \diff^{d+1} x \, \sqrt{|g|}\, (R-2\Lambda) - \frac{1}{8\pi G}  \int_{\partial \mathcal{M}} \diff^{d} x \, \sqrt{-\gamma}\, K 
\end{equation}
where $\gamma$ is the induced metric at the boundary and $K = -\gamma^{\mu \nu} \Gamma^{\alpha}_{\mu \nu} n_\alpha$ is the trace of the extrinsic curvature ($n^\mu$ being the normal vector to $\partial \mathcal{M}$). 
\\Even after adding $\mathcal{S}_{\text{GHY}}$ the action still diverges if the boundary is located at infinity, as is the case for (asymptotically) AdS spaces. As we are going to see, one adds counterterms absorbing the divergences of the action, or subtract the action of a reference background. 
\subsubsection{Example: Schwarzschild black hole}
Let us see how this works concretely for a Schwarzschild AdS$_{d+1}$ geometry \cite{zaanen}. At the boundary $r\to \infty$ the geometry becomes AdS and so the scalar curvature is given by (\ref{ricciads}):
\begin{equation}\label{ricci}
\lim_{r\to \infty}R = -\frac{d(d+1)}{L^2}
\end{equation}
While the cosmological constant appearing in the EH action (\ref{EHaction}) is 
\begin{equation}
\Lambda = -\frac{d(d-1)}{2L^2}
\end{equation}
Wick rotating the metric ($\tau = it$) we obtain
\begin{equation}\label{wickadsschw}
\diff s_E^2 = \frac{r^2}{L^2} \left(f(r)\diff \tau^2 +  \diff \mathbf{x}^2\right) + \frac{L^2}{r^2f(r)} \diff r^2
\end{equation}
With $\lim_{r \to \infty} f(r)=1$. The determinant of the metric reads: 
\begin{equation}
\sqrt{g_E} = \left( \frac{r}{L}\right)^{d-1}
\end{equation}
We now have to determine the normal vector at the boundary, defined by sending $r \to \infty$. It points in radial direction; since it has to be normed, it is given by
\begin{equation}
n^\mu = \frac{r\sqrt{f(r)}}{L} \, \delta^\mu_r
\end{equation}
The induced metric $\gamma_{\alpha \beta}$ is the simply the original metric, minus the radial component, so that its determinant amounts to:
\begin{equation}
\sqrt{|\gamma_E|} = \left( \frac{r}{L}\right)^d \sqrt{f(r)}  \, \xrightarrow[r\to \infty] \, \left(\frac{r}{L} \right)^d
\end{equation}
One can now compute the trace of the extrinsic curvature $K= \gamma_E^{\alpha \beta} \nabla_\alpha n_\beta=- \gamma_E^{\alpha \beta}\,\Gamma^{\mu}_{\alpha \beta}\,n_\mu$. Since $n^\mu$ points in radial direction and $\gamma_{\alpha \beta}$ has no radial component we only need $\Gamma^{r}_{\mu \nu}$ with $\mu, \nu \neq r$. The only non-vanishing Christoffel symbols of interest are then (denoting with $x^i$ all coordinates different from $r$ and $\tau$):
\begin{align}
\begin{split}
 \Gamma^{r}_{x^i x^i}& = -\frac{r^3f(r)}{L^4} \qquad \qquad  \Gamma^{r}_{\tau \tau} = -\frac{r^3f(r)\big(2f(r)+rf'(r)\big)}{2L^4} \\
\end{split}
\end{align}
Therefore one obtains:
\begin{equation}
K = \frac{\sqrt{f(r)}}{L} \left( d + \frac{rf'(r)}{2f(r)}\right)  \, \xrightarrow[r\to \infty] \, \frac{d}{L} \
\end{equation}
We can now compute the counterterm, at leading order in $1/r$. For this we only need the asymptotic value of the Ricci scalar curvature (\ref{ricci}). Integrating the EH action over $\diff r$ we find (to leading order in $r$):
\begin{equation}
\mathcal{S}_{\text{bdy}} = -\frac{d-1}{8 \pi G} \int_{\partial \mathcal{M}} \diff \tau \diff \mathbf{x} \, \frac{r^d}{L^{d+1}} = (\text{finite terms}) -\lim_{\epsilon \to 0} \, \frac{d-1}{8\pi G} \int^{\beta}_0 \diff \tau \int_{\mathbb{R}^{d-1}} \diff \mathbf{x}\, \frac{\epsilon^{-d}}{L^{d+1}} \quad \text{,}
\end{equation}
where we have recalled that the limits of integration for Euclidean time are $0$ and $T=1/\beta$ and we have parametrized the UV boundary by setting $r=\epsilon^{-1}$, with $\epsilon \to 0$. We can absorb this divergence by adding a corresponding counterterm to the boundary action\footnote{Note that for $d>2$ this is not the only needed correction. Further corrections come in powers of the curvature of the boundary metric \cite{haroskend, bianchirenorm}. For the purpose of illustrating the computation of the entropy, though, in this chapter we are restricting ourselves to the leading term.} 
\begin{equation}
\mathcal{S}_{\text{ct}} \equiv \frac{d-1}{8\pi G L} \int_{r \to 0}  \diff \tau \diff \mathbf{x} \, \sqrt{|\gamma_E|}
\end{equation}
With this counterterm we can finally perform the integration; recall that the lower integration limit on $r$ is the horizon radius $r_h$. We get \cite{zaanen}: 
\begin{equation}
\mathcal{S}^* = -\frac{1}{4G}\left(\frac{r_h}{L}\right)^{d-1} \text{Vol}_{d-1}=- \frac{1}{4G}\frac{(4\pi)^{d-1}L^{d-1}T^{d-1}}{d^{d}} \text{Vol}_{d-1}\quad \text{,}
\end{equation}
where $ \text{Vol}_{d-1}$ is the volume enclosed by the $d-1$ spatial coordinates $\mathbf{x}$.
\\So, with (\ref{freeen}) and (\ref{Tss}), the saddle point approximation for the free energy is: 
\begin{equation}
F \simeq T \mathcal{S}^* =  -\frac{T^d}{4Gd}\left(\frac{4\pi L}{d}\right)^{d-1}\text{Vol}_{d-1}
\end{equation}
We can then compute the entropy:
\begin{equation}
S = -\frac{\partial F}{\partial T} = \frac{1}{4G}\text{Vol}_{d-1} \left( \frac{4\pi L T}{d}\right)^{d-1} = \frac{1}{4G}\left( \frac{r_h}{L}\right)^{d-1} \text{Vol}_{d-1}
\end{equation}
With (\ref{wickadsschw}) we recognize that we can rewrite this last expression in terms of the horizon area:
\begin{equation}
S = \frac{A_{\text{horizon}}}{4G}
\end{equation}
This is the famous result predicted by Bekenstein and Hawking \cite{bekenstein}. 
\subsubsection{Example: Reissner-Nordstr\"om black hole}
Note: from now on we are going to set $16\pi G=L=1$ for simplicity's sake. 
\bigskip
\\By coupling a gauge field to the metric one obtains the so-called \textit{Reissner-Nordstr\"om} (RN) black hole. In an asymptotically AdS$_{d+1}$ geometry its bulk action is given by the Einstein-Maxwell action, which  (in Lorentzian signature) reads:
\be
\mathcal{S}_{\text{EM}} = \int \diff^{d+1}x \sqrt{-g} \, \left[ \big(R + d(d-1) \big)-\frac{1}{4e^2}F_{MN}F^{MN}\right]
\ee
where $F_{MN} = \partial_M A_N -\partial_N A_M$ is the field strength tensor associated to the gauge field. 
\\Assuming the gauge field to be timelike (that is, only $A_t \neq 0$), the equations of motion obtained by extremizing this action are solved by \cite{zaanen}:
\begin{equation}\label{rnmetric}
\diff s^2 = \frac{r^2}{L^2} \left(-f(r)\diff t^2 +  \diff \mathbf{x}^2\right) + \frac{L^2}{r^2f(r)} \diff r^2 \qquad \text{with} \qquad f(r)=1+\frac{Q^2}{r^{2d-2}} - \frac{M}{r^d}
\end{equation}
\begin{equation}\label{rngauge}
A_t = \mu \left( 1 -\frac{r_h^{d-2}}{r^{d-2}}\right)
\end{equation}
The chemical potential $\mu$ and $Q$ are related by:
\be
Q = \sqrt{\frac{(d-2)}{2(d-1)e^2}}r_h^{d-2} \mu
\ee
Note that in the limit $e \to \infty$ we recover the Schwarzschild black hole solution. This is the so-called ``probe limit''. $r_h$ denotes the larger root of $f(r)$ and is called \textit{outer horizon radius}. It is related to the charge $Q$ and the the mass $M$ of the black hole by:
\be \label{Mrn}
M = r_h^d + Q^2 \, r_h^{2-d}
\ee
The temperature reads:
\be \label{Trn}
T = \frac{d\, r_h}{4\pi} - \frac{(d-2)^2\, \mu^2}{8\pi(d-1)\,e^2\,r_h}
\ee
With a similar procedure to the Schwarzschild case one can compute the free energy:
\be \label{Frn}
F = -T \ln \mathcal{Z} \approx T \mathcal{S}^* = - \frac{V_{d-1}}{d}\left( 4\pi  r_h^{d-1}T + \frac{(d-2)\mu^2}{e^2} r_h^{d-2} \right)
\ee
We can then compute the entropy density and verify the Bekenstein-Hawking conjecture:
\be
s\equiv -\frac{\partial F}{\partial T}\bigg|_\mu = 4\pi r_h ^{d-1}
\ee
Note that this solution displays finite entropy at zero temperature.
\\Similarly, the charge density associated with the conserved current of the boundary field theory is given by: 
\be \label{rncharge}
n\equiv -\frac{\partial F}{\partial \mu}\bigg|_T = \frac{d-2}{e^2}  r_h^{d-2}\mu
\ee
Note that the charge density obtained in this way is proportional to $\partial_r A_t$ on the boundary, as prescribed by the holographic dictionary that we are going to introduce in the upcoming sections.
\bigskip
\\Having obtained analytical expressions for both $s$ and $\rho$, together with their sources $T$ and $\mu$, we can now compute the related susceptibilities. Recall the thermodynamical definition of susceptibilities: given an ensemble $\phi_a$, $a=1,\cdots, n$ and its sources $s_a$ the susceptibility $\chi_{ab}$ is the variation of the vacuum expectation value $\phi_a$ with respect to the source of $\phi_b$, keeping all other sources constant:
\be
\chi_{ab} = \frac{\partial \phi_a}{\partial s_b}\bigg|_{s_{c\neq b}}
\ee
Here we find:
\be
\chi_{nn} = \frac{r_h^{d-2}}{e^2}\frac{\left(2d(d-2)(d-1)\,e^2r_h^2+(d-2)^3(2d-3)\,\mu^2\right)}{2d(d-1)\,e^2r_h^2 + (d-2)^2\, \mu^2}
\ee
\be
\chi_{ss}=  \frac{32(d-1)^2\,\pi e^2 r_h^d}{2d(d-1)\,e^2r_h^2 + (d-2)^2\,\mu^2}
\ee
\be
\chi_{ns}=\chi_{sn} = \frac{8(d-2)^2(d-1)\,\pi \mu r_h^{d-1}}{2d(d-1)\,e^2r_h^2 + (d-2)^2\, \mu^2}
\ee
One can see that the susceptibility matrix is symmetric: $\chi_{s n}=\chi_{n s}$, and that the diagonal susceptibilities $\chi_{nn}$ and $\chi_{ss}$ are positive, as required by thermodynamic equilibrium. 
\bigskip
\\The energy density and the pressure can be extracted from the boundary behaviour of the metric, which is dual to the stress-energy tensor of the field theory. In Fefferman-Graham coordinates, where $g_{rm}=r^{-2}\delta_{rm}$, the subleading behaviour of the metric at the boundary gives the expectation value of the stress-energy tensor in the dual field theory \cite{haroskend}. 
\\In particular, as we are going to see in the upcoming chapter, the boundary behaviour of the metric in $d+1$ dimensions is\footnote{If $d$ is odd logarithmic terms appear in the expansion \cite{Skenderis:2002wp}}:
\be
\lim_{r\to \infty} = r^2\left[ g_{(0)\mu \nu}\left(1+\mathcal{O}(r^{-2})\right) + r^{-d}g_{(d)\mu \nu}\left(1+\mathcal{O}(r^{-2})\right) \right]
\ee
For example for a Reissner-Nordstr\"om metric we find\cite{zaanen}:
\begin{align}
\epsilon&= d\langle T^{00}\rangle =d\, g_{(d)}^{tt} = dM \\
p&= d\langle T^{xx}\rangle =M
\end{align}
Notice that $\langle T^\mu_\mu\rangle=0$, as required for a conformal field theory. 
\\Finally, with equations (\ref{Mrn}, \ref{Trn}, \ref{Frn}, \ref{rncharge}) one can then check the Smarr law:
\be
\epsilon+ p  = sT + n \mu 
\ee

\section{Definition of propagators and quasinormal modes}
From now on we are going to set $L=16\pi G=1$. Moreover, for the sake of consistency with the upcoming chapters, we are going to use the coordinate $u\equiv 1/r$, so that the AdS boundary is located at $u\to 0$.   Finally, we assume to be working in AdS$_{d+1}$ with $d$ odd, as for even $d$ logarithmic terms arise in the boundary expansion of the fields \cite{Skenderis:2002wp}.

\subsection{Asymptotic behaviour of the fields}
In this section we examine the asymptotic behaviour of scalar and gauge field, assuming that the metric is asymptotically AdS (\ref{ppatchu}):
\begin{equation}\label{ads}
\lim_{u\to 0} \diff s^2 = \frac{1}{u^2}\left( \diff u^2+ \eta_{\mu \nu}\, \diff x^\mu \diff x^\nu \right)
\end{equation}
So that the square root of the metric asymptotes to:
\begin{equation}
\lim_{u\to 0}  \sqrt{-g} = u^{-(d+1)}
\end{equation}
Let us start with the simple example of a scalar field $\phi$ whose bulk action reads \cite{mcgreevy}:
\begin{equation}
\mathcal{S}_\phi = \int \diff^{d+1} x\, \sqrt{-g} \left( m^2\phi^2 + \partial_M \phi \partial^M \phi\right)
\end{equation}
The resulting equation of motion is \cite{Herzog:2010vz}: 
\begin{equation}\label{eqscalarfield}
\frac{1}{\sqrt{-g}}\partial_M \left( \sqrt{-g} \partial^M \right)\phi = m^2 \phi
\end{equation}
We assume radial symmetry (that is, the fields only depend on $u$). On the boundary equation (\ref{eqscalarfield}) then becomes:
\begin{equation}
\left(u^{d+1}\partial_u u^{1-d} \partial_u - m^2\right)\phi =u^2 \partial_u^2\phi + (1-d)u\,\partial_u \phi  - m^2\phi=0
\end{equation}
Assuming an asymptotic behaviour $\phi \sim u^{\Delta}$ we obtain an equation for the falloff exponent $\Delta$:
\begin{equation}
\Delta(\Delta-d)=m^2
\end{equation}
which is solved by
\begin{equation}\label{deltascalar} 
\Delta =\Delta_\pm \equiv\frac{1}{2}\left(  d \pm \sqrt{d^2+4m^2}\right) \, \text{,}
\end{equation}
so that we can write:
\be
\phi = \phi_+ u^{\Delta_+}+ \phi_- u^{\Delta_-}
\ee
for two coefficients $\phi_+$ and $\phi_-$, which in general are functions of the non-radial coordinates $x^\mu$.
\\Notice that under dilatations $u \mapsto \lambda u$ the scalar field is invariant. Hence the coefficients transform as:
\be
\phi_\pm \mapsto \lambda^{-\Delta_\pm} \phi_\pm
\ee
In particular, as we are going to see in the coming sections, $\phi_+$ gives the vacuum expectation value of the dual field theory operator, which has \textit{scaling dimension} $\Delta_+$ \cite{mcgreevy, zaanen}, determined by $m^2$.
\bigskip
\\One can see that a negative mass does not lead to an instability (that is, $\phi$ gets a complex scaling dimension) as long as:
\begin{equation} \label{BFbound}
m^2>m_{\text{BF}}^2 \equiv -\frac{d^2}{4}
\end{equation}
This is the so-called Breitenlohner-Freedman (BF) bound \cite{bf1,bf2}. 
\bigskip
\\Let us now add a gauge field to the picture. Considering a complex scalar together with a gauge field as in \cite{Hartnoll:2008kx} the simplest form the bulk part of the action is given by \cite{Herzog:2010vz,Bhattacharya:2011eea}:
\begin{equation}\label{action}
\mathcal{S}_{A_\mu, \psi} = \int \diff^{d+1} x\, \sqrt{-g} \left( \frac{1}{4e^2} F_{M N}F^{M N} +m^2|\psi|^2 +|\partial \psi- iA\psi|^2|\right)
\end{equation}
From this action one can derive the equations of motion for the scalar and the gauge field \cite{Herzog:2010vz}: 
\begin{equation}\label{complexscal}
\frac{1}{\sqrt{-g}}(\partial_M-iA_M) \sqrt{-g}(\partial^M \psi - iA^M\psi) = m^2 \psi^*
\end{equation}
\begin{equation}\label{maxwell}
\partial_M\left( \sqrt{-g}F^{M N}\right) =e^2 g^{NK}J_K
\end{equation}
With the field strength tensor $F_{MN}= \partial_M F_N - \partial_N F_M$ and the current
\begin{equation}
J_M = i\left( \psi^*(\partial_M - iA_M )\psi - \psi (\partial_M + iA_M)\psi^*\right) = 2A_M |\psi|^2 + i\left( \psi^* \partial_M \psi - \psi \partial_M \psi^*\right)
\end{equation}
We work in the so-called radial gauge, where $A^u=0$. Then the above equations of motion (\ref{complexscal},\ref{maxwell}) become:
\begin{equation}\label{eqforpsi}
u^2 \partial_u^2\psi - (d+1)u\,\partial_u \psi - A_\mu A^\mu \psi - m^2\psi^*= 0
\end{equation}
\begin{equation}\label{eqforA}
\partial_u (u^{3-d}\,\partial_u A_\mu)- 2u^2e^2  |\psi|^2A_\mu = 0
\end{equation}
Assuming the same asymptotic behaviour for $\psi$ and $\psi^*$: $\psi \sim \psi^* \sim u^{\Delta}$ with $\Delta$ as in (\ref{deltascalar}) we see that (\ref{eqforpsi}) is satisfied if the leading behaviour for the gauge field at infinity is just a constant. If we then consider the second equation (\ref{eqforA}) without the mixing term we see that it is indeed solved by 
\begin{equation}\label{asymptA}
A_\mu \sim \alpha  + \beta u^{d-2}
\end{equation}
We can now plug this expansion back into (\ref{eqforA}) and see that the mixing term is subleading in $u$ if $m^2<d+1$: 
\begin{equation}
u^2 |\psi|^2A_\mu \sim u^{2+d-\sqrt{d^2+4m^2}} 
\end{equation}
So for $m^2<0$, as will be the our case in the rest of this thesis, we can retain the asymptotic behaviours given by (\ref{deltascalar}) and (\ref{asymptA}). 
\\Also, note that a positive squared mass gives the scalar field a diverging boundary behaviour. Its dual will then be an irrelevant operator, which, if sourced (i.e. if it gets a non-vanishing leading coefficient),  alters the UV behaviour of the theory, driving the theory to a different UV fixed point and endangering the asymptotic ``AdS-ness'' of the metric \cite{mcgreevy}. 
\bigskip
\\As we are going to soon see in better detail, the subleading component of the fields at the boundary give the vacuum expectation values of the operators in the dual conformal field theory, while the leading components are their sources. For instance the timelike component $A_t$ of the gauge field delivers the conserved charge and its associated chemical potential in the field theory: 
\be
A_t \sim \mu -\rho \, u^{d-2}
\ee 
Notice that the Reissner-Nordstr\"om solution (\ref{rngauge}) for $A_t$ shows this behaviour, with $\rho = r_h^{d-2} \mu$.
\\Similarly, having a finite background superfluid velocity $\sim \xi$ means \cite{Herzog:2010vz}:
\be
A_x = \xi -\frac{e^2}{2}\big<J_x\big> \,u^{d-2} + \mathcal{O}(r^3)
\ee
\subsection{Quasinormal modes and retarded functions}
From a practical point of view, in order to make use of gauge/gravity duality one has to compute the quasinormal modes (QNMs) and the propagators of the operators on the gravity side. 
\bigskip
\\In finite temperature field theory there are two ways of computing response functions. The first option is to perform a Wick rotation to Euclidean time, compute the response functions for complex frequencies, and then analytically continue them to real ones. This, however, is a rarely viable strategy: one needs to know the correlators at extremely high precision (basically, analytically) in order for the analytical continuation to deliver acceptable results \cite{zaanen}.
\\Alternatively, one can directly work in Lorentzian signature. It turns out that, given a bulk theory with a black hole, one can holographically reproduce the Schwinger-Keldysh structure of the retarded and advanced propagators of the dual thermal field theory. This is done by selecting infalling and outgoing solutions to the equations of motion at the black hole horizon. Here we are not going to prove this construction; instead, we will limit ourselves to outlining the practical procedure than one has to follow in order to compute the correlators.
\bigskip
\\The first step consists in deriving the equations of motion for the background by minimizing the bulk action. Once one has solved for the background (usually numerically, although an analytical approach is possible around $T_c$ \cite{Herzog:2011ec, Bhattacharya:2011eea} or, more generally, in the long wavelength limit \cite{donos1,donos2,donos3,donos4}) one can compute the QNMs by allowing for small fluctuations of the fields around their background values. The standard procedure is the following (note that for simplicity we are setting the horizon radius $r_h=1$) \cite{Amado:2009ts}: 
\begin{enumerate}
\item One uses a plane wave ansatz for the perturbations to the background fields $\Lambda_i$. :
\be \label{planewave}
\delta \Lambda_i = \lambda_i(u)\, e^{-i\omega t + i kx}
\ee
Note that one is allowed to use this ansatz if one considers flat spatial sections for the metric; this will be the case in this thesis.
\\After imposing (\ref{planewave}) one linearizes the bulk action in the fluctuations  and derives the corresponding equations of motion for $\lambda(z)$.
\item One computes the boundary action up to quadratic order in the fluctuations. The action carries diverging contributions, that need regularizing. This is done by subtracting the diverging part of the action, as we will see in the upcoming section with the concrete example of the holographic dual to a superfluid system in AdS$_4$.
\item One then writes the quadratic part of the renormalized boundary action as 
\be \label{SB}
S_B = \int_B c_i(-\omega,-k) \, \mathcal{F}^{ij}(\omega,k)\, c_j(\omega,k)
\ee
Recall the definition of the holographic Green's functions with the GKPW rule (\ref{npointholo}):
\be \label{npointholo}
\left \langle \mathcal{O}(x_1) \cdots \mathcal{O}(x_n) \right \rangle = (-1)^{n+1} \frac{\delta \mathcal{S}}{\delta \phi_0(x_1)\cdots \delta \phi_0(x_n)}\bigg|_{\phi_0 = 0}
\ee
This formalism, generalized to multiple fields and naively applied to the boundary action (\ref{SB}) (forgetting about the boundary terms originating from the lower integral bound, which is the horizon radius), gives \footnote{Note that in the presence of a black hole the procedure outlined here is non-trivial and relies on the definition of Schwinger-Keldysh correlators at finite temperature. In particular, one takes into account a second boundary, corresponding to a region of spacetime containing a white hole in the infinite past. After a careful comparison with the construction of Schwinger-Keldysh correlators in finite-temperature field theory one can show that (\ref{quadraticac}) holds. \cite{Son:2002sd,herzogschwinger, zaanen, erdmenger}}  \cite{Son:2002sd,Amado:2009ts, kaminskiland}: 
\be \label{quadraticac}
G^{ij}(\omega,k) = -2\lim_{z \to 0}\mathcal{F}^{ij}(\omega,k)
\ee
From this expression we can also obtain the quasinormal modes $\omega(k)$, as the poles of the propagators  \cite{kaminskiland, poli2}. As such, they are going to be the poles of $G^{ij}(\omega,k)$.
\item In order to obtain the propagators one evaluates the quadratic action (\ref{quadraticac}) over the solutions to the equations of motion with infalling or outgoing boundary conditions at the horizon \cite{policastro, kovstar, Son:2002sd}:\be\label{infalling}
\begin{split}
\lim_{u\to 1}\lambda_i &= (1-u)^{i\nu \omega} \tilde{\lambda}_i(1-u) \,
\end{split}
\ee
where the $\lambda_i(1-u)$'s are regular functions of $1-u$. In most cases $\lim_{u\to1} \tilde{\lambda}_i(1-u)=\text{const}$, but for some perturbations (notably the timelike components of the metric and gauge field) $\tilde{\lambda}_i$ vanishes at the horizon. 
\\In general the coefficients in the Taylor expansions of the $\tilde{\lambda}_i$'s in the vicinity of the horizon are not independent. One can pick the value of a subset of the leading coefficients; all the others will be fixed via  Robin boundary conditions  obtained by expanding the equations of motion in the vicinity of $u=1$. The number of available linearly independent solutions is equal to the amount of free coefficients.
\bigskip
\\The coefficient $\nu$ can be determined by studying the leading terms in the expansion of the equations of motion around the horizon. It is proportional to the inverse Hawking temperature \cite{erdmenger, natsuume, mcgreevy}:
\be
\nu = \pm \frac{1}{4\pi T}
\ee
In the probe limit, for example, where the metric is that of an asymptotically AdS$_{d+1}$ Schwarzschild black hole, $\nu = \pm L^2/(d)$ (with unit horizon radius).
\bigskip
\\A negative coefficient describes an infalling solution; a positive one, on the other hand, an outgoing solution. Infalling solutions yield retarded Green's functions; outgoing ones give advanced propagators. We can see this by defining a new coordinate  \cite{erdmenger, natsuume} (for a Schwarzschild geometry, this amounts to choosing Eddington-Finkelstein coordinates):
\be
\tilde{u} \equiv \frac{\ln(1-u)}{4\pi T}
\ee
Notice that the horizon is located at $\tilde{u} \to - \infty$ and the boundary at $\tilde{u}\to 0$.
\\Using the redefined coordinate the near-horizon behaviour (\ref{infalling}), multiplied with the plane wave factor (\ref{planewave}), becomes:
\be
\delta \Lambda_i \sim e^{-i\omega(t\mp \tilde{u})}
\ee
A perturbation with a negative exponent will fall into the black hole; conversely, a positive exponent will describe propagation out of the black hole. It is then intuitively clear that a negative exponent is necessary for the computation of retarded functions: it guarantees that causality is not violated. In other words, the absorption of perturbations by the black hole is dual to dissipation in the field theory.
\item 
Different solutions to the equations of motion are determined by different boundary conditions on the AdS boundary. Physically, this means turning on different sources.
\\In order to find a solution to the equations of motions with given sources one goes through the following steps: 
\begin{itemize}
\item Imposing one boundary condition  for every field would result in redundant boundary conditions because of gauge- and diffeomorphism invariance. Instead, one has to impose gauge-invariant boundary conditions.  For instance, the gauge-invariant boundary condition for the gauge field sets the boundary value of the electric field. 
\\Alternatively, as we saw above one can obtain linearly independent solutions by picking the values of a subset of the leading coefficients of the fields at the horizon. 
\\The two approaches are equivalent. However, note that the number of independent solutions obtained in this way will be smaller than that of the fields, for there will be fewer boundary conditions than there are fields. Hence the need to make use of pure gauge solutions: these are gauge transformations of the trivial solution.  
\\For instance, for a geometry coupled to a gauge field and a complex scalar field (as will be the case in this thesis) the pure gauge solutions can be parametrized via a scalar field $\theta$ and a vector field $\beta^M$: 
\be \label{puregaugebackreacted}
\begin{split}
\delta \psi &= i\psi \theta + \beta^M \partial_M \psi  \qquad \delta \psi^* =- i\psi \theta + \beta^M \partial_M \psi  \\
\delta A_M &= \partial_M \theta +  \left(\mathcal{L}_\beta A\right)_M  \\
\delta g_{MN} &=  \left(\mathcal{L}_\beta g\right)_{MN} = \nabla_M \beta_N + \nabla_N \beta_M
\end{split}
\ee
In the probe limit, where one neglects backreaction between the metric and the matter fields, one can take $\beta=0$ and $\theta = e^{-i\omega t +ikx}$ and the relevant pure gauge solution takes the following form \cite{Amado:2009ts}:
\be \label{puregaugeprobe}
a_t = -i\omega \qquad a_x = ik \qquad \delta \psi = i \psi \qquad \delta \psi^* =- i \psi 
\ee
\item A general solution will be a linear combination of the above solutions, including the pure gauge ones. Labelling the source of fluctuation $j$ for the $i$-th solution as $\ell_i^j(\omega, k)$ the source of the general solution will be given by $\ell_{i,\text{gen}}(\omega, k)= \ell_i^j (\omega, k) c_j$ for some coefficients $c_j$. 
\\In particular, one can set one source to a finite value and all the others to 0. As we are going to see in detail in the upcoming chapter, the one-point functions are given by the subleading terms in the horizon expansion of the fields. One can then variate them with respect to the nonzero source and compute the corresponding Green's functions as in (\ref{quadraticac}).
\item One can set all sources to 0 and maintain a non-trivial general solution if \cite{Amado:2009ts}
\be \label{eqformodes}
\det  \ell_i^j(\omega, k)  = 0
\ee
One computes the modes $\omega(k)$ by solving for this equation in powers of $k$ (the intuition being that the Green's functions are given by the variation of the fields with respect to the sources, at general $k$ and $\omega$ - if all sources vanish they diverge). The obtained modes are gauge invariant \cite{Amado:2009ts}; 
this is due to the form of the pure gauge solution. 
\end{itemize}
Note: If one only wishes to compute the modes one does not have to look for independent solutions. 
One defines a grid $x_i$ and writes the equations of motion as a linear problem for the vector $\mathbf{f}$ containing the values of the fields at the grid points. Formally:
\be
\tilde{D}(\omega, k) \mathbf{f}= \boldsymbol{\lambda}
\ee
where $\boldsymbol{\lambda}$ contains the boundary conditions. The modes $\omega(k)$ are given by the solution to $\det \tilde{D}(\omega(k),k)=0$. More details are given in the appendix.

\end{enumerate}

\chapter{Holographic Superfluids}
The analysis of superfluids in the context of gauge gravity duality (commonly denoted as ``holographic superfluids'') was laid out in 2008 by Hartnoll, Herzog and Horowitz \cite{HHHprobe, Hartnoll:2008kx} and has been used extensively over the years \cite{Herzog:2008he,Sonner:2010yx,Arean:2010wu,Bhattacharya:2011eea,Herzog:2009md, Amado:2009ts, Herzog:2010vz,Herzog:2011ec, Amado:2013aea,Gouteraux:2019kuy, Arean:2021tks }. 
\bigskip
\\The Lagrangian that we are going to employ in this thesis was first introduced in 2008 by Gubser \cite{Gubser:2008px} and  Hartnoll, Herzog and Horowitz  \cite{HHHprobe, Hartnoll:2008kx}. It involves a metric tensor $g_{MN}$ coupled to a gauge field $A_M$ and a complex scalar $\psi$. The Lorentzian action is:
\begin{equation} \label{HHHaction}
\mathcal{S} = \frac{1}{16\pi G}\int \diff^{d+1}x \sqrt{|g|} \,\left[ R - 2\Lambda - \frac{1}{4e^2}F_{MN}F^{MN} - |D\psi|^2 - \frac{m^2}{L^2}|\psi|^2\right]
\end{equation}
As one can see, the first part is the Einstein-Hilbert action (\ref{EHaction}).  From now on we are going to set the AdS radius $L$ to 1 for simplicity. Furthermore, we set $m^2=-2$: in this case the behaviour of the scalar field at the boundary (\ref{deltascalar}) will simply involve a linear and a quadratic term. 
This Lagrangian can be modified in various ways, for instance by adding a quartic potential for the scalar field \cite{Herzog:2010vz}. 
\\The equations of motion obtained by extremizing the action (\ref{HHHaction}) can be written as \cite{Bhattacharya:2011eea, Hartnoll:2008kx,Herzog:2010vz}: 
\begin{align} \label{theeoms1}
\frac{1}{\sqrt{-g}} \left( \partial_M - iA_M\right) \sqrt{-g}\left( \partial^M - iA^M\right) \psi &= m^2\psi \\
\label{theeoms2}\frac{1}{\sqrt{-g}} \partial_M \left( \sqrt{-g}F^{MN}\right) &= e^2 J^N \\
\label{theeoms3} R_{MN}- \frac{R}{2}g_{MN} + \Lambda g_{MN}&=\frac{1}{e^2} T_{MN}
\end{align}
with the current:
\be
J_M \equiv i\left(\bar{\psi}(\partial_M -iA_M)\psi -\psi(\partial_M +iA_M)\bar{\psi}\right)
\ee
and the stress tensor:
\begin{equation}
\begin{split}
T_{MN} \equiv& -\frac12 \left( F_{MJ}F^J_N +\frac14 g_{MN}\,F_{JK}F^{JK}\right) \\
&+ \frac14 \left( (\partial_M -iA_M)\psi \, (\partial_N +iA_N)\bar{\psi}  + M \leftrightarrow N\right) -\frac{1}{4}g_{MN}\left( |\partial_J\psi - iA_J\psi|^2 + 2|\psi|^2\right) \,.
\end{split}
\end{equation}
Below a critical temperature $T_c$ the scalar field condensates, leading to spontaneous symmetry breaking.  We interpret the condensate as the conjugated charge to the chemical potential $\mu$. 
\\Notably, the conductivity shows the same features that we outlined at the end of chapter 2. In particular, $\text{Im}\,\sigma(\omega)$ features a pole at $\omega = 0$ even without translation invariance \cite{ HHHprobe}, as one can see in Figure \ref{fig:cond}. Furthermore, at $\omega \to 0$ the deviation from the normal fluid behaviour $\text{Re}\,\sigma(\omega)=1$ becomes more and more pronounced at lower temperatures, eventually reaching $\text{Re}\,\sigma(\omega)=0$ at small frequencies, which signals a nondissipative flow \cite{Hartnoll:2008kx}.
\begin{figure}[tp]
    \centering
    \includegraphics[scale=0.48]{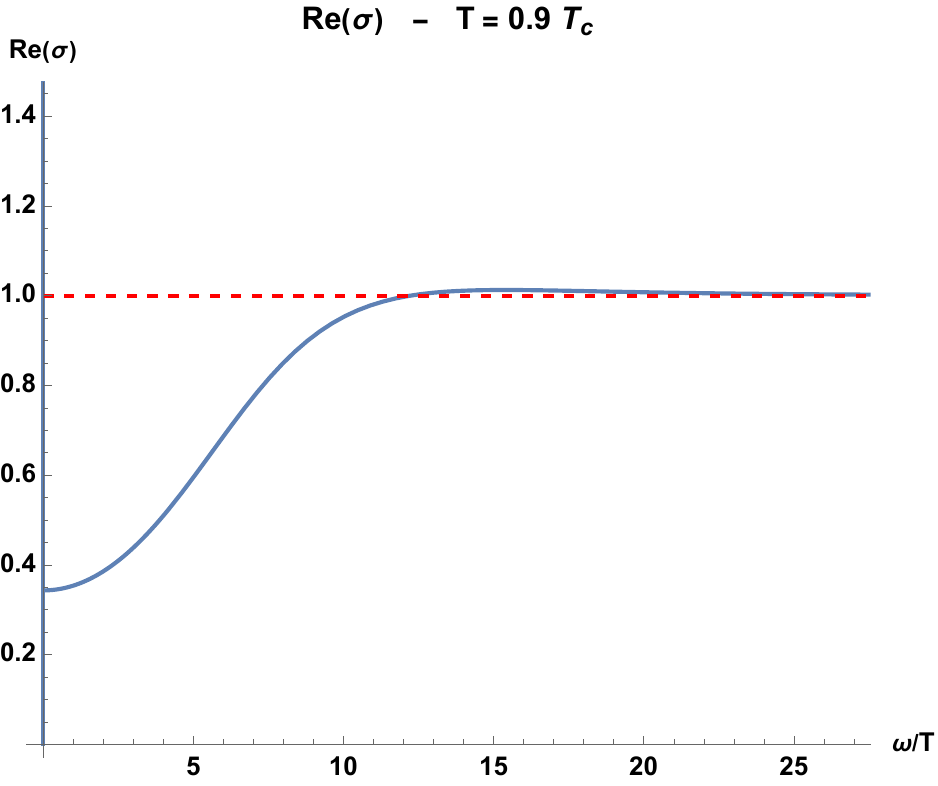}
    \hspace*{.4cm}
     \includegraphics[scale=0.48]{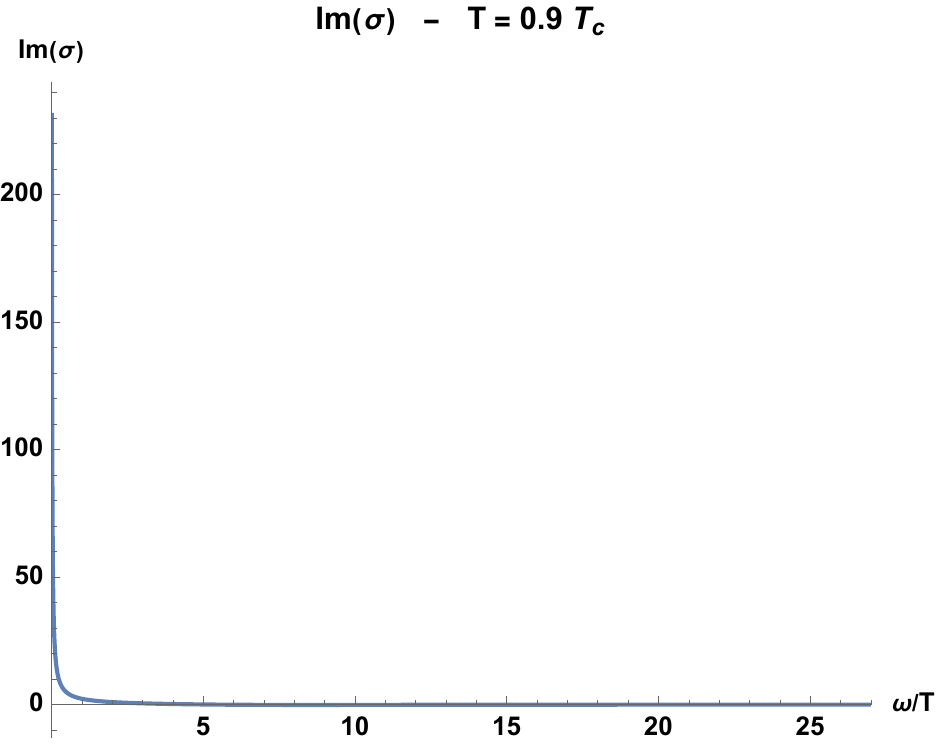}
    \caption{The real and imaginary parts of the conductivity in the probe limit (i.e. neglecting backreaction between the metric and the matter fields, and thus not imposing translation invariance) for $T = 0.9 \,T_c$. The background scalar field is set to behave as $u^2$ near the boundary.}
    \label{fig:cond}
\end{figure}
\bigskip
\\The equations of motion (\ref{theeoms1},\ref{theeoms2},\ref{theeoms3}) can be solved either analytically or numerically. The analytical approach is possible in some limited cases; however, in order to extensively study the behaviour of such a theory one therefore usually has to resort to a numerical approach.
\\As we are going to see, a first simplification is the so-called ``probe limit'', that is the large charge limit: $e \to \infty$. In this case the gauge and scalar field decouple from the metric and one obtains a Schwarzschild black hole. This limit corresponds to freezing temperature and momentum fluctuations on the field theory side. Although quantitatively not necessarily accurate, it allows to get a first qualitative understanding of the physics described by the chosen Lagrangian, especially close to the critical temperature. 
\section{The holographic dictionary for holographic superfluids}
In order to compute the vacuum expectation values from (\ref{npointholo}), one realizes that both sides of the GKPW prescription can be organized into representations of the conformal group. To wit, the fields appearing in action (\ref{HHHaction}) can be related to their dual operators by simply looking at their transformation properties. Thus varying the renormalized action with respect to the metric, the gauge field and the phase of the complex scalar field will respectively yield the vacuum expectation value of stress-energy tensor, the current and the Goldstone boson on the field theory side. In particular one has \cite{Skenderis:2002wp,bianchirenorm}: 
\be \label{1ptscal}
\langle \mathcal{O}\rangle = \frac{1}{\sqrt{-\gamma_{(0)}}}\frac{\delta S_\text{ren}}{\delta \psi_{(0)}}
\ee
\be\label{1ptgauge}
\langle J_\mu \rangle = \frac{1}{\sqrt{-\gamma_{(0)}}}\frac{\delta S_\text{ren}}{\delta A_{\mu(0)}}
\ee
\be\label{1ptmet}
\langle T_{\mu \nu}\rangle = \frac{2}{\sqrt{-\gamma_{(0)}}}\frac{\delta S_\text{ren}}{\delta \gamma_{\mu \nu(0)}} \,,
\ee
where $\gamma_{\mu \nu}$ is the boundary metric and the $_{(0)}$ subscript denotes the leading boundary behaviour of a given field, stripped of its factors of $u$, i.e. the source of its dual operator.
\bigskip
\\We are now going to lay out the computation of the renormalized boundary action for holographic superfluids. In particular, we are going to see how the various counterterms arise: this will allow us to compute the vacuum expectation values of the fields and, consequently, their propagators.
\section{The renormalized boundary action}
For simplicity's sake we are going to stick to AdS$_4$ and keep the AdS radius $L=1$, since this is the configuration in which we are going to work within this thesis. Furthermore, we set $16\pi G=1$.
\bigskip
\\The AdS boundary $\partial \Sigma$ is defined by setting $u = \epsilon$ with $\epsilon \to 0$. We work in the radial gauge ($A_r=0$) and in the Fefferman-Graham gauge, where in the vicinity of the boundary the metric is parametrized as  \cite{erdmenger, Skenderis:2002wp}:
\be \label{metricexpansion}
\diff s^2 = \frac{\diff u^2}{u^2} + \frac{g_{\mu \nu}(u^2,x)}{u^2}\diff x^\mu \diff x^\nu
\ee \footnote{Note that here Greek indices denote the $d$ boundary coordinates, while latin indices also include $u$.}
The outward pointing normal vector to the boundary is given by:
\be \label{normalvec}
n^M = -\frac{u}{L}\delta^M_u
\ee
The induced metric on the boundary is simply $\gamma_{\mu \nu} =g_{\mu \nu}/u^2$. One can show that $g_{\mu \nu}$ can be expanded in powers of $u$ as follows \cite{skendstress, haroskend}: \footnote{In odd bulk dimensions logarithmic terms in $u$ are also present in the expansion \cite{Skenderis:2002wp}; we are, however, working in AdS$_4$ so we do not need to take them into account.} 
\be \label{metricexpansion}
g_{\mu \nu} = g_{(0)\mu \nu}+ g_{(0)\mu \nu} \, u^2 + \cdots + g_{(d)\mu \nu}u^d
\ee
Notice that the only odd term in $u$ is the last one (in even bulk dimensions). The other subleading terms can be recursively determined starting from $g_{(0)}$ \cite{haroskend,erdmenger}.
\\The determinant of the induced metric asymptotes to:
\be \label{detg}
\sqrt{|\gamma|} =\sqrt{-\gamma}= \left(\frac{L}{u}\right)^{d} + \mathcal{O}(u^{2-d})
\ee
We start from action (\ref{HHHaction}), setting for simplicity $e=L=1$ and $d=3$:
\begin{equation}
\mathcal{S} = \int \diff^{d+1}x \sqrt{|g|} \, \left(R - 6 - \frac{1}{4}F_{M N}F^{MN} - |D\psi|^2 +2|\psi|^2 \right)
\end{equation}
From this action one gets diverging boundary terms. Boundary counterterms are added in order to cure the divergences without altering the bulk equations of motion.
\bigskip
\\One goes though the following steps:
\begin{itemize}
\item \textbf{The contribution of the gauge field to the action is finite}
\\From (\ref{asymptA}) we know that near the boundary:
\be \label{behgauge}
A_M = \alpha + \beta u \qquad \Longrightarrow \qquad F_{M \nu} \sim u^0 \quad \text{and}\quad  F^{M \nu} \sim u^4
\ee
After partial integration we find:
\be \label{gaugeboundarycontrib}
 -\frac{1}{4}\int \diff^{d+1}x \sqrt{|g|} \, F_{MN}F^{MN} \longrightarrow  -\frac{1}{2} \int_{\partial \Sigma} \diff^dx \, \sqrt{-\gamma}\, n_M A_\nu F^{M\nu} \sim \epsilon^0
\ee
This term behaves as $\epsilon^0$ and is therefore finite in the $\epsilon \to 0$ limit and does not need renormalizing.
\item \textbf{Renormalization of the scalar field part}
\\We know from (\ref{deltascalar}) the boundary behaviour of the scalar field. With $d=3$ and $m^2=-2$ it becomes:
\be \label{scalarHHH}
\psi = \alpha u + \beta  u^{2}
\ee
The variation of the contribution to the action proportional to $|\psi|^2$ vanishes for solutions to the equations of motion; the part proportional to $|D\psi|^2$, however, does produce a boundary term after partial integration. One has:
\be
\begin{split}
 |D\psi|^2& =g^{M N}\left(\partial_M \psi - iA_M \psi\right) \left(\partial_N \psi^* + iA_N \psi^* \right) \\
&=A^M A_M |\psi|^2+ \partial_M \psi \partial^M \psi^* +iA^M\left(\psi^* \partial_M \psi  - \psi \partial_M \psi^* \right)
\end{split}
\ee
The first term contains no derivatives and thus will not produce any boundary terms after partial integration and its variation will vanish if the equations of motion are satisfied. The boundary contribution obtained from partially integrating the third term vanishes altogether. The second term, however, gives rise to a nontrivial boundary contribution \cite{erdmenger}:
\be
S_{\text{bdy}, |D\psi|^2} = - \int_{\partial \Sigma} \diff^dx\, \sqrt{-\gamma}\,\psi^* n^M \partial_M \psi
\ee
With (\ref{normalvec}), (\ref{detg}) and (\ref{scalarHHH}) we see that the integrand behaves as 
\be
\sqrt{-\gamma}\,\psi^* n^M \partial_M \psi = \frac{|\alpha|^2}{u} + \mathcal{O}(u^0)
\ee
We can absorb this divergence by adding the following counterterm:
\be
S_{\text{ct}, |D\psi|^2} \equiv - \int_{\partial \Sigma} \diff^dx \, \sqrt{-\gamma}\,|\psi|^2 
\ee
\item \textbf{Renormalization of the metric part}
\\First, in order for the variational problem to be well-defined, one has to subtract the so-called Gibbons-Hawking-York (GHY) term \cite{gibbonshawking}:
\be \label{GHYterm}
\mathcal{S}_\text{GHY} = 2 \int_{\partial \Sigma} \diff^d x\,\sqrt{-\gamma}\, K
\ee
Here $K \equiv-\gamma^{\mu \nu}\Gamma^{\rho}_{\mu \nu}n_\rho$ is the trace of the extrinsic curvature. 
This correction ensures that the boundary action will not depend on the derivatives of the metric at the boundary. This requirement is necessary since we are going to impose Dirichlet boundary conditions on the metric. 
\bigskip
\\We can now examine the diverging parts of the boundary action. 
\\First we want to study the boundary behaviour of the bulk Ricci scalar curvature. With (\ref{metricexpansion}) one has
\be
R_{\mu \nu} = -d \, \frac{g_{(0),\mu \nu}}{u^2} + \mathcal{O}(u^0)
\ee
Then, on inserting this into the Einstein equations of motion $R_{\mu \nu}- Rg_{\mu \nu}/2 + \Lambda g_{\mu \nu}$, one finds at leading order in $u$ (recall that $\Lambda = -d(d-1)/2$):
\be
-dg_{(0),\mu \nu}- \frac{1}{2}Rg_{(0),\mu \nu} -\frac{d(d-1)}{2} g_{(0),\mu \nu}= 0  \quad \Longrightarrow \quad R = -d(d+1) +\mathcal{O}(u^2)
\ee
Furthermore, at leading order the trace of the extrinsic curvature behaves as:
\be
K = d +\mathcal{O}(u^2)
\ee
Thus, integrating the leading behaviour of $\sqrt{|g|}(R-2\Lambda)$ over $\diff u$, and adding the leading contribution of the GHY term, one finds the following divergence:
\be
\mathcal{S}_{\text{bdy}}^{d} = \epsilon^{-d}\,{2(d-1)} \intd \sqrt{-\gamma} 
\ee
Hence the corresponding counterterm:
\be
\mathcal{S}_{\text{ct},g_{(0)}} \equiv -2(d-1) \intd \sqrt{-\gamma} \,  \stackrel{d=3}{=}  \,  -4 \intd \sqrt{-\gamma} 
\ee
In $d=3$ boundary dimensions, however, the next-to-leading contribution is proportional to $\epsilon^{-d+2}=\epsilon^{-1}$: thus it also diverges and needs to be taken care of.
The next diverging contribution (and, in AdS$_4$, the last one), is given by \cite{haroskend}:
\be \label{metricnext}
S_{\text{bdy}}^{d-2}  = -\frac{(d-4)(d-1)}{d-2}\epsilon^{-d+2} \intd \sqrt{-\gamma} \,\text{Tr} g_{(2)}  \,  \stackrel{d=3}{=}  \,  2 \intd \sqrt{-\gamma} \,\text{Tr} g_{(2)}
\ee
One can show  \cite{haroskend} that:
\be
R_{(\gamma)} = 2(d-1)\epsilon^2 \,\text{Tr} g_{(2)} + \mathcal{O}(\epsilon^4)
\ee
Inserting this relation into (\ref{metricnext}) we find that in $d=3$ boundary dimensions the next counterterm reads: 
\be
\mathcal{S}_{\text{ct},g_{(2)}} \equiv - \int \text{d}^3x \sqrt{-\gamma} \, R_{(\gamma)}
\ee
In $d=3$ these are all the needed counterterms; in higher dimensions one would have contributions proportional to higher powers of $R_{(\gamma),\mu \nu}$, too \cite{skenderis, erdmenger, Andrade:2013gsa}.
\end{itemize}
We therefore obtain the following counterterms \cite{erdmenger, Emparan:1999pm, Arean:2021tks}:
\be \label{HHHcounterterms}
\mathcal{S}_\text{ct} = \int_{\partial \Sigma}  \diff^3x \sqrt{-\gamma} \, \left( 2K - 4 - |\psi|^2 - R_{(\gamma)}\right)
\ee
And with it the renormalized boundary action is defined as:
\be \label{HHHSren}
\mathcal{S}_\text{ren} = \mathcal{S}_\text{bdy} + \mathcal{S}_\text{ct}
\ee
\\Now we can compute the variation of this boundary metric under a linear perturbation of the background fields. As we saw in the previous section, this is necessary in order to compute the vacuum expectation values and the Green's functions of the dual boundary theory. 
\begin{itemize}
\item First, let us consider the part of the Lagrangian solely depending on the gauge field. Varying the boundary action (\ref{gaugeboundarycontrib}) we find:
\be
\delta \mathcal{S}_{F^2} =- \int \diff^3 x \, \sqrt{-\gamma}\, F^{MN}n_{M}\, \delta A_{M}
\ee
\item Now let us variate the complex scalar field. From the counterterm we get:
\be
\delta \mathcal{S}_{|\psi|^2, \text{ct}} =-  \int \diff^3 x \, \sqrt{-\gamma}\, \left( \psi^*\, \delta \psi + \psi\, \delta \psi^*\right)
\ee
After integrating by parts the $\sqrt{-g}|D\psi|^2$ term in the boundary action we find: 
\be
\delta \mathcal{S}_{|D\psi|^2} =- \int \diff^3 x \, \sqrt{-\gamma}\, g^{MN}n_M \Big( A_N\left( \psi^* \, \delta \psi - \psi \, \delta \psi^*\right) +\left(\,\delta \psi^*\partial_N \psi  -  \delta \psi\,\partial_N \psi^*\right) \Big)
\ee
\item Let us now variate the boundary metric. Recall the identities:
\begin{subequations}
\begin{gather}
\delta \sqrt{-\gamma} = -\frac12 \sqrt{-\gamma} \, \gamma_{\mu \nu}\, \delta \gamma^{\mu \nu}  \\
R_{(\gamma)}\equiv \gamma^{\mu \nu} R_{(\gamma),\mu \nu}  
\end{gather}
\end{subequations}
Furthermore, one can show that $\delta K =\gamma^{\rho \sigma}n^\mu\partial_\mu  \delta g_{\rho \sigma}/2 +\delta \gamma^{\mu \nu} \, K_{\mu \nu}/2$. The first term exactly cancels the contribution of $\partial g$ to the boundary action, obtained after integration by parts; the second one will contribute to the one-point functions. 
Then the variation of the boundary action due to the variation of the boundary metric $\gamma_{\mu \nu}$ amounts to \cite{Herzog:2009md, Arean:2021tks}:
\be
\delta \mathcal{S}_{\gamma} =  \int \diff^3 x \, \sqrt{-\gamma}\, \delta\gamma^{\mu \nu}\left(K_{\mu \nu} - R_{(\gamma),\mu \nu} - \frac12(2K-4+|\psi|^2+R_{(\gamma)})\,\gamma_{\mu \nu} \right)
\ee
\end{itemize}
Summing all the variations together we write:
\be \label{HHHvariation}
\delta \mathcal{S}_\text{ren} =  \delta \mathcal{S}_{F^2} + \delta \mathcal{S}_{|\psi|^2, \text{ct}} + \delta \mathcal{S}_{|D\psi|^2} + \delta \mathcal{S}_{\gamma}
\ee
Varying this expression with respect to the fluctuations of the fields we can now obtain the vacuum expectation values and the propagators of their dual operators.

\section{One- and two-point functions}
First, recall the boundary expansions of the matter fields (the hat denoting the leading mode of the background solutions):
\begin{align}
\psi &= u^2\hat{\psi} + u\psi^{(1)} + u^2\psi^{(2) }  \\
A_\mu &= \hat{A}_\mu + A_\mu^{(0)} + uA_\mu^{(1) }
\end{align}
We denote $\mu = \hat{A}_t^{(0)}$ and  $n = -\hat{A}_t^{(1)}$, as well as $\xi = -\hat{A}_x^{(0)}$.
\bigskip
\\Let us start by computing the vacuum expectation value of the scalar field $\langle \psi \rangle$:
\be
\begin{split}
\langle \psi \rangle& \equiv \frac{1}{\sqrt{-\gamma^{(0)}}} \frac{\delta \mathcal{S}_{\text{ren}}}{\delta {\psi}^{*(1)}} \\
&=-\frac{2}{u^2}\left(\psi + n_Mg^{MN} \partial_N \psi \right) \\
&= -\frac{2}{u^2}\left( \hat{\psi} -u \,\partial_u \hat{\psi} + u\,\psi^{(1)}+ u^2\psi^{(2)}- u\,\psi^{(1)}-2u^2\psi^{(2)} \right) \\
& \stackrel{u\to 0}{=} 2\, \left(\hat{\psi}^{(2)} + \psi^{(2)}\right) \quad \text{,}
\end{split}
\ee
where we have exploited the fact that $n^M$ points in radial direction, that the radial component of the gauge field is set to zero and that we have chosen the boundary behaviour of the background scalar field to be quadratic in $u$. Notice that, as expected, the counterterms cancel the diverging terms due to $\psi^{(1)}$, proportional to $u^{-1}$. 
\\Similarly:
\be
\langle \psi^* \rangle \equiv \frac{1}{\sqrt{-{\gamma}^{(0)}}} \frac{\delta \mathcal{S}_{\text{ren}}}{\delta {\psi}^{(1)}}= 2\hat{\psi}^* + 2\psi^{*(2)}
\ee
varying the action with respect to the leading behaviour of the gauge field we get:
\be
\langle J^\mu \rangle \equiv \frac{1}{\sqrt{-{\gamma}^{(0)}}} \frac{\delta \mathcal{S}_{\text{ren}}}{\delta {A}_\mu^{(0)}}= \frac{1}{u^4}\,n_N F^{N\mu}\Big|_{u\to 0} = \eta^{\mu \nu}A_\nu^{(1)}
\ee
And in particular:
\be
\langle J^t \rangle = n - A_t^{(1)}
\ee
We see that the subleading coefficients in the boundary expansion of the scalar and gauge fields give the expectation values of their dual operators.
\bigskip
\\Similarly, we obtain the expectation value of the stress-energy tensor by varying the action with respect to the boundary behaviour of the induced metric. In particular, Balasubramanian and Kraus showed in \cite{Balasubramanian:1999re} that, given the counterterms introduced above, in an asymptotically AdS$_4$ geometry:
\begin{equation*}
\langle T^{\mu \nu}\rangle\equiv\frac{2}{\sqrt{-{\gamma}^{(0)}}} \frac{\delta \mathcal{S}_{\text{ren}}}{\delta {\gamma}_{\mu \nu}^{(0)}}  = 2\left[ K^{\mu \nu}-\left(K + 2 + \frac{R_{(\gamma)}}{2}\right)\gamma^{\mu \nu} - R_{(\gamma)}^{\mu \nu} \right]
\end{equation*}
Adding the contribution from the scalar field to the boundary action (\ref{HHHvariation}) we have:
\begin{equation}
\langle T^{\mu \nu}\rangle = 2\left[ K^{\mu \nu}-\left(K +2 + \frac{R_{(\gamma)}}{2}+\frac{\psi^2}{2}\right)\gamma^{\mu \nu} - R_{(\gamma)}^{\mu \nu} \right]
\end{equation}
In the Fefferman-Graham gauge (\ref{metricexpansion}), by direct evaluation of this tensor, one finds that, up to the contribution of the scalar field, it is proportional to the coefficients of the subleading terms in the boundary expansion of the metric \cite{haroskend, skendstress}. Hence the main advantage of the Fefferman-Graham gauge: it allows to treat the metric like the gauge and scalar field. \footnote{In odd bulk dimensions one has additional contributions to $\langle T^{\mu \nu}\rangle$, which depend on $g^{(n)}$, with $n<d$ \cite{haroskend}. }
\\For instance, in this gauge the $tt$-component of the stress-energy tensor reads: 
\be
\langle T^{tt} \rangle = 3\left(\hat{\gamma}^{tt}_{(3)}+\gamma_{(3)}^{tt}\right)\equiv \epsilon+3\gamma_{(3)}^{tt}  \,
\ee
where we define the energy density $\epsilon \equiv  \hat{\gamma}^{tt}/3$.
\\As we saw in the last chapter, in a conformally invariant theory the trace of the stress-energy tensor vanishes: $\langle T^{\mu}_\mu\rangle=0$. This relation holds here, too. In fact, by inserting the boundary metric expansion (\ref{metricexpansion}) into the equations of motion for the fluctuations, one can show that $\gamma_\mu^{(3),\mu}=0$, whence $\langle T^{\mu}_\mu\rangle=0$.
\bigskip
\\As we previously saw, further differentiating the one-point functions with respect to the source we can obtain the propagators \cite{Skenderis:2002wp}. For example, let us consider the gauge field. If we turn off all the sources, except for $a_x^{(0)}$, we simply have, up to contact terms \cite{Arean:2021tks}: 
\be
G^R_{J^xJ^x} = \langle J^x J^x \rangle =\frac{a_x^{(1)}}{a_x^{(0)}}\, \text{,}\qquad \qquad G^R_{J^tJ^x} = \langle J^t J^x \rangle =-\frac{a_t^{(1)}}{a_x^{(0)}}
\ee
We see that the Ward identity $k G_{J^xJ^x}+\omega G_{J^xJ^t} = 0$ is satisfied if $ka_x^{(1)} + \omega a_t^{(1)}=0$. This does indeed hold, owing to the Maxwell equations of motion (\ref{complexscal}). In fact, to leading order in $u$ one gets $\partial_c F^{cu}=0$. But then, recalling that $A_u=0$, this gives $\partial_x \partial_u A_x- \partial_t \partial_u A_t = 0$. Inserting the boundary behaviour of the gauge field (\ref{behgauge}), together with the plane wave ansatz, one recovers $ka_x^{(1)} + \omega a_t^{(1)}=0$. 
\\In an analogous way one can compute the other propagators by turning on the corresponding sources. The Ward identities for the propagators related to the stress-energy tensor hold due to relations between the subleading components of the metric, in the same fashion as what we just saw for the gauge field. In general, the Ward identities of the boundary theory are related to the gauge symmetry of the bulk theory \cite{bianchirenorm}.
\bigskip
\\From the Green's functions one can then extract the dissipative coefficients through Kubo relations such as those we listed in (\ref{kubo}). A famous result is the so-called Kovtun-Son-Starinets (KSS) bound for the ratio of shear viscosity to entropy density for isotropic boundary field theories dual to the classical Einstein-Hilbert action. With:
\be
\eta = \lim_{\omega \to 0} \frac{1}{\omega} \, G_{T^{xy}T^{xy}}^R(\mathbf{k}=0, \omega )
\ee
In natural units one finds \cite{kss, buchelliu, erdmenger, zaanen}:
\be
\frac{\eta}{s} = \frac{1}{4\pi}
\ee
Note that in general $\eta$ can be smaller than this ``bound'' in presence of higher derivative corrections \cite{cremoninikss} or explicit breaking translational or rotational invariance. This has been shown in a range of different setups; for a review of the literature we refer the reader to \cite{baggiolikss}.

\chapter{Probe limit}
\section{Hydrodynamics}
A first simplified framework is the so-called ``probe-limit'', where the fluctuations of the normal fluid velocity and temperature are frozen. In other words, the only fluctuating currents are the electric current $j^\mu$ and the Goldstone field $\phi$, with external source $s_\phi$. In the presence of a background gauge field we define the gauge invariant combination \begin{align}
\xi_{\nu} \equiv \partial_\nu\phi - A_\nu
\end{align}
and its transverse part $\zeta^\mu \equiv \Delta^{\mu}_\nu \xi^\nu$.
The equations of motion are current conservation and the Josephson relation: 
\be
\begin{split}
\partial_\nu j^\nu &= s_\phi \\
u^\nu\xi_\nu  +\mu +\mu_{\text{diss}} &=0.
\end{split}
\ee
In the following we are going to work in the collinear limit, where the spacelike component of $\xi^\mu$ points in the same direction as the wavevector. 
\bigskip
\\The constitutive relation for the current up to first order in gradients makes use of two independent dissipative coefficients. Keeping the same notation as in (\ref{disscolllim}) we write:
\be \label{jnuprobe}
j^\mu = n u^\mu + \frac{n_s}{\mu}\xi^\mu - \sigma_0 \Delta^{\mu\nu}\partial_\nu \mu - \zeta_2 \xi^\mu \partial_\nu \left(\frac{n_s}{\mu}\xi^\nu\right )
\ee
Similarly, we parametrize the dissipative contribution to the Josephson relation as:
\be
\mu_{\text{diss}}  = -\zeta_3\partial_\nu\left(\frac{n_s}{\mu}\xi^\nu\right) - \zeta_2\xi^\nu\partial_\nu \mu
\ee
 The superfluid density $n_s$ depends on $\mu$ and $\xi$. We write its variation as:
\begin{align}
\delta n_s = \frac{\partial n_s}{\partial \mu}\biggr|_{\xi} \delta \mu + \frac{\partial n_s}{\partial \xi}\biggr|_\mu \delta \xi.
\end{align}
Considering the susceptibilities (\ref{chicomplete}) and retricting ourselves to the $(n,\xi)$ sector, we find that  the static susceptibility matrix reads:
\begin{align}
\chi = \left(\begin{array}{cc} \chi_{nn}+\chi_{n h_\xi}^2\chi_{\xi\xi}&\chi_{n h_\xi}\chi_{\xi\xi}\\
\chi_{n h_\xi}\chi_{\xi\xi} & \chi_{\xi\xi}\end{array}\right).
\end{align}
Recall that:
\be 
\chi_{\xi \xi} = \frac{\mu}{n_s + \xi\, \partial_{\xi}n_s|_{T,\mu, u_x}} \, .
\ee
\bigskip
\\We can cast the equations of motion in the following form:
\begin{equation}
-i\omega \, \begin{pmatrix} \delta n \\ \delta \xi  \end{pmatrix} + \tilde{M} \cdot \begin{pmatrix} \delta \mu \\ \delta h_\xi \end{pmatrix} = 0
\end{equation}
with
\begin{equation}
\tilde{M}=\begin{pmatrix} \sigma_0\,k^2 & ik +k^2\zeta_2 \xi \\
ik +k^2\zeta_2 \xi & \zeta_3\,k^2
\end{pmatrix} 
\end{equation}
This matrix being symmetric, we see that Onsager reciprocity is satisfied.
\bigskip
\\The spectrum of linearized fluctuations at small $k$ is given by two sound modes with dispersion relations:
\begin{align}
\label{modesprobe}
\omega_{\pm} = v_\pm k +i\frac{\Gamma_\pm}{2}k^2+ \mathcal{O}(k^3)
\end{align}
where
\be
\label{modesprobe2}
v_{\pm} =\frac{-\mu \chi_{nh_\xi} \pm \sqrt{\mu(\mu \chi_{nh_\xi}^2+\xi\partial_\xi n_s \chi_{nn} +n_s\chi_{nn}) }}{\mu \chi_{nn}}
\ee
\be
\Gamma_{\pm} = \mp \left( \pm 2\xi\, \zeta_2 + \frac{\sqrt{\mu(\mu \chi_{nh_\xi}^2+\xi\partial_\xi n_s \chi_{nn} +n_s\chi_{nn}) }}{\mu} \, \zeta_3 +\frac{\mu}{\sqrt{\mu(\mu \chi_{nh_\xi}^2+\xi\partial_\xi n_s \chi_{nn} +n_s\chi_{nn}) }}\, \sigma_0 \right)v_\pm 
\ee
Some comments:
\begin{itemize}
\item The non-diagonal susceptibility $\chi_{nh_\xi}$ is, by construction, odd under time-reversal. This implies that it can only depend on odd powers of $\xi$ and so that it vanishes at zero background superfluid velocity: $\chi_{nh_\xi}(\xi=0)=0$. In this case the speeds of sound and attenuation constants reduce to: 
\be
v_\pm(\xi=0) = \pm \sqrt{\frac{n_s}{\mu^2 \chi_{nn}}}
\ee
\be
\Gamma_\pm(\xi=0) =-\frac{n_s}{\mu^2}\, \zeta_3 - \frac{1}{\chi_{nn}}\,\sigma_0
\ee
\item Recall that in order for entropy production to be nonnegative it holds that $\zeta_3,\sigma_0\geq0$. Thus at zero background superfluid velocity $\xi=0$ both attenuation constants are negative: $\Gamma_{\pm}(\xi=0)\leq 0$. This ensures hydrodynamic stability, since $\text{Im}\,\omega_\pm \leq 0$.
\end{itemize}
Notably, $v_- <0$ when $\xi = 0$. As we increase $\xi$, $v_-$ increases. Most importantly, $v_- = 0$ when:
\begin{align} \label{thecriterion}
\boxed{\partial_\xi(\xi n_s) = 0 \quad\quad \Longleftrightarrow \quad\quad \chi_{\xi\xi}\to\infty.}
\end{align}
For larger $\xi$, $v_->0$ and $\Gamma_->0$. This signals a dynamical instability. At the same time we see that the instability is also thermodynamic, for the diagonal susceptibility $\chi_{\xi \xi}$ diverges and changes sign. 
\\Note that a similar criterion can be used to determine the critical current in superconductors \cite{bardeen}. Furthermore, the double nature (dynamical and thermodynamic, or ``energetic'') of the instability was already pointed out in \cite{schmitt2}.

\section{Holographic approach}
The holographic equivalent of the probe limit is obtained by neglecting the backreaction of the matter fields (in our case, the gauge field $A_M$ and the scalar field $\psi$) on the metric. This approach is qualitatively valid close to the critical temperature. Indeed, a low temperature is equivalent to a large chemical potential and gives rise to a larger condensate, so that the assumption of negligible backreaction between the matter fields and the metric becomes less justified. Moreover, at $T\to 0$ the condensate was found to diverge if backreaction was not taken into consideration \cite{Hartnoll:2008kx,Sonner:2010yx}.
\bigskip
\\Formally, we consider the action (\ref{HHHaction}) together with the following rescaling:
\begin{align}
A_M \to \frac{A_M}{e}\,,\quad \psi \to  \frac{\psi}{e}\,
\end{align}
while taking the large charge limit $e\to \infty$. In this limit the complex scalar and gauge fields do not backreact on the geometry, so that we consider their propagation on a fixed background. 
\\We work in four bulk dimensions ($d=3$). The background metric is given by the Schwarzschild black hole metric ( we set the horizon radius at $z=1$, as well as $M=1$) \cite{erdmenger, zaanen, HHHprobe}:
\be
\diff s^2 = -f(r)\, \diff t^2 + f(r)^{-1}\, \diff r^2 + r^2 \, \diff \mathbf{x}^2
\ee
With 
\be
f(r) = r^2 - \frac1r
\ee
In order to numerically solve the equations of motion we choose a coordinate system where the AdS boundary lies at $u \equiv 1/r \to 0$. In this case the metric reads:
\be \label{probemet}
\diff s^2 = \frac{1}{u^2}\left( \left(-1+u^3\right)\,\diff t^2 + \frac{1}{1-u^3} \, \diff u^2 + \diff \mathbf{x}^2\right)
\ee
As we already saw, in the $u\to 0$ limit the boundary behaviour of the gauge field is given by: 
\be \label{asymptA}
\lim_{u\to 0}A_\mu = -\xi_{\mu} - \langle J_\mu \rangle \, u+...
\ee
Furthermore, choosing $m^2=-2$ the scalar field has a boundary expansion in integer powers of $u$:
\be
\lim_{z\to 0}\psi = \psi_s \,u + \langle \mathcal{O} \rangle \,u^{2}+...
\ee
The coefficient $\psi_s$ is interpreted as a source for an operator $\langle \mathcal{O}\rangle$ with dimension 2 \cite{HHHprobe}. Setting $\psi_s$ to zero, a solution with $\langle \mathcal{O}\rangle \neq 0$ signals spontaneous condensation of the order parameter $\mathcal{O}$. 
\bigskip
\\The sign convention of (\ref{asymptA}) is due to the fact that in hydrodynamics the phase $\phi$ of the complex order parameter must appear in the gauge invariant combination
\begin{align}
\xi_\mu =\partial_\mu\phi - A_\mu
\end{align}
so that our choice of a real scalar is equivalent to fixing a gauge in which the chemical potential and background superfluid velocity is:
\begin{align}
\label{BC2}
\mu = \lim_{u\to0}A_t,\quad \xi_a = -\lim_{u\to0}A_a.
\end{align}
Recalling (\ref{Tschwarzschild}) we see that, setting the horizon radius to 1, the Schwarzschild temperature is fixed:
\be
T = \frac{3}{4\pi}
\ee
The physically relevant parameter is therefore the temperature-to-chemical potential ratio. In other words, we measure the temperature in units of $\mu$.

\subsection{Background equations of motion and fields}
%
Assuming the background scalar to be real ($\psi = \bar{\psi}$), with the Schwarzschild metric (\ref{probemet}) the Maxwell and scalar equations of motion (\ref{theeoms1},\ref{theeoms2}) for the background fields take a rather simple form: 
\begin{align}
0&=2u^2\Psi^2 A_t  - f(u)\,\partial_u^2 A_t  \\
0&=2u^2\Psi^2 A_x + 3u^2\partial_u A_x -f(u) \,\partial_u^2 A_x\\
0&=u\left(  A_t^2 -f(u)(4u+A_x^2) \right)\, \psi + (2-7u^3+5u^6)\,\partial_u \Psi + uf(u)^2 \,\partial_u^2\Psi  
\end{align}
with $f(u)\equiv 1-u^3$ and the redefined scalar field $\Psi(u) \equiv u^2\psi(u)$. Furthermore, the Maxwell equation for the radial component of the gauge field gives $\psi^2 A_u=0$, so that we set $A_u = 0$ \cite{Herzog:2010vz} (i.e. we work in the radial gauge).
\\Finally, we set the timelike component $A_t$ of the gauge field to $0$ at the horizon, so that $A_\mu A^\mu$ stays finite (and that the Wilson loop associated to the gauge field does not diverge).

\subsection{Equations of motion for the fluctuations}
We perturb the background fields with fluctuations following a plane wave ansatz:
\be
\begin{split}
A_t \to A_t + a_t(u)e^{-i(\omega t-kx)} \qquad A_x \to A_x+ a_x(u)e^{-i(\omega t-kx)} \\
\psi \to \psi+ \left[\sigma(u) + i \bar{\sigma}(u)\right]e^{-i(\omega t-kx)} \qquad \psi^* \to \psi^*+ \left[\sigma(u) -i \bar{\sigma}(u)\right]e^{-i(\omega t-kx)}
\end{split}
\ee
Linearizing the equations of motion (\ref{theeoms1},\ref{theeoms2}) one obtains the equations of motion for the fluctuations. Notably, one finds two dynamical, second order equations for the scalar field and its conjugate, and one dynamical equation for each non-zero component of the gauge field.
\\One can, however, trade one dynamical Maxwell equation for the first-order constraint obtained from the $u$-component of the Maxwell's equations.  One can show that a set of fields satisfying  the two scalar equations, one dynamical Maxwell's equation and the constraint automatically solves the other dynamical equation. From a practical point of view this approach reduces the amount of boundary conditions needed and can guarantee better numerical stability. It is also used when searching for analytical solutions, as we are going to see in the coming section. When numerically computing the quasinormal modes, however, using the constraint produces a larger amount of spurious modes (i.e. numerical artifacts due to the finite grid size and precision). The modes  presented in the rest of this chapter were computed using the dynamical equations.
\bigskip
\\As explained in the last chapters, we impose infalling boundary conditions at the horizon:
\begin{align}\label{infallingbcprobenum}
a_t = (1-u)^{1-i\omega/3}\left[a_t^{(0)}+a_t^{(1)}(1-u)+...\right]\nonumber\\
a_x = (1-u)^{-i\omega/3}\left[a_x^{(0)}+a_x^{(1)}(1-u)+...\right]\nonumber\\
\sigma = (1-u)^{-i\omega/3}\left[\sigma^{(0)}+\sigma^{(1)}(1-u)+...\right]\nonumber\\
\bar{\sigma} = (1-u)^{-i\omega/3}\left[\bar{\sigma}^{(0)}+\bar{\sigma}^{(1)}(1-u)+...\right]
\end{align}
Note the additional power of $(1-u)$ in the horizon expansion of $a_t$.
\\The leading terms are not independent. To wit, the constraint equation evaluated at the horizon implies:
\be \label{robinprobe}
3k a_x^{(0)} + 6i \psi \sigma^{(0)} - (3i+\omega)a_t^{(0)} = 0
\ee
Further boundary conditions involving the subleading terms can be derived by expanding the dynamical equations of motion around $u=1$. These are mixed boundary conditions between the leading values and the derivatives of the fields at the horizon, so that one usually refers to them as Robin boundary conditions.
\\At the boundary we impose gauge invariant boundary conditions. In particular, we can independently fix the leading behaviour of $\sigma$ and $\bar{\sigma}$, but not of the individual components of the gauge field. Instead, we fix the electric field at the boundary $E_x \equiv ka_t +\omega a_x \big|_{u=0}$.
\section{Results}
\subsection{Numerical analysis}
\subsubsection{Location of the instability}
Recalling the constitutive relation (\ref{jnuprobe}) and the boundary behaviour of the gauge field (\ref{asymptA}) we find:
\be
\xi =- \lim_{u\to 0}  A_x \qquad \qquad \frac{\xi n_\text{s}}{\mu} = -\lim_{u\to 0} \partial_u A_x 
 \ee
 These simple relations allow us to plot $\xi n_s$ as a function of the background superfluid velocity $\xi$ for different values of the chemical potential, as we did in Figure \ref{fig:nsvs}. For any $\mu$, the value of $\xi$ where this quantity attains its maximum determines the position of the instability (\ref{thecriterion}).
 \begin{figure}[h]
    \centering
    \includegraphics[scale=0.52]{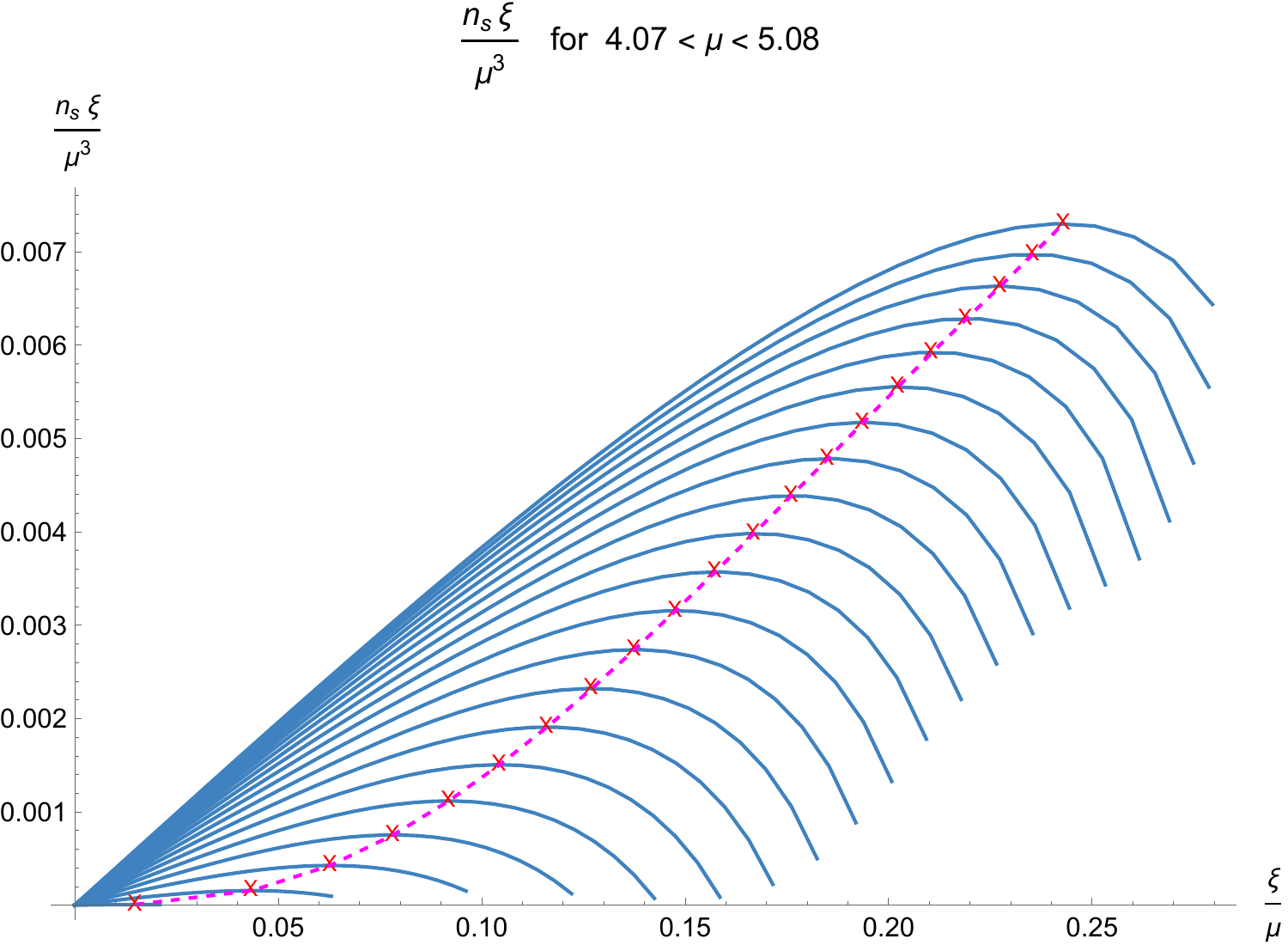}
    \caption{$n_\text{s}\xi/\mu^3$ for a range of 21 values of $\mu$ between 4.07 and 5.08, corresponding to $\frac{T}{\mu}$ between $0.998 \frac{T}{\mu_c}$ (lowest curve) and $0.8 \frac{T}{\mu_c}$ (highest curve). The red crosses indicate the position of the maximum of each curve, estimated via a polynomial regression. Data computed on a grid of 16 Chebyshev points, at MachinePrecision. We plot $n_\text{s}\xi/\mu^3$ instead of $n_\text{s}\xi$ because it is a dimensionless, scale invariant quantity. }
   \label{fig:nsvs}
\end{figure}
\begin{figure}[tp]
    \centering
    \includegraphics[scale=0.54]{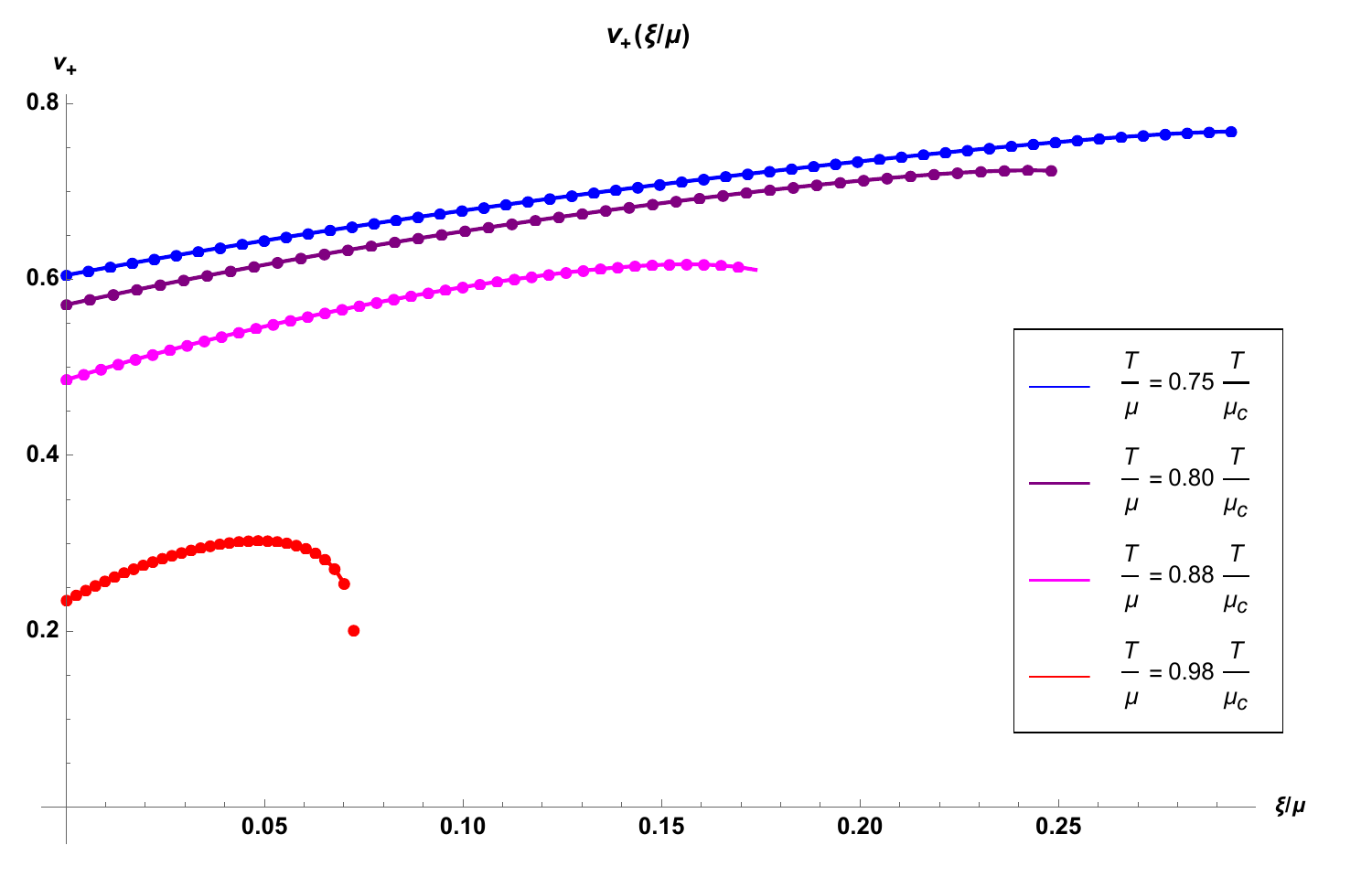}
      \includegraphics[scale=0.54]{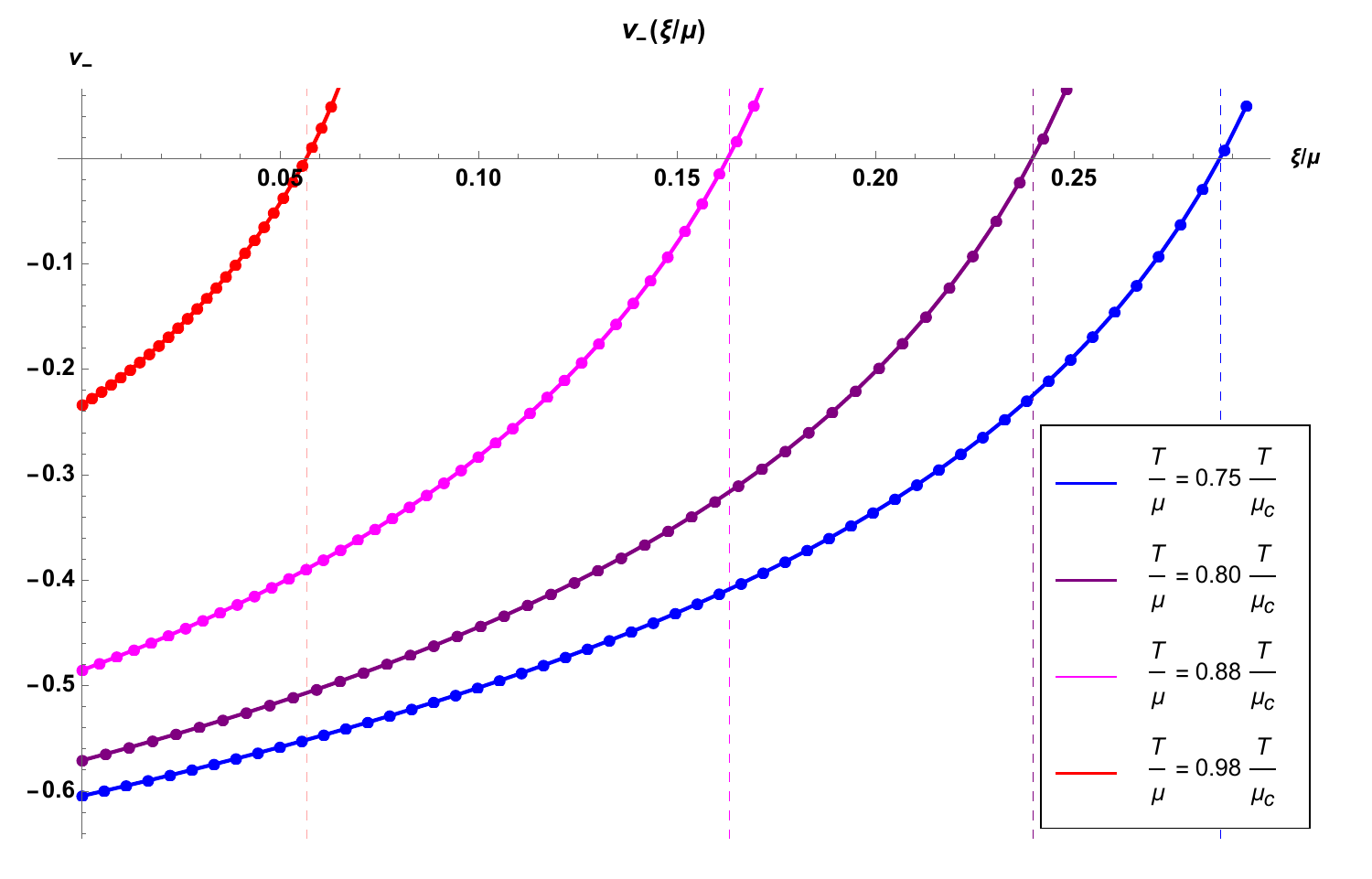}
    \caption{$v_\pm$ as a function of the background superfluid velocity for four different chemical potentials. The dots are obtained from a fit of the QNMs (computed with 24 Chebyshev grid points and working precision 30); the lines correspond to the hydrodynamic dispersion relation (\ref{modesprobe}), obtained from backgrounds with 24 grid points (precision 60). The consistency between the two sets of data is quite good already at such small grid sizes and precisions, with a maximum relative error generally below $10^{-6}$ in the range displayed in this graph. The vertical lines signal, for every value of $\mu$, the value of $\xi$ where $\xi n_s$ attains its maximum. As one can see at this critical value $v_-$ vanishes and changes sign, validating the criterion (\ref{thecriterion}).}
    \label{fig:vplus}
\end{figure}
\begin{figure}[tp]
    \centering
    \includegraphics[scale=0.54]{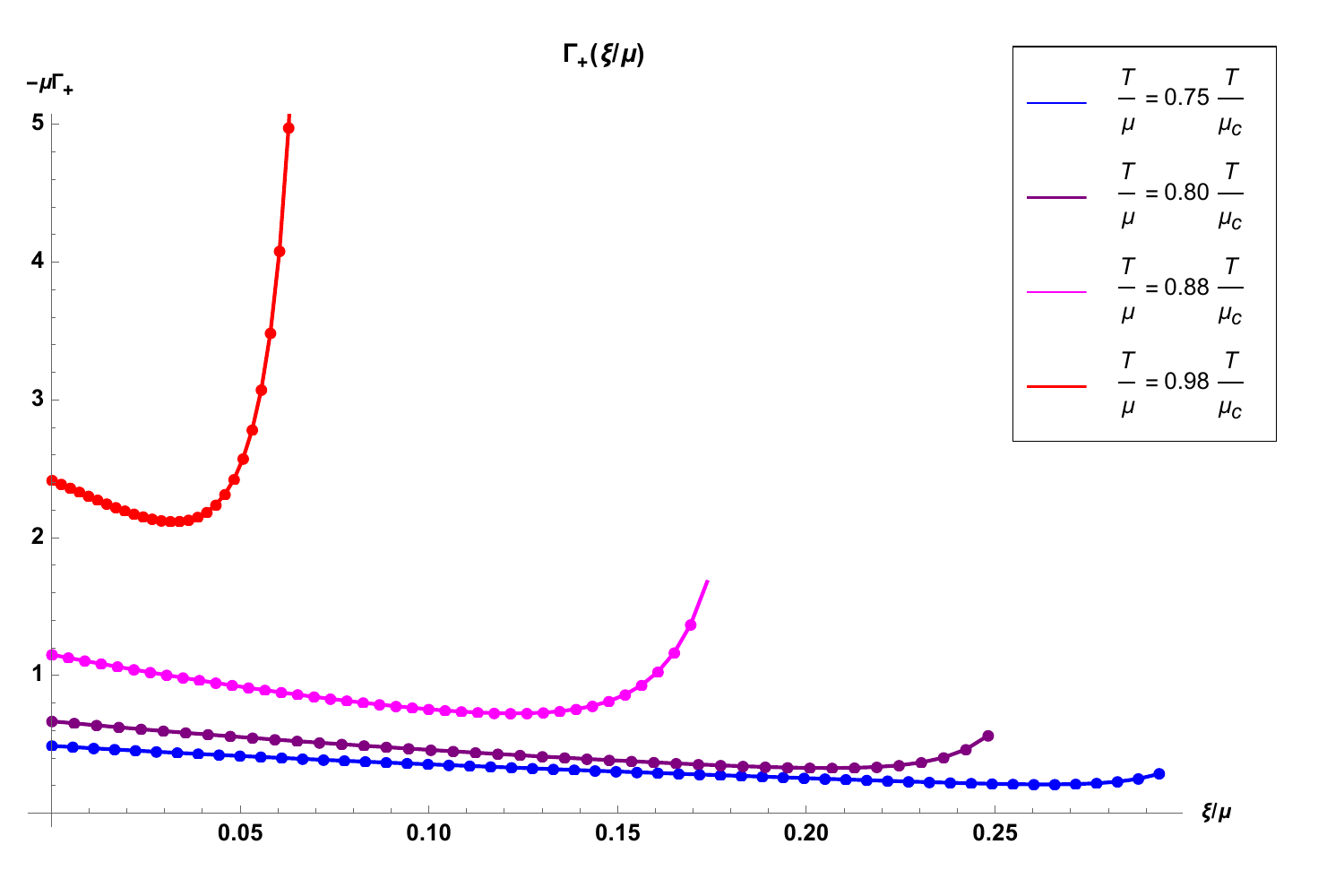}
        \includegraphics[scale=0.54]{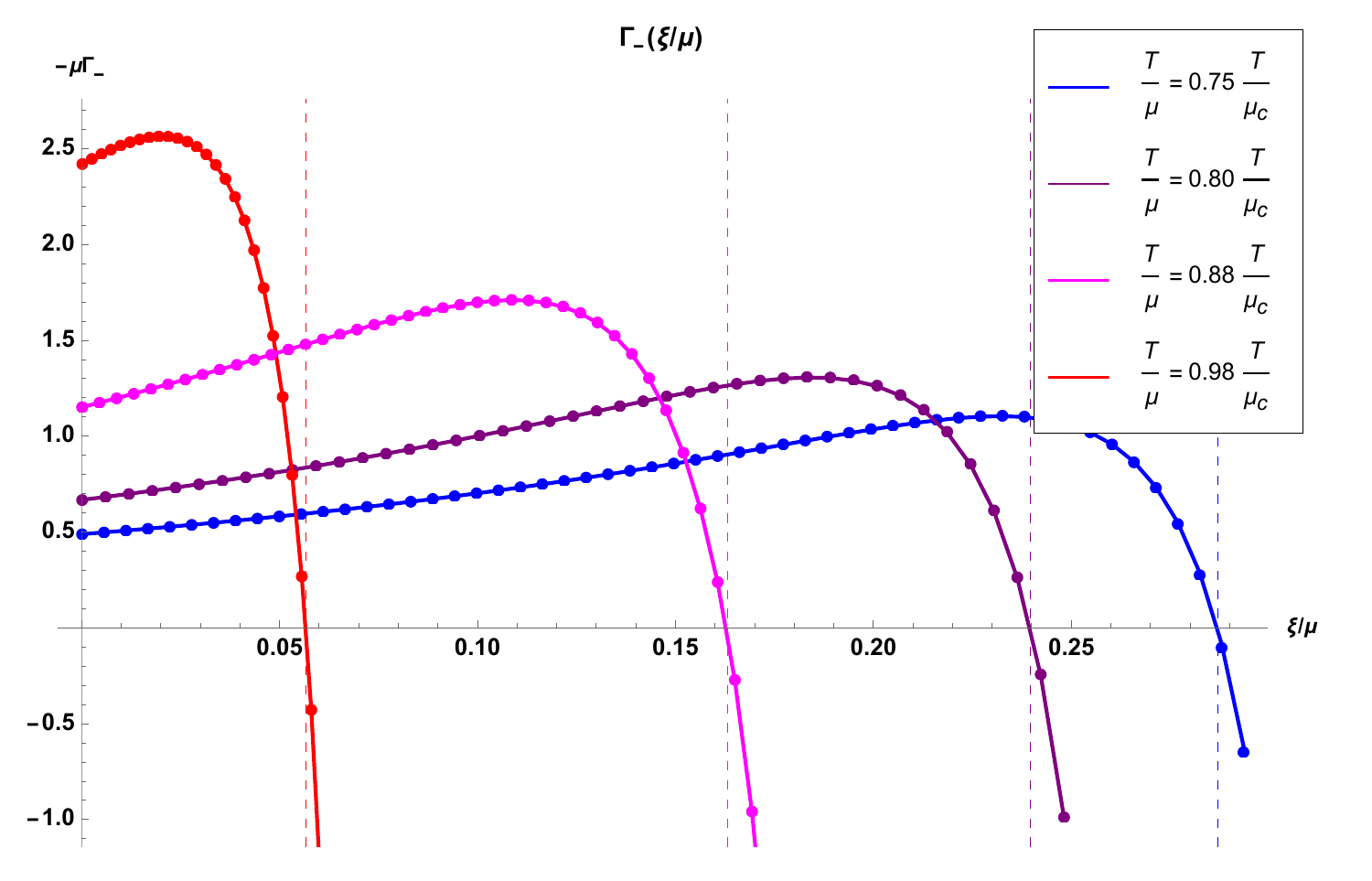}
    \caption{The dots correspond to the attenuation constants $\Gamma_\pm$ as a function of the background superfluid velocity for four different chemical potentials, obtained from a fit of the QNMs (computed with 24 Chebyshev grid points and working precision 30). The lines correspond to the hydrodynamical prediction (\ref{modesprobe}). The relative discrepancy between the two sets of data is below $0.1\%$, except around the phase transition, where it lies below $1\%$. Even more consistent results are achieved at larger grid sizes and precisions.}
    \label{fig:gammaplus}
\end{figure}
\bigskip
 \\The next step consists in computing the QNMs and fit them to get the speeds of sound and the attenuation constants, as we did in Figures \ref{fig:vplus} and \ref{fig:gammaplus}. This data can then be compared with (\ref{modesprobe2}), after computing the susceptibilities from the background data \footnote{More details are given in the coming chapter.} and extracting the dissipative coefficients from the $k,\omega \to 0$ limit of the correlators through Kubo relations. 
\bigskip
 \\In order to compute the susceptibilities we generated multiple backgrounds, close in parameter space, and estimated the thermodynamic derivatives by finite differences. We found a very good agreement between the QNMs and the matched data throughout the ranges of $\xi$ and $\mu$ that we considered.
 \bigskip
 \\Finally, in Figure \ref{fig:gammaplus} we can see that the attenuation changes sign together with the speed of sound, validating the criterion. Moreover, this happens at the same value of $\xi$ for which $n_\text{s}\xi$ attains its maximum, thus validating our criterion (\ref{thecriterion}).

\subsubsection{Phase space}
One can also plot the phase space of the holographic hydrodynamical theory, as we did in Figure \ref{fig:phasespace}. To wit, one can distinguish two stable regions: a stable superfluid region (coloured in red in the graph), where both attenuation constants in (\ref{modesprobe}) $\Gamma_\pm$ are negative and a region (in white) where the scalar field does not condense, so that the superfluid component vanishes. The part of phase space between them is the unstable region and can be divided in two sections. With growing $\xi/\mu$, one first lands in the region (in blue) where both speeds of sound are positive. At even larger background superfluid velocity, one then enters the region (in grey) where the argument of the square root in (\ref{modesprobe2}) becomes negative, so that they feature an imaginary part which is linear in the wavevector $k$.
\bigskip
\\As we saw above, the first instability corresponds to the divergence of $\chi_{\xi \xi}$. Furthermore, recalling the susceptibility matrix (\ref{chicomplete}) we see that the argument of the square root in  (\ref{modesprobe2}) is simply $\mu \chi_{nn}^{h_\xi}$, so that we can identify a similar mechanism to the first one. This kind of instability, signalled by the insurgence of a complex speed of sound, has been interpreted as a ``two-stream'' instability, appearing beyond the Landau critical velocity \cite{schmitt1,schmitt2,schmitt3}.

\begin{figure}[h]
    \centering
    \includegraphics[scale=0.45]{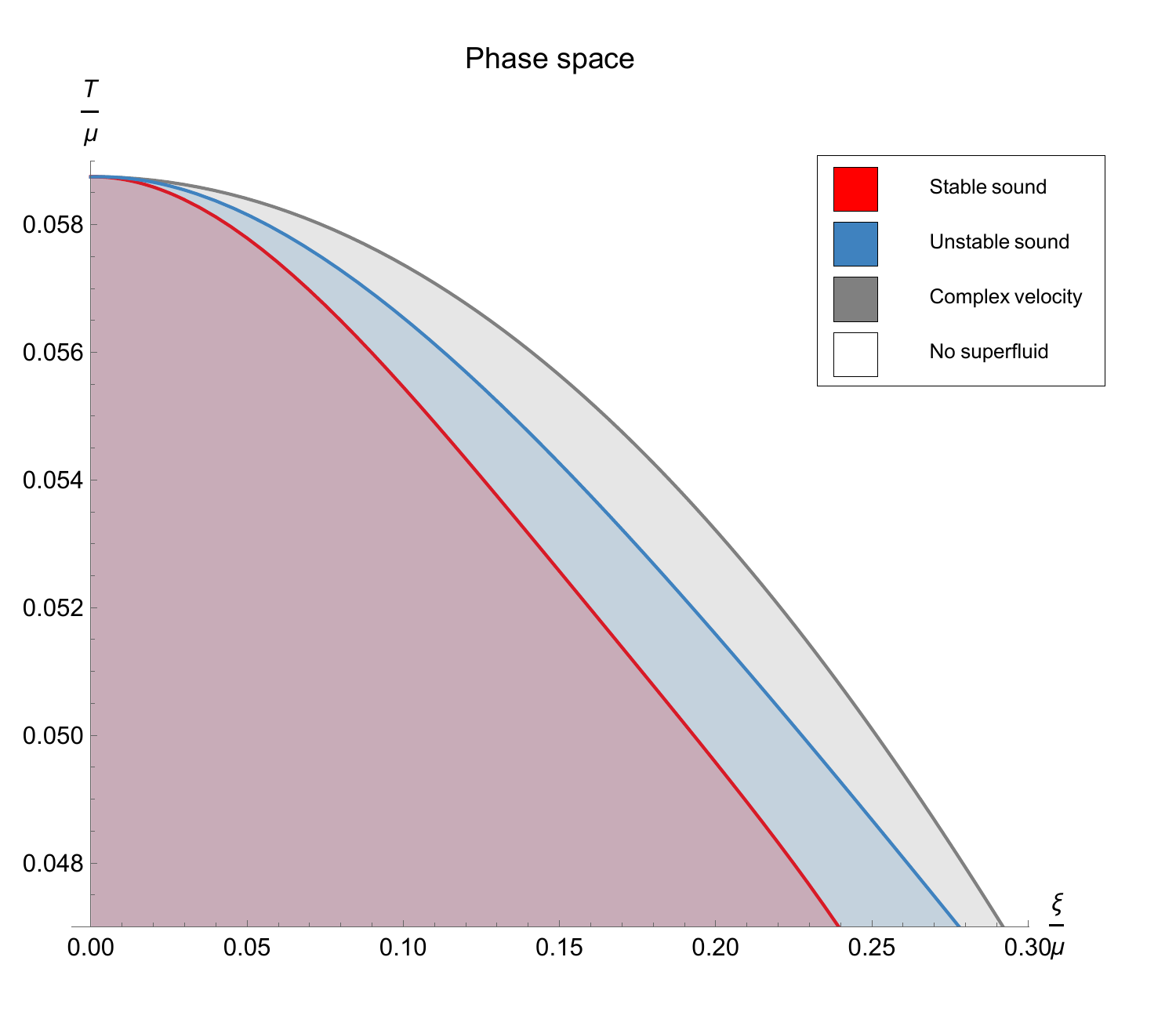}
    \caption{The phase space. The red area features two stable sound modes; beyond the red line, in the blue area, $v_-$ and $\Gamma_-$ change sign, leading to a dynamical instability.  In the grey region $\text{Im}\omega$ acquires a linear contribution in $k$. In the white region the scalar field does not condense and the system reduces to a normal fluid. The curves are obtained from fits of background data,  namely the maxima of the curves plotted in Figure \ref{fig:nsvs} for the red line and the speeds of sound computed with the matching procedure described in the previous section for the blue line. The grey line shows, for a given $T/\mu$ ratio, the value of $\xi/\mu$ that gives $\psi(0)=10^{-7}$.}
    \label{fig:phasespace}
\end{figure}
\subsection{Analytical approach near $T_c$}
In the small condensate limit an analytical approach to holographic superfluid hydrodynamics is possible. It was first introduced by Herzog in \cite{Herzog:2010vz} in the probe limit, and later generalized in  \cite{Bhattacharya:2011eea} by  Bhattacharya et al., including backreaction and a background superfluid velocity. 
\\Although only applicable near the critical temperature, this strategy allows to explicitly check the AdS/CFT correspondence. Also, it allows us to achieve a more precise understanding of the behaviour of thermodynamical quantities such as susceptibilities near the phase transition.
\bigskip
\\In this section we work in an asymptotically AdS$_5$ geometry for consistency with previous literature. The AdS radius is still set to the unity for simplicity. We choose $m^2=-4$; this value of the mass saturates the BF bound (\ref{BFbound}). 
\bigskip
\\In this configuration the leading and subleading modes of the scalar field differ by a factor of $\log u$ \cite{Skenderis:2002wp}: 
\be
\psi = u^2\log(u/\delta)\,\psi_s -u^2\, \langle \mathcal{O}\rangle +...  \,
\ee
where $\delta$ is a cutoff that we introduce by hand. 
\\The boundary behaviour of the gauge field is given by (\ref{asymptA}):
\be
A_\mu =- \xi_\mu - e^2\frac{u^2}{2}\,\langle J_\mu\rangle +...
\ee
We fix the background metric to be that of a Schwarzschild black hole:
\begin{align}
\diff s^2 = \frac{1}{u^2}\left(-f(u)dt^2+\frac{du^2}{f(u)} + dx^2+dy^2+dz^2\right)
\end{align}
with $f(u)=1-u^4$ (setting the horizon radius at $u_h=1$).
\bigskip
\\We choose a superfluid velocity pointing along the $x$-direction and define:
\be
\epsilon \equiv  -\sqrt{2}\langle\mathcal{O} \rangle \qquad \qquad \xi \equiv \xi_x
\ee
Owing to the five-dimensional bulk geometry, the temperature now reads:
\be
T = \frac{1}{\pi}
\ee
We assume $\epsilon$ and $\xi$ to come at the same order in perturbation theory. For notational compactness we set $e=1$ and consider $\mathcal{O}(\iota)\equiv \mathcal{O}(\epsilon)=\mathcal{O}(\xi)$. Furthermore, we impose $\mu = 2 + \mathcal{O}(\iota)$. 
\bigskip
\\To leading order in $\epsilon$ and $\xi$ the gauge and scalar fields read:
\begin{align}
A_t &= \frac{2}{1+u^2} + \frac{(1-5u^2)\,\epsilon^2 +24(1+u^2)\,\xi^2}{48(1+u^2)^2} +\mathcal{O}(\iota^4)\\
A_x &= -\xi + \frac{\xi \epsilon^2}{8(1+u^2)}u^2 +\mathcal{O}(\iota^4)\\
\psi &=  \frac{\epsilon \, u^2}{1+u^2} + \epsilon^3 \, \frac{u^2(1+u^2)\log(1+u^2)-2u^4}{48(1+u^2)^2} - \epsilon \xi^2 \, \frac{\log(1+u^2)}{4(1+u^2)} +\mathcal{O}(\iota^4)
\end{align}
Hence we can extract (obtaining $n_s$ from (\ref{jnuprobe})):
\begin{align}
\mu &= \lim_{u\to 0} A_t = 2 + \frac{\xi^2}{2}+ \frac{\epsilon^2}{48} +\mathcal{O}(\iota^4) \\
n &=2 \lim_{u\to 0} \partial_u^2 A_t = 4+\frac{7}{24}\epsilon^2 + \xi^2 +\mathcal{O}(\iota^4) \\
n_s &= -\frac{\mu}{\xi}\langle J_x\rangle = \frac{\epsilon^2}{2}+ \mathcal{O}(\iota^2)
\end{align}
Now we can compute the susceptibilities. For instance to leading order we get $\delta n = 2\xi \delta \xi + \frac{7}{12}\epsilon \delta \epsilon$ and $\delta \mu = \xi \delta \xi + \frac{1}{24}\epsilon \delta \epsilon$. Hence:
\be \label{suscan1}
\chi_{nh_\xi} \equiv \pdertmedium{n}{\xi}{\delta \mu=0} = -12\xi
\ee
Similarly:
\be \label{suscan2}
\chi_{nn} = 14 \qquad \partial_\xi n_s|_\mu = -24\xi
\ee
\\Next, we consider perturbations to this background of the following form:
\begin{align}
\label{perturbationsprobe}
A_t &= A_{t,0}(u)+a_t(r)e^{-i\omega t+ikx},\quad A_x = A_{x,0}(u)+a_x(u)e^{-i\omega t+ikx},\quad \nonumber\\
 \Psi &= \psi(u)+[\sigma(u)+i\bar{\sigma}(u)]e^{-i\omega t+ikx}, \quad \Psi^* =\psi(r)+ [\sigma(u)-i\bar{\sigma}(u)]e^{-i\omega t+ikx}
\end{align}
where we have chosen the wavevector to point along the superfluid velocity. Importantly, while we can use a gauge transformation to set $\Psi=\Psi^*$ in the background, the fluctuations of the scalar field and its complex conjugate will be independent. 
\\As $u\to 0$, the fluctuations can be expanded as:
\begin{align}
a_t &\to \delta A_t^{(0)} + \frac{\delta J_t}{2}u^2+...\nonumber\\
a_x &\to \delta A_x^{(0)}+\frac{\delta J_x}{2}u^2+...\nonumber\\
\sigma&\to \Sigma^{(0)}u^2\log(u/\delta) + \Sigma^{(1)}u^2+...\nonumber\\
\bar{\sigma}&\to \bar{\Sigma}^{(0)}u^2\log(u/\delta) + \bar{\Sigma}^{(1)}u^2+...
\end{align}
As we did before, at the horizon we impose ingoing boundary conditions (the infalling exponent being now $1/4$ instead of $1/3$ as in (\ref{infallingbcprobenum}) because we are working in AdS$_5$):
\begin{align}
a_t = (1-u)^{1-i\omega/4}\left[a_t^{(0)}+a_t^{(1)}(1-u)+...\right]\nonumber\\
a_x = (1-u)^{-i\omega/4}\left[a_x^{(0)}+a_x^{(1)}(1-u)+...\right]\nonumber\\
\sigma = (1-u)^{-i\omega/4}\left[\sigma^{(0)}+\sigma^{(1)}(1-u)+...\right]\nonumber\\
\bar{\sigma} = (1-u)^{-i\omega/4}\left[\bar{\sigma}^{(0)}+\bar{\sigma}^{(1)}(1-u)+...\right]
\end{align}
As we saw above, the equations of motion for the fluctuations are three second-order equations for $a_x, \sigma,$ and $\bar{\sigma}$ and one first-order equation for $a_t$. As such, in addition to fixing the four ingoing boundary conditions, we are also able to fix three of the leading behaviours at the horizon, though not all four. In the following, we are going to pick the horizon values of $a_x^{(0)}, \sigma^{(0)}$ and $\bar{\sigma}^{(0)}$ at the horizon and determine $a_t^{(0)}$ as their function \cite{Amado:2009ts}. To leading order:
\be
a_t^{(0)} = - \frac{96a_x^{(0)} + i\sqrt{2}\sigma_i^{(0)}\big(48 + \epsilon^2(\log2-1)-12\xi^2\log 2\big)}{24(4i+\omega)} + ...
\ee
\\For simplicity we choose the three following linearly independent sets of boundary conditions:
\begin{align}
\{a_x^{(0)},\sigma^{(0)},\bar{\sigma}^{(0)}\}^\text{I} = \{1,0,0\},\quad \{a_x^{(0)},\sigma^{(0)},\bar{\sigma}^{(0)}\}^{\text{II}} = \{0,1,0\},\quad \{a_x^{(0)},\sigma^{(0)},\bar{\sigma}^{(0)}\}^{\text{III}} = \{0,0,1\} \,,
\end{align}
which lead to three linearly independent solutions $X^i = \{a_t, a_x, \sigma, \bar{\sigma}\}^i$ for $i=\{\text{I},\text{II}, \text{III}\}$. 
\bigskip
\\The Green's functions are obtained in the standard way from linear response \cite{KadanoffandMartin}. Given two currents $J^a$ and $J^b$ with sources $s_a$ and $s_b$ we write: 
\begin{align}
\delta J^a = G^R_{J^aJ^b}(\omega,k)\delta s^b.
\end{align}
Here the sources are the UV boundary values of the fluctuations, $\delta A_t^{(0)}, \delta A_x^{(0)}, \delta \Sigma^{(0)}$ and $\delta \bar{\Sigma}^{(0)}$ (we write them in capital letters to distinguish them from the horizon behaviours). As we saw in the last chapter, in order to independently vary each source we need to supplement our three solution with the pure gauge solution, which produces the following sources:
\begin{align}
X^{\text{IV}} = \{-i\omega, ik, 0, \psi\}.
\end{align}
With this solution, any solution for the fluctuating fields can be written as:
\begin{align} \label{gensol}
X = c_\text{I} X^\text{I} + c_{\text{II}}X^{\text{II}} + c_{\text{III}}X^{\text{III}} + c_{\text{IV}}X^{\text{IV}}
\end{align}
In particular we can find the solution corresponding to a particular choice of $\delta A_t^{(0)}, \delta A_x^{(0)}, \Sigma^{(0)},$ and $\bar{\Sigma}^{(0)}$ via this method. In the end we find that, turning on the sources $\delta A_t^{(0)}$ and $\delta A_x^{(0)}$ of the gauge field, the subleading modes are given by:
\begin{align}
\delta A_t^{(1)} &= 2k\frac{k\delta A_t^{(0)} +\omega \delta A_x^{(0)}}{\mathcal{P}}\biggl[-480\omega^2-4i\omega(37\epsilon^2+24k(k+6i\xi))\nonumber\\
&\quad\quad\quad\quad\quad\quad\quad\quad\quad\quad\quad\quad\quad+(\epsilon^2+4k^2)(7\epsilon^2+12[k^2-4\xi^2])+\mathcal{O}(\iota^4)\biggr]\nonumber\\
\delta A_x^{(1)} &=- 2\omega\frac{k\delta A_t^{(0)} + \omega \delta A_x^{(0)}}{\mathcal{P}}\biggl[-480\omega^2-4i\omega(37\epsilon^2+24k(k+6i\xi))\nonumber\\
&\quad\quad\quad\quad\quad\quad\quad\quad\quad\quad\quad\quad\quad+(\epsilon^2+4k^2)(7\epsilon^2+12[k^2-4\xi^2])+\mathcal{O}(\iota^4)\biggr]\nonumber\\
\Sigma^{(1)} &= -24\sqrt{2}\epsilon\frac{k\delta A_t^{(0)}+\omega\delta A_x^{(0)}}{\mathcal{P}}\left[k(\epsilon^2+4k^2) - 8\omega(\xi+2ik)+\mathcal{O}(\iota^4)\right]\nonumber\\
\bar{\Sigma}^{(1)} &=\frac{i\sqrt{2}\epsilon\delta A_x^{(0)}}{\mathcal{P}}\biggl[-288\omega^2(k+2i\xi)+k(\epsilon^2+4k^2)(\epsilon^2+12[k^2-4\xi^2])\nonumber\\
&\quad\quad\quad\quad\quad\quad-12i\omega(\epsilon^2(3k+4i\xi)+16k[k^2+2ik\xi-2\xi^2])+\mathcal{O}(\iota^4)\biggr]\nonumber\\
&+\frac{ i\sqrt{2}\epsilon\delta A_t^{(0)}}{\mathcal{P}}\left[-960 i\omega^2+8\omega(7\epsilon^2+24k[2k+3i\xi])-48k(\epsilon^2+4k^2)\xi+\mathcal{O}(\iota^4)\right]
\end{align}
where $\mathcal{P}$ is the determinant of the $4\times 4$ matrix whose entries are the values of the four boundary sources for the four solutions in (\ref{gensol}):
\begin{align}
\mathcal{P} &=-1920i\omega^3 +16\omega^2(7\epsilon^2+12k(7k+12i\xi))+24ik\omega(16k(k+i\xi)(k+2i\xi)+\epsilon^2[3k+8i\xi])\nonumber\\
&\quad-2k^2(\epsilon^2+4k^2)(\epsilon^2+12[k^2-4\xi^2]).
\end{align}
\\The modes are given by the zeros of $\mathcal{P}$. In the limit $k\ll \epsilon, \xi\ll1$ we find two sound modes $\omega_\pm$ and a gapped mode $\omega_g$:
\begin{align}
\label{modesprobeanalytic}
\omega_\pm &= v_\pm k+i\frac{\Gamma_\pm}{2}k^2+ \mathcal{O}(k^3)\nonumber\\
\omega_g &= -i\frac{7}{120}\epsilon^2 +\frac{18}{35}k\xi -ik^2\left(\frac{89}{245}+\mathcal{D}_g\right)+ \mathcal{O}(k^3)
\end{align}
where 
\begin{align}
\label{modesprobeanalytic2}
v_\pm = \frac{6\xi}{7} \pm\frac{\sqrt{14\epsilon^2-96\xi^2}}{28},\quad \qquad \Gamma_\pm =  \mp v_\pm \frac{462\epsilon^2+48\xi(8\xi\mp9\sqrt{14\epsilon^2-96\xi^2})}{49\epsilon^2\sqrt{14\epsilon^2-96\xi^2}}.
\end{align}
The expression for $\mathcal{D}_g$ is rather involved. However, in the limit $\xi\ll \epsilon$,
\be
\mathcal{D}_g =  \frac{2496}{343}\frac{\xi^2}{\epsilon^2}.
\ee
We can compare the sound modes with (\ref{modesprobe2}). With relations (\ref{suscan1}) and (\ref{suscan2}) we immediately recover the speeds of sound. In order to compute the attenuations $\Gamma_{\pm}$ we also need to extract the dissipative coefficients via the Kubo relations (\ref{kubo}) (using $\xi = ik\phi$): 
\begin{equation}
\begin{split}
\sigma_0 &= -\lim_{\omega \to 0, k\to 0} \, \frac{\omega}{k^2}\, \text{Im} \,G^R_{J^tJ^t}(\omega, k) \\
\xi \zeta_2 &= -\lim_{\omega \to 0, k\to 0} \, \omega\, \text{Re} \, G^R_{J^t \phi}(\omega, k)\\
\zeta_3 &=- \lim_{\omega \to 0, k\to 0} \, \omega\,  \text{Im} \,G^R_{\phi \phi }(\omega, k) \\
\end{split}
\end{equation}
To leading order in $\epsilon$ and $\xi$ we get: 
\be
\zeta_2 =-\frac{216}{49 \epsilon^2}, \qquad \zeta_3 = \frac{52}{49\epsilon^2},\qquad \sigma_0 = 1+\frac{1440}{49}\frac{\xi^2}{\epsilon^2}.
\ee
Using these expressions we recover the imaginary parts of  (\ref{modesprobe2}).
\bigskip
\\We are now able to study the instabilities of these modes. In particular, we can see that when 
\begin{align}
\xi=\xi_c \equiv \frac{\epsilon}{4\sqrt{3}}
\end{align}
the velocity $v_- = 0$. Taking a look at the attenuation constant we also see that for $\xi<\xi_c$, $\Gamma_- <0$, whereas for $\xi>\xi_x$, $\Gamma_->0$: this signals a dynamical instability. Furthermore, it is easily checked that at $\xi=\xi_c$:
\be
\partial_\xi(\xi n_s)\big|_{\xi=\xi_c} = 0  
\ee
which is the criterion introduced above.

\chapter{The full picture}
\section{Hydrodynamics}
\subsection{Intermediate approach: breaking translations}
In the last chapter we considered the probe limit, where we froze the fluctuations of both the temperature and the normal fluid velocity. As a first step we can turn on $\delta T$, but not $\delta \mathbf{u}$. In this limit we can describe so-called ``dirty'' superfluids, where spatial translations are broken. Such systems where studied in \cite{blaisedelac2016} for zero background superfluid velocity.
\bigskip
\\The equations of motion are described by the matrix given in (\ref{eomscollinear}) without the third column and row. To wit:
\begin{equation} \label{eoms3x3}
 \tilde{M}=\begin{pmatrix} \sigma_0\,k^2 & \frac{ -\xi^2 n_\text{s} \, \sigma_0 + n_\text{s} \xi^2 \zeta_2}{T \mu}\, k^2  & ik -k^2\zeta_2 \xi \\
\frac{ -\xi^2 n_\text{s} \, \sigma_0 + n_\text{s} \xi^2 \zeta_2}{T \mu}\, k^2 & \frac{n_\text{s}\xi^2(n_\text{s} \zeta_3 - 2\mu^2 \zeta_2)+ \mu^4\sigma}{T^2 \mu^2}\, k^2  & \frac{k^2\xi\left( -n_\text{s}\zeta_3 + \mu^2\zeta_2 \right)}{T \mu} \\
ik -k^2\zeta_2 \xi& \frac{k^2\xi\left( -n_\text{s}\zeta_3 + \mu^2\zeta_2 \right)}{T \mu}  & \zeta_3\,k^2
\end{pmatrix} 
\end{equation}
The resulting modes are two sound modes and one diffusive mode. Notably, it is the latter which now signals the instability. Indeed, we find that it is given by $\omega = i\Gamma k^2/2$ with:
\be \label{diffusive3x3}
\Gamma = -2\frac{\ns + \xi \, \partial_\xi \ns}{\mu^2T^2\left(\mu \chi_{sh_\xi}^2+\chi_{ss}(\ns + \xi \, \partial_\xi \ns)\right)}\,\left( \ns^2\xi^2 \, \zeta_3 - 2\ns\mu^2  \xi^2 \,\zeta_2 +\mu^4\, \sigma_0\right)
\ee
We immediately see that $\Gamma$ vanishes if $\ns + \xi \, \partial_\xi \ns=0$.
Furthermore, recall that positivity of entropy production implies that $\sigma_0, \zeta_3 \geq 0$ and $\sigma_0 \zeta_3 \geq \zeta_2^2\xi^2$. This gives:
\be
\ns^2\xi^2 \, \zeta_3 - 2\ns \mu^2 \xi^2 \,\zeta_2 +\mu^4\, \sigma_0 \geq \left( \ns \xi \sqrt{\zeta_3}-\mu^2 \sqrt{\sigma_0}\right)^2 \geq 0
\ee
Therefore, denoting $\xi_c \equiv -\ns/\partial_\xi \ns$ as in (\ref{thecriterion}) and setting:
\be
\xi = \xi_c+ \delta \xi + \mathcal{O}(\delta \xi^2)
\ee
We find:
\be
\Gamma = 2\frac{\ns}{\xi_c\, \mu^3T^2 \chi_{sh_\xi}^2}  \left( \ns^2\xi_c^2 \, \zeta_3 - 2\ns \mu \xi_c^2 \,\zeta_2 +\mu^4\, \sigma_0\right) \, \delta \xi  + \mathcal{O}(\delta \xi^2) \, \text{,}
\ee
which implies that $\text{Im}(\omega)$ becomes positive for $\delta \xi > 0$, that is, for $\xi > \xi_c$. 
\bigskip
\\Analogously to what we saw in the probe limit, then, we are facing here a dynamical instability, which is also a thermodynamical instability since $\chi_{\xi \xi}= (\ns + \xi \, \partial_\xi \ns)/\mu$. Now, however, the instability is captured by a diffusive mode and not by a sound mode.
\subsection{``Full'' superfluid hydrodynamics}
As we saw at the end of chapter 2 in the $\xi = 0$ limit, adding the fluctuation $\delta \mathbf{u}$ of the normal fluid velocity to the picture we obtain two pairs of sound modes, together with one diffusive mode governed by the shear diffusivity $\eta$.
\\In general, with a non-vanishing background superfluid velocity the modes become much too involved to be written down. Similarly to what we did in the previous section, however, one can compute the modes around the critical value found in (\ref{thecriterion}) and check whether one still finds a vanishing mode. 
\bigskip
\\As expected, one does see an instability at $\xi =\xi_c \equiv -\ns/\partial_\xi \ns$. 
\\In particular, setting $\xi = \xi_c+ \delta \xi + \mathcal{O}(\delta \xi^2)$ as before, we find a sound mode $\omega = vk +i\Gamma k^2/2$ which is linear in $\delta \xi$:
\be \label{soundinst}
v =\nu \, \delta \xi + \mathcal{O}(\delta \xi^2)
\ee
\be 
\nu =\frac{\ns \, s^2}{2\, \mu \, \xi_c \left( -s^2\,\chi_{nh_\xi} + s\, \chi_{sh_\xi}(n + \xi_c \chi_{nh_\xi})-\xi_c \,\nn \,\chi_{sh_\xi}^2\right)} 
\ee
The attenuation constant $\Gamma$ can be written as:
\be \label{gammainst}
\Gamma = \frac{2\mu \,  \xi_c^2\, \nu^2}{\ns^2 \,T^2\,s^4}\, \gamma_A \gamma_BM^{AB} \, \delta \xi + \mathcal{O}(\delta \xi^2) \, \text{,}
\ee 
with
\be
M^{AB} = \begin{pmatrix} \sigma_0 & \xi_c  \,\zeta_6 & \xi_c \, \zeta_2 \\
\xi_c \, \zeta_6 & \eta & \zeta_1\\
\xi_c \, \zeta_2 & \zeta_1 & \zeta_3
  \end{pmatrix}
\ee
and
\be
\gamma_A = \begin{pmatrix} 
s\,\left(sT +\mu (n + \xi_c\, \chi_{nh_\xi}) \right) + \xi_c \,\chi_{sh_\xi} \left(sT + 2\mu\, \nn \right) \\
sT\, \chi_{sh_\xi} \\
(s^2\,\mu\, T \,\chi_{nh_\xi} + \ns s\xi_c (n + \xi_c\, \chi_{nh_\xi}) - 2\ns \nn \,\xi_c^2\,  \chi_{sh_\xi} -s\,T\,\mu\, n \,\chi_{sh_\xi})/\mu
\end{pmatrix} \, \text{.}
\ee
Most importantly, the matrix $M^{AB}$ is (semi)positive definite. This follows from positivity of entropy production, applied to matrix (\ref{dissmat}) in the collinear limit. Thus we see that for $\delta \xi >0$ (i.e. $\xi>\xi_c$), the imaginary part of this mode becomes positive, therewith signalling a dynamical instability.
\bigskip
\\In the rest of this chapter we are going to test this result on a holographic superfluid. In contrast to the last chapter, we are now going to include backreaction. This will allow us to account for the interplay between the fluctuations of the temperature and the normal fluid velocity and those of the Goldstone field and the chemical potential.
\section{Holographic check}
In the last chapter, in order to study the physics described by the holographic superfluid action (\ref{HHHaction}), we used the Schwarzschild metric: we were neglecting the backreaction between the metric and the matter fields. Now, however, we want to account for it, so, assuming that the background superfluid velocity points in $x$-direction, we write the background metric as:
\be \label{backreactedmetricansatz}
\diff s^2 = \frac{1}{u^2} \left(  -  \chi(u)f(u) \diff t^2  + b(u)\diff x^2 + \diff y^2 - 2c(u)\, \diff t \diff x+ \frac{1}{f(u)}\diff u^2  \right)
\ee
Note that in writing the ansatz in this manner we have chosen the normal fluid rest frame, in which the normal velocity, $u^\mu$, points only along the time direction. At the boundary we want to recover the AdS$_4$ metric (\ref{ppatchu}):
\be
\lim_{u\to 0} \diff s^2 = \frac{1}{u^2}\left(-\diff t^2+\diff \mathbf{x}^2+\diff u^2\right)
\ee
In particular, this means that:
\be
\lim_{u\to 0}c(u)=0 \qquad \qquad\lim_{u\to 0}f(u)=\lim_{u\to 0}\chi(u)=\lim_{u\to 0}b(u)=1
\ee
For what concerns the gauge field, as we did in the probe limit we require:
\be
\lim_{u\to 0}A_t = \mu \qquad \lim_{u\to 0}A_x= -\xi \, \text{,}
\ee
while we set the leading behaviour of the other components of $A_\mu$ to 0. Furthermore, as in the probe limit, we set the source of the background scalar field to zero, i.e. we impose:
\be
\lim_{u\to 0}\partial_u \psi = 0
\ee
As we did in the probe limit, we set the horizon radius $u_h$ to 1. This choice is convenient since it simplifies the notation and requires use to vary one less parameter when producing the numerical background solutions. However, it obscures the dimensionality of the fields, so that we have to make sure that one is always working with dimensionless variables, as we are going to see shortly.
\\At the horizon we want to obtain an asymptotically Schwarzschild geometry:
\be
\lim_{u\to 1}c(u)=\lim_{u\to 1}f(u)=0 \qquad \qquad\lim_{u\to 1}\chi(u)=\lim_{u\to 0}b(u)=1
\ee
As we saw in the probe limit, at the horizon we also need $\lim_{u\to 1} A_t = 0$ in order for $A_M A^M$ to stay finite.
\\Note that we recover the $3+1$-dimensional Schwarzschild metric with $c=0$, $b=\chi=1$ and $f(u)=1-u^3$.
\bigskip
\\With this ansatz the temperature (\ref{Tss}) of the system amounts to:
\be \label{thetemp}
T =  \frac{f'(1)\sqrt{\chi(1)}}{4\pi}
\ee
At the horizon $c(1)$ vanishes, so that the Bekenstein-Hawking entropy density reads:
\be
s = 4\pi \sqrt{b(1)}
\ee 
We work in the collinear limit, where the wavevector $\mathbf{k}$ points in the same direction as the background superfluid velocity: $\mathbf{k}=k \hat{x}$ \footnote{This choice is motivated by practicality and by the fact that one finds the smallest value for the critical velocity when the background superfluid velocity and the wavevector are antiparallel, see \cite{Schmitt:2014eka, Amado:2013aea}. Also recall that in the derivation of the Landau criterion (\ref{landaucrit}) one finds the critical velocity by assuming that $\mathbf{v}_\text{s}$ and $\mathbf{p}$ be antiparallel.}. Working in a gauge where $A_u = 0$ and  $\delta g_{Mu} = 0$, we perturb the fields in the following fashion \footnote{As we saw in the probe limit, the Maxwell equation of motion for $A_u$ is simply $\psi^2A_u=0$, which is trivially solved by $A_u=0$.}:
\begin{align}
\psi &\to \psi + [\sigma+i\bar{\sigma}]e^{-i(\omega t-kx)} \qquad \psi^* \to \psi + [\sigma-i\bar{\sigma}]e^{-i(\omega t-kx)} \nonumber\\
A_\mu &\to A_{\mu} + a_\mu e^{-i(\omega t-kx)} \,\,\,\quad \qquad g_{\mu \nu} \to g_{\mu \nu}+ \frac{1}{u^2}h_{\mu \nu}e^{-i(\omega t-kx)} 
\end{align}
In the collinear limit we can consistently set all fluctuations to zero apart from $a_t, a_x, h_{tt},h_{tx}$, $h_{xx}$ and $h_{yy}$. This follows from symmetry considerations: for instance the spatial part of the vector $h_{t\mu}$ is odd under parity, so in the equations of motion it will couple to the wavevector and the spatial part of the gauge field. But since both of them point in $x$-direction, $h_{xy}$ decouples from the other fluctuations.
\\In general, with the above ansatz the linearized equations of motion are easily found, but their expressions are too cumbersome to be shown here. 
\bigskip
\\Since we wish to compute retarded Green's functions, as we saw in Chapter 3 we have to impose infalling boundary conditions \cite{Son:2002sd}, imposing that in the vicinity of the horizon the fields behave as $(1-u)^{-i\omega/4\pi T}$. Some fields require one or two additional powers of $(1-u)$. In particular, we find:
\begin{align}
a_t(u) &= (1-u)^{1-i\omega/4\pi T}\left[a_{t}^{(0)}+a_{t}^{(1)}(1-u)+\cdots \right]\nonumber\\
a_x(u) &= (1-u)^{-i\omega/4\pi T}\left[a_x^{(0)}+a_{x}^{(1)}(1-u)+\cdots \right]\nonumber\\
\sigma(u) &=(1-u)^{-i\omega/4\pi T}\left[\sigma_r^{(0)}+\sigma^{(1)}(1-u)+\cdots \right]\nonumber\\
\bar{\sigma}(u) &=(1-u)^{-i\omega/4\pi T}\left[\sigma_i^{(0)}+\bar{\sigma}^{(1)}(1-u)+\cdots \right]\nonumber\\
h_{tt}(u) &=(1-u)^{2-i\omega/4\pi T}\left[h_{tt}^{(0)}+h_{tt}^{(1)}(1-u)+\cdots \right]\nonumber\\
h_{tx}(u) &=(1-u)^{1-i\omega/4\pi T}\left[h_{tx}^{(0)}+h_{tx}^{(1)}(1-u)+\cdots \right]\nonumber\\
h_{xx}(u) &=(1-u)^{-i\omega/4\pi T}\left[h_{xx}^{(0)}+h_{xx}^{(1)}(1-u)+\cdots \right]\nonumber\\
h_{yy}(u) &=(1-u)^{-i\omega/4\pi T}\left[h_{yy}^{(0)}+h_{yy}^{(1)}(1-u)+\cdots \right] \,.
\end{align}
Inserting this expansion into the equations of motion (\ref{theeoms1},\ref{theeoms2},\ref{theeoms3}) we obtain eight dynamical second-order equations, one for every field. Similarly to what we saw in the probe limit, however, we can trade four dynamical equations for four first-order constraint equations. In particular, one can show that a set of fields solving the dynamical equations and satisfying the constraint equations at one value of $u$ will be a solution of the latter in the whole space \cite{gauntlettdonos}. 
\\In contrast to what we did in the probe limit, the numerical results presented here are obtained by replacing four dynamical equations by four constraint equations, the main practical advantage being the reduced number of boundary condition needed. We explicitly checked numerically that the two sets of equations deliver the same results. 
\bigskip
\\In the probe limit we had one boundary condition (\ref{robinprobe}) between the leading behaviours of the fields at the horizon. Now we have four. They are derived by expanding the equations of motion in the vicinity of $u=1$; setting the charge $e$ to 1 they read:
\be
h_{xx}^{(0)}+b(1)\,h_{yy}^{(0)}=0
\ee
\be
a_t^{(0)} + i\, \frac{ c'(1)\,\omega - 4\pi T \sqrt{\chi(1)}\,k}{(4\pi T -i\omega)\,b(1)}\, a_x^{(0)} +\frac{8\pi T\sqrt{\chi(1)}\,\psi(1)}{4\pi T -i\omega}\, \bar{\sigma}^{(0)}=0
\ee
\be
h_{tx}^{(0)} - \frac{8\pi T \sqrt{\chi(1)}\,\psi(1)\,A_x(1)}{4\pi T -i\omega} \, \bar{\sigma}^{(0)}  -\frac{4\pi T \, A_t'(1) }{4\pi T -i\omega} \, a_x^{(0)} + i\frac{4\pi T \sqrt{\chi(1)}\, k -\omega c'(1)}{4\pi T-i\omega} h_{yy}^{(0)}=0
\ee
\be
\begin{split}
h_{tt}^{(0)} &+ \frac{8\pi T\Big(4\pi TA_x(1)^2 + i\omega \left(A_x(1)^2-2b(1)\right)\Big)\,\chi(1)\psi(1)}{(4\pi T -i\omega)(6\pi T-i\omega)\, b(1)} \, {\sigma}^{(0)}  \\
&+ \frac{8i\pi T\Big( 4\pi T A_x(1)\sqrt{\chi(1)}k +i(8\pi T -i\omega) b(1)A_t'(1)-\omega \,c'(1)A_x(1)\Big)\,\sqrt{\chi(1)}\psi(1)}{(4\pi T -i\omega)(6\pi T-i\omega)\, b(1)} \, \bar{\sigma}^{(0)} \\
&+ \frac{8\pi T\Big( 4\pi T \left( \chi(1)\,A_x(1)\,\psi(1)^2-k\sqrt{\chi(1)}\,A_t'(1)\right) +i\omega\, A_t'(1)\,c'(1)\Big)}{(4\pi T -i\omega)(6\pi T-i\omega)\, b(1)} \, {a_x}^{(0)} \\
&+ \frac{4\pi T\chi(1)  \left( -4\pi T k^2 +(4\pi T + i\omega)A_x(1)^2 \psi(1)^2 \right) + 8\pi T k\omega \, \sqrt{\chi(1)}c'(1)-\omega^2 c'(1)^2}{(4\pi T -i\omega)(6\pi T-i\omega)\, b(1)} \, {h_{yy}}^{(0)}  =0
\end{split}
\ee
\\As we explained at the end of Chapter 3, at the boundary we impose gauge-invariant boundary conditions. We have a total of four such conditions. We already encountered three of them in the probe limit: they set the boundary value of the electric field and of the scalar sources. Notably, they do not change when taking the backreaction between the matter and the metric fields into consideration. The fourth one involves the metric tensor:
\be
2\omega k \, h_{tx} + \omega^2\,(h_{xx}-h_{yy}) + k^2(h_{yy}-h_{tt}) \bigg|_{u=0}
\ee
\subsubsection{Using scale-invariant quantities}
In order to compute the modes in hydrodynamics we start by writing the equations of motion as in (\ref{eomscollinear}) and build the matrix $K$ as in (\ref{Kdef}). The modes $\omega(k)$ are the solutions of $\text{det}(K)=0$, since the propagators are built with $K^{-1}$. 
\\A meaningful test of AdS-CFT correspondence is the comparison between the quasinormal modes and the hydrodynamic modes of the boundary theory, the so-called ``matching'' of the modes. To do so, one has to extract from the boundary theory the thermodynamic data entering the hydrodynamic equations of motion. In particular, one has to compute thermodynamic derivatives. In principle, the estimation of derivatives is not complicated: one can simply resort to computing finite differences. For instance, let us assume that one wants to compute the derivative of a vacuum expectation value $\alpha$ with respect to the source $\beta$ in the vicinity of $\beta=\beta_0$. One simply generates a first background solution with $\beta =\beta_0$, and then another one with $\beta =\beta_0 + \epsilon$ (where $\epsilon \ll \beta_0$). Then one can estimate:
\be \label{findiff}
\frac{\partial \alpha}{\partial \beta} \approx \frac{\alpha(\beta_0 + \epsilon)-\alpha(\beta_0)}{\epsilon}
\ee
For a better estimation one can generate multiple backgrounds at various values of $\beta$ and apply Runge-Kutta methods, or even produce a whole set of background on a set of values of $\beta$ sitting on a Chebyshev grid (see the Appendix for more details). In any case, the basic idea is still encoded in (\ref{findiff}).
\bigskip
\\There is, however, one subtlety compared to the probe limit. Thermodynamic derivatives with respect to one source are computed keeping all other sources constant. In the probe limit the hydrodynamics of the system depended on two fluctuations: that of the chemical potential $\delta \mu$ and that of the superfluid velocity $\delta \xi$. Now, however, we also have to keep track of the fluctuation of the temperature $\delta T$ (as we saw above, we generate the background solutions in the normal fluid's rest frame, so $\delta \mathbf{u}=0$, and in any case we already know the susceptibilities of the $\boldsymbol{\pi}$-sector (\ref{chicomplete})). The temperature (\ref{thetemp}) depends on the horizon radius $r_h$ and the boundary data $\xi$ and $\mu$. So, in order to vary one of the three fluctuations independently we should generate solutions with different horizon radii. This is possible to do, but proves technically cumbersome.
\bigskip
\\Fortunately, in holography it is only meaningful to measure dimensionless ratios (in fact, the boundary theory is conformal, so it is invariant under rescalings). 
We can build two dimensionless variables out of $\mu$, $\xi$ and $T$: for instance the ratios $\tilde{\xi}\equiv \xi/\mu$ and $\tilde{T}\equiv T/\mu$. Any dimensionless quantity can be written as a function of these two ratios alone. In particular, in this work we used the \textit{reduced entropy} $\sigma \equiv s/n$ \cite{Arean:2021tks} and the \textit{superfluid fraction} $r_\text{s}\equiv n_\text{s}/n$. Moreover, we make use of the fact that the boundary theory is conformal, giving rise to relations between the susceptibilities as we saw in (\ref{thsusc2}) for vanishing background superfluid velocity.
\\Specifically, in a two-dimensional conformal field theory one finds that:
\begin{align}
\begin{split}
\chi_{ss} &= \frac{\xi^2 (\ns -\xi\partial_\xi \ns ) + \mu \left(2sT +\mu (\mu \chi_{nn}-2n + 2\xi \chi_{nh_\xi})\right)}{\mu T^2} \\
\chi_{sn} &= \frac{\mu \chi_{nn}-2n +\xi \chi_{nh_\xi}}{T} \\
\chi_{sh_\xi} &= \frac{\xi^2\partial_\xi \ns -\xi \ns + \mu^2 \chi_{n h_\xi}}{\mu T}
\end{split}
\end{align}
With:
\be
\begin{split}
\chi_{nn} = \frac{1     }{\mu^2(sT+n \mu)(sT\mu + n\mu^2-\ns \xi^2)} \bigg[&\mu \xi^3 n^3 \,\pdertmedium{r_\text{s}}{\tilde{\xi}}{\tilde{T}} + n^2T^2(sT\mu +n\mu^2 -\ns \xi^2)\,\pdertmedium{\sigma}{\tilde{T}}{\tilde{\xi}}  \\
&  +   Tn^2\xi (sT\mu + 2n\mu^2-\ns\xi)\,\pdertmedium{\sigma}{\tilde{\xi}}{\tilde{T}}  \\
&\quad + n^2\mu^2\left( 2\mu(sT+n\mu)-\ns \xi^2\right)  \bigg]
\end{split}
\ee
\be
\chi_{nh_\xi} = -\frac{1}{\mu(sT\mu +n\mu^2-\ns \xi^2)}\bigg[ n^2T\mu \,\pdertmedium{\sigma}{\tilde{\xi}} {\tilde{T}}-n^2\xi^2\,\pdertmedium{r_\text{s}}{\tilde{\xi}}{\tilde{T}}  +\ns n \mu\, \xi\bigg]
\ee
\be
\partial_\xi \ns = \frac{1}{sT\mu +n\mu^2-\ns \xi^2}\bigg[ \ns n T \,\pdertmedium{\sigma}{\tilde{\xi}} {\tilde{T}}-n(sT+n\mu)\,\pdertmedium{r_\text{s}}{\tilde{\xi}}{\tilde{T}}  +\ns^2 \, \xi \bigg]
\ee
In this way we can write any thermodynamic derivative appearing in (\ref{eomscollinear}) in terms of derivatives of scale-invariant quantities in a two-parameter space parametrized by the dimensionless ratios $\tilde{\xi}$ and $\tilde{T}$ (even if the closed-form expressions for the modes are much too cumbersome to be written down).
\subsection{Results}
As we did in the probe limit, we first plotted $\ns \xi/\mu^3$ for a range of temperatures (Figure \ref{fig:nsvsbr}). Note the $\mu^3$ factor, which guarantees that we are plotting a dimensionless ratio. Compared to the probe limit, which is a valid approximation around $T_c$, it is now physically meaningful to investigate lower temperatures, so went we went down to $\sim 0.14 T_c$. For every temperature we find the maximum of the curve, which gives the critical velocity as in (\ref{thecriterion}).
 \begin{figure}[h]
    \centering
    \includegraphics[scale=0.52]{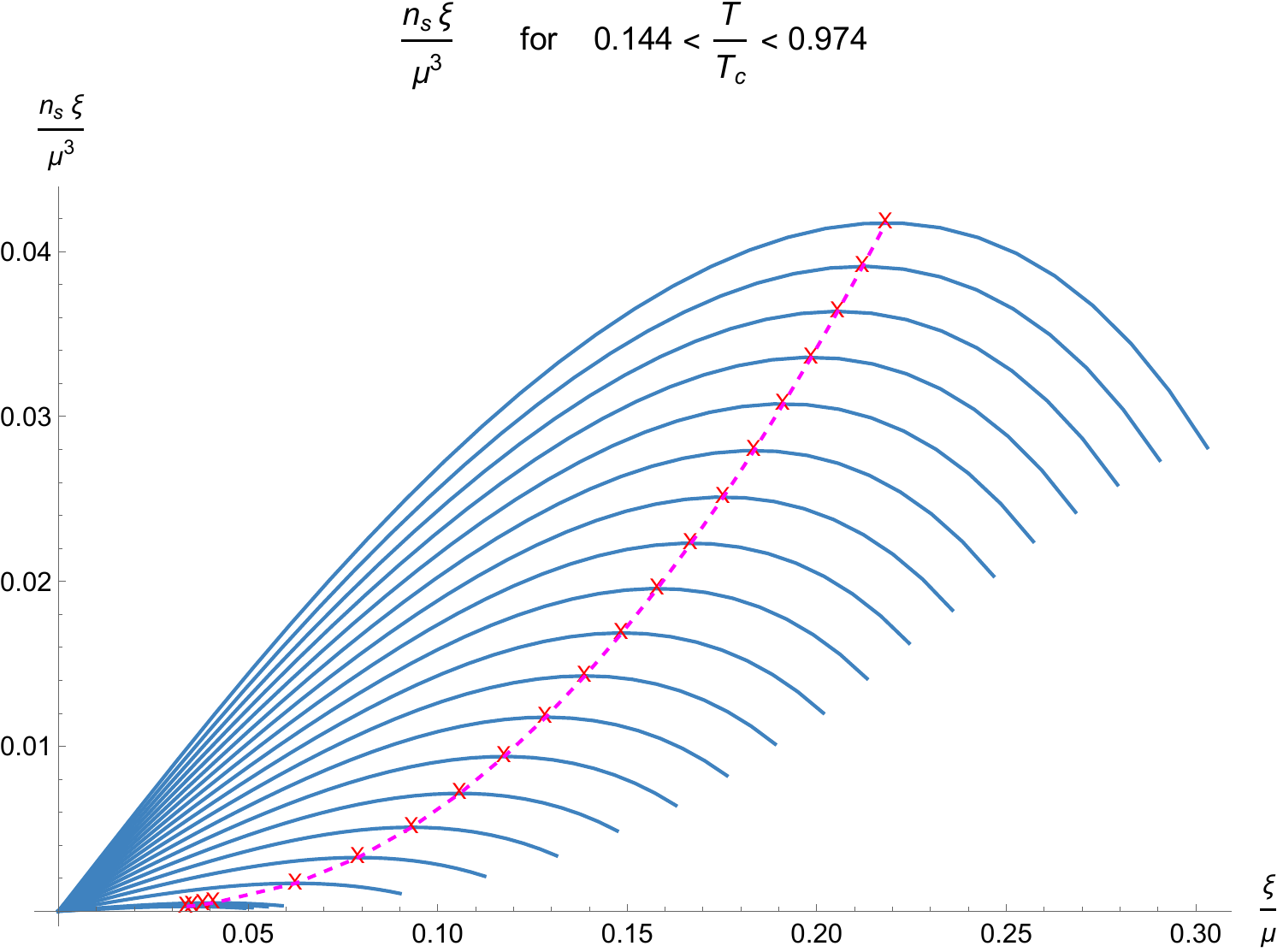}
    \caption{$n_\text{s}\xi/\mu^3$ for a range of 20 values of temperatures between 0.144$T_c$ and 0.97$T_c$. The red crosses indicate the position of the maximum of each curve, estimated via a polynomial regression. $T_c$ is the critical temperature at zero background superfluid velocity; more precisely, it is the $T/\mu$ ratio for a solution close to the phase transition to the normal phase. Data computed on a grid of 30 Chebyshev points, at precision 30. }
   \label{fig:nsvsbr}
\end{figure}
\bigskip
\\We then computed the quasinormal modes for a range of temperatures. As one can see in Figures \ref{fig:vplusbr} and \ref{fig:gammaplusbr}, at the values of $\xi/\mu$ where $\ns \xi/\mu^3$ attains its maximum (signalled by the dashed vertical lines), one second sound mode does indeed show an instability: both its real and imaginary part change sign. This happens consistently throughout the range of temperatures that we took into consideration. \\At the same time, the other modes do not show any unstable behaviour, as it can be seen in Figures \ref{fig:vminusbr}, \ref{fig:gammaminusbr},  \ref{fig:vplusfirstbr}, \ref{fig:gammaplusfirstbr},  \ref{fig:vminusfirstbr} and \ref{fig:gammaminusfirstbr}.
\\In Figures \ref{fig:vplusbr},  \ref{fig:gammaplusbr}, \ref{fig:vminusbr}, \ref{fig:gammaminusbr}, \ref{fig:vplusfirstbr},   \ref{fig:gammaplusfirstbr}, \ref{fig:vminusfirstbr}, \ref{fig:gammaminusfirstbr}  we also plotted the speeds of sound and the attenuations, as explained in the chapter on hydrodynamics (\ref{modesformula}), computed using scale-invariant quantities as we saw in the last section (in the plots, the dots correspond to the QNM data, while the lines represent the speeds of sound obtained by matching). We observe a very good agreement between the two sets of data; as expected, we found the real part of the QNMs to be linear in the wavevector $k$, while the imaginary part has a quadratic behaviour (at least in the stable region). 
\\As we saw in the probe limit, for a comparison of the attenuation constants it is necessary to compute the dissipative coefficients through the Kubo relations (\ref{kubo}). We do so by taking the $\omega \to 0$ limit of the holographic Green's functions of spacelike quantities, evaluated at $k=0$.
\\As long as the background is determined up to a very high precision, the speeds of sound and attenuation constants extrapolated from the QNMs are found to agree very well with those expected from the boundary hydrodynamics.
\begin{figure}[h]
    \centering
    \includegraphics[scale=0.59]{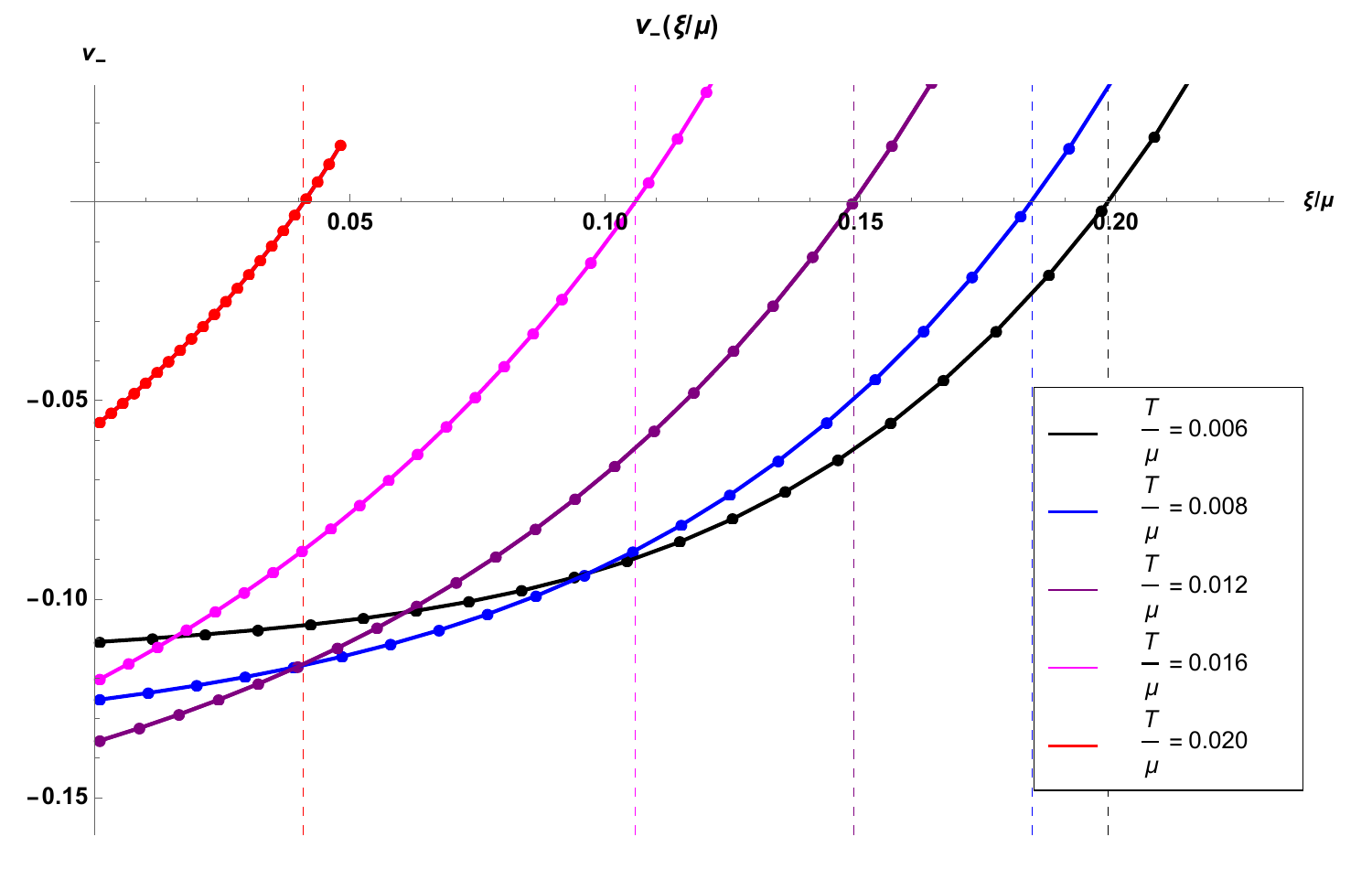}
    \caption{Second sound: the speed of sound of the mode signalling the instability, as a function of $\xi/\mu$, for four different temperatures ($T\approx 0.96 T_c$, $T\approx 0.77 T_c$, $T\approx 0.58 T_c$, $T\approx 0.39 T_c$, $T\approx 0.29 T_c$). The dots are obtained from $\text{Re}(\omega)$ at $k=10^{-3}$ (computed with 50 Chebyshev grid points and working precision 50, except for the lower temperature, where we used 70 grid points at working precision 70). The lines correspond to the hydrodynamic dispersion relations solving (\ref{modesformula}), obtained from backgrounds with 110 points for the three lowest temperatures, while for the two highest temperatures we used 70 grid points to evaluate the Kubo formulae and 50 for the thermodynamic derivatives. The consistency between the two sets of data is quite good, with a maximum relative error around $10^{-6}$ in the range displayed in this graph, for the four hydrodynamic modes. The vertical lines signal, for every value of $T/\mu$, the value of $\xi$ where $\xi \ns/\mu^3$ attains its maximum. As one can see at this critical value $v_-$ vanishes and changes sign, thus validating the criterion (\ref{thecriterion})}
    \label{fig:vplusbr}
\end{figure}
\bigskip
\\Finally we can plot the phase space, see Figure \ref{fig:phasespacebr}. This plot is obtained from background data, but as we just saw we checked the agreement with the QNM is very good. 
\\The stable region is plotted in red, while in the white region the condensate vanishes, leaving us with a normal fluid. The transition to the first unstable region, in blue, is signalled by our criterion being fulfilled (\ref{thecriterion}). A second instability (grey region) is signalled by the speeds of second acquiring an imaginary part. As we mentioned in the case of the probe limit, this corresponds to the ``two-stream instability'' discussed in \cite{schmitt1, schmitt2, schmitt3}.
\\Note that for $T/\mu \lesssim 0.0139$ we find a second region with unstable sound, but real sound speeds, beyond the area with complex speed of sound. 
\begin{figure}[h]
    \centering
    \includegraphics[scale=0.59]{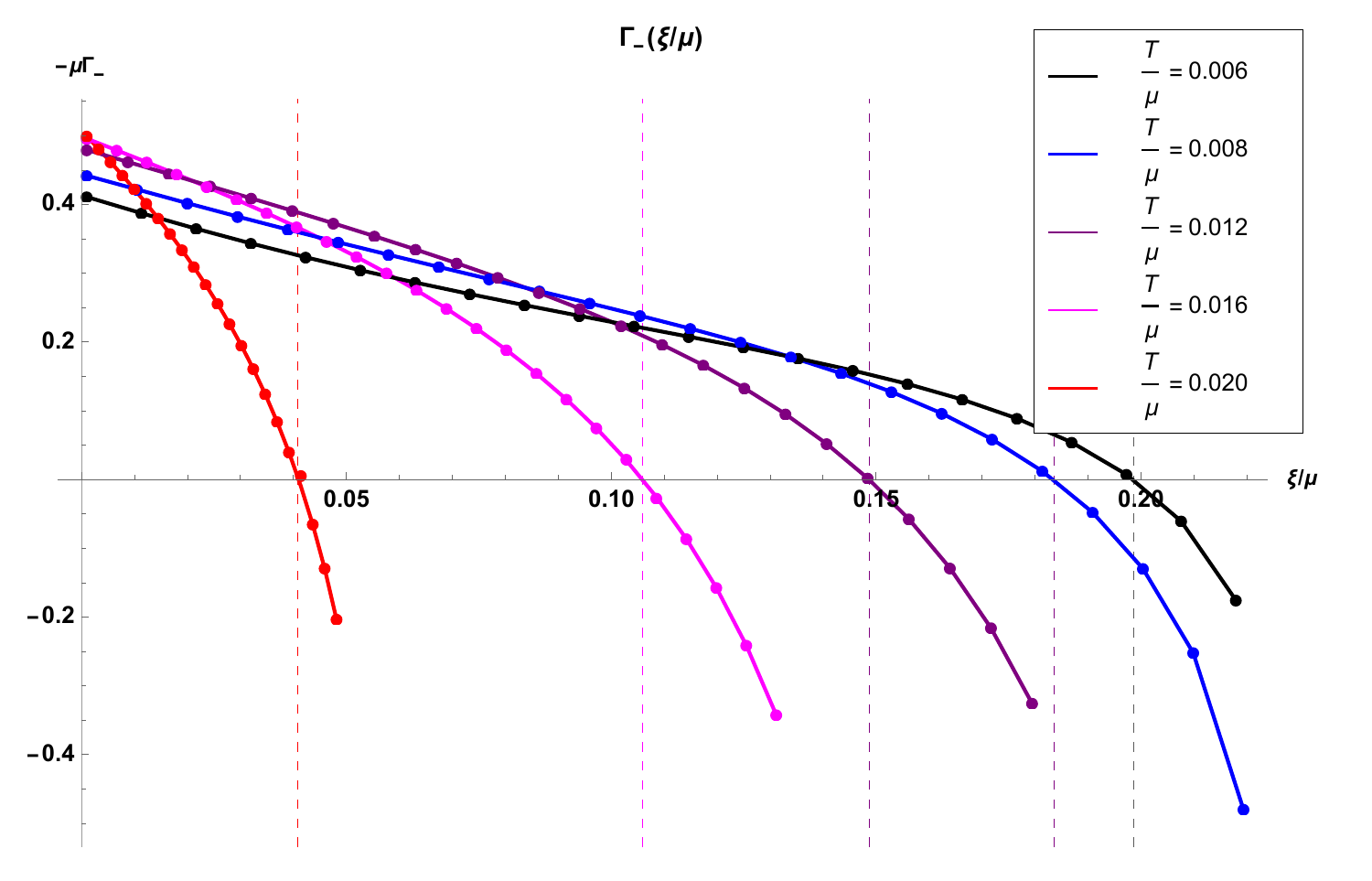}
    \caption{Second sound: the attenuation constant $\Gamma$ of the mode signalling the instability, as a function of $\xi/\mu$, obtained from the same set of data as Figure \ref{fig:vplusbr}. As one can see at the critical value of $\xi$ also $\Gamma_-$ vanishes and changes sign, validating the criterion (\ref{thecriterion}). As before, the dots correspond to the QNMs, the lines to the hydrodynamic dispersion relation (\ref{modesformula}) evaluated with the boundary data. The relative error on the imaginary parts of the modes is generally around $10^{-4}$ on all of the four hydrodynamic modes, as long as the background is determined to a very high precision.}
    \label{fig:gammaplusbr}
\end{figure}

\begin{figure}[h]
    \centering
    \includegraphics[scale=0.59]{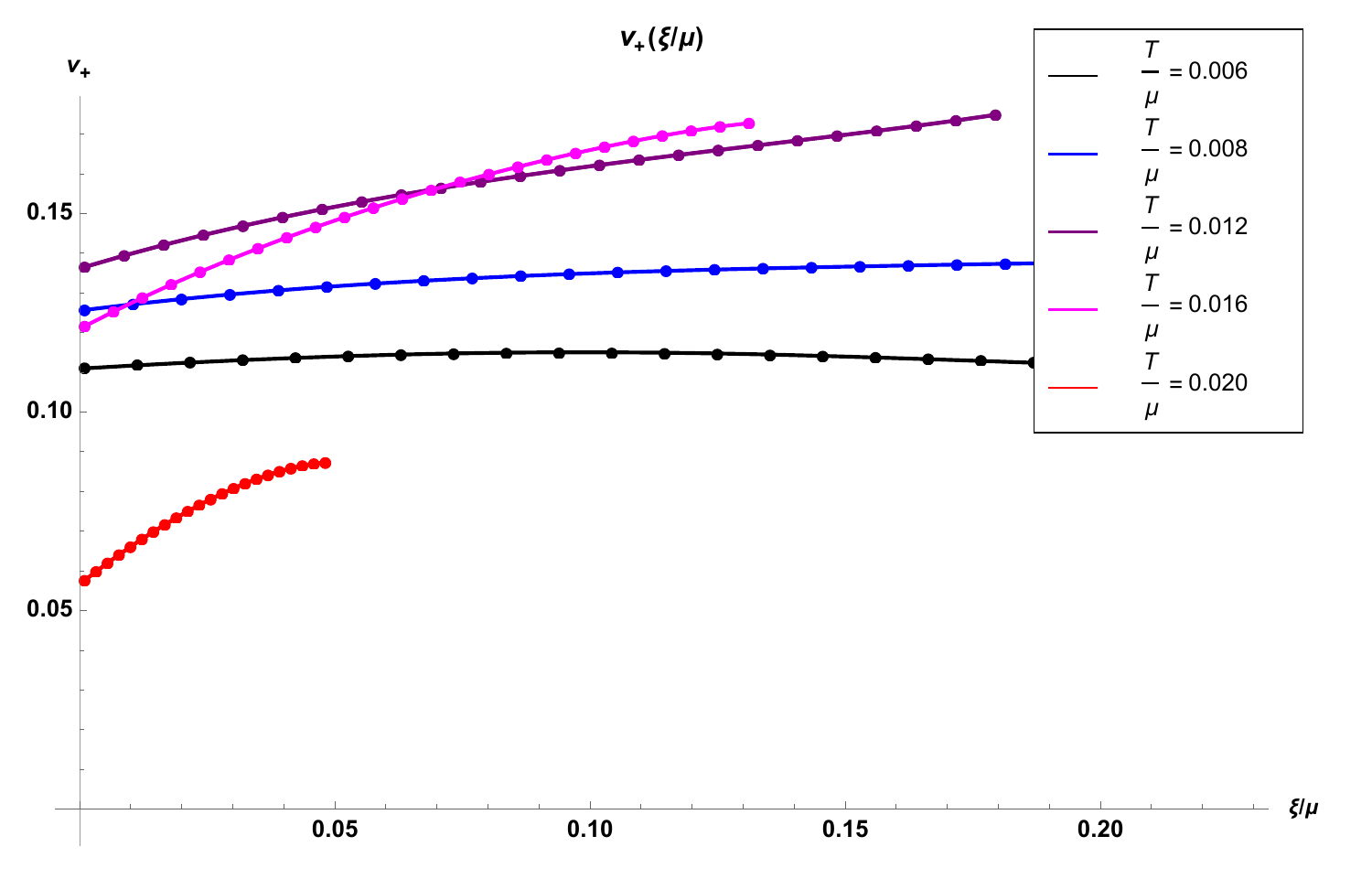}
    \caption{Second sound: the speed of sound of the stable mode as a function of $\xi/\mu$, obtained from the same set of data as Figure \ref{fig:vplusbr}.}
    \label{fig:vminusbr}
\end{figure}

\begin{figure}[h]
    \centering
    \includegraphics[scale=0.59]{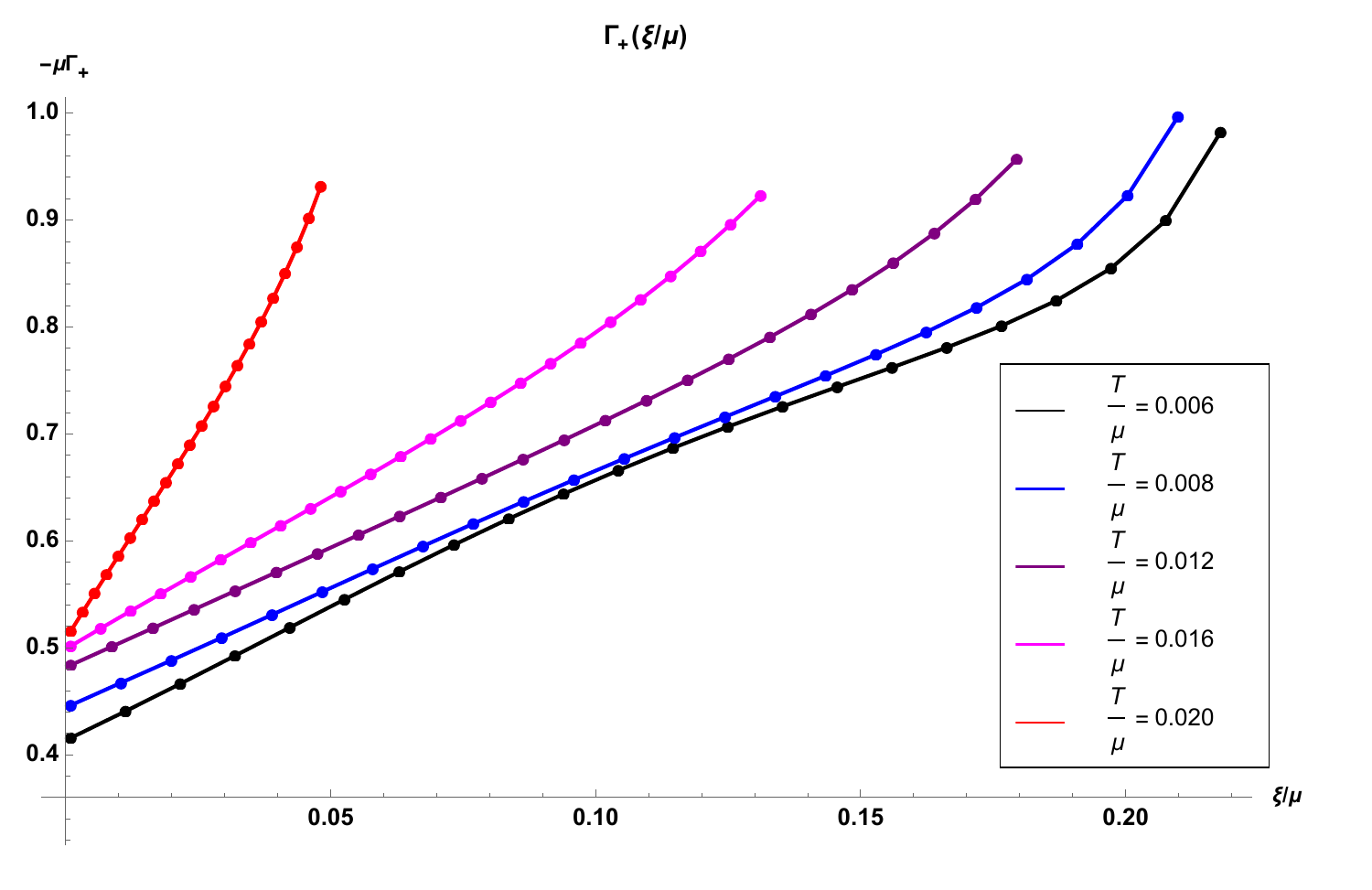}
    \caption{Second sound: the attenuation constant $\Gamma$ of the stable mode as a function of $\xi/\mu$, obtained from the same set of data as Figure \ref{fig:vplusbr}.}
    \label{fig:gammaminusbr}
\end{figure}

\begin{figure}[h]
    \centering
    \includegraphics[scale=0.59]{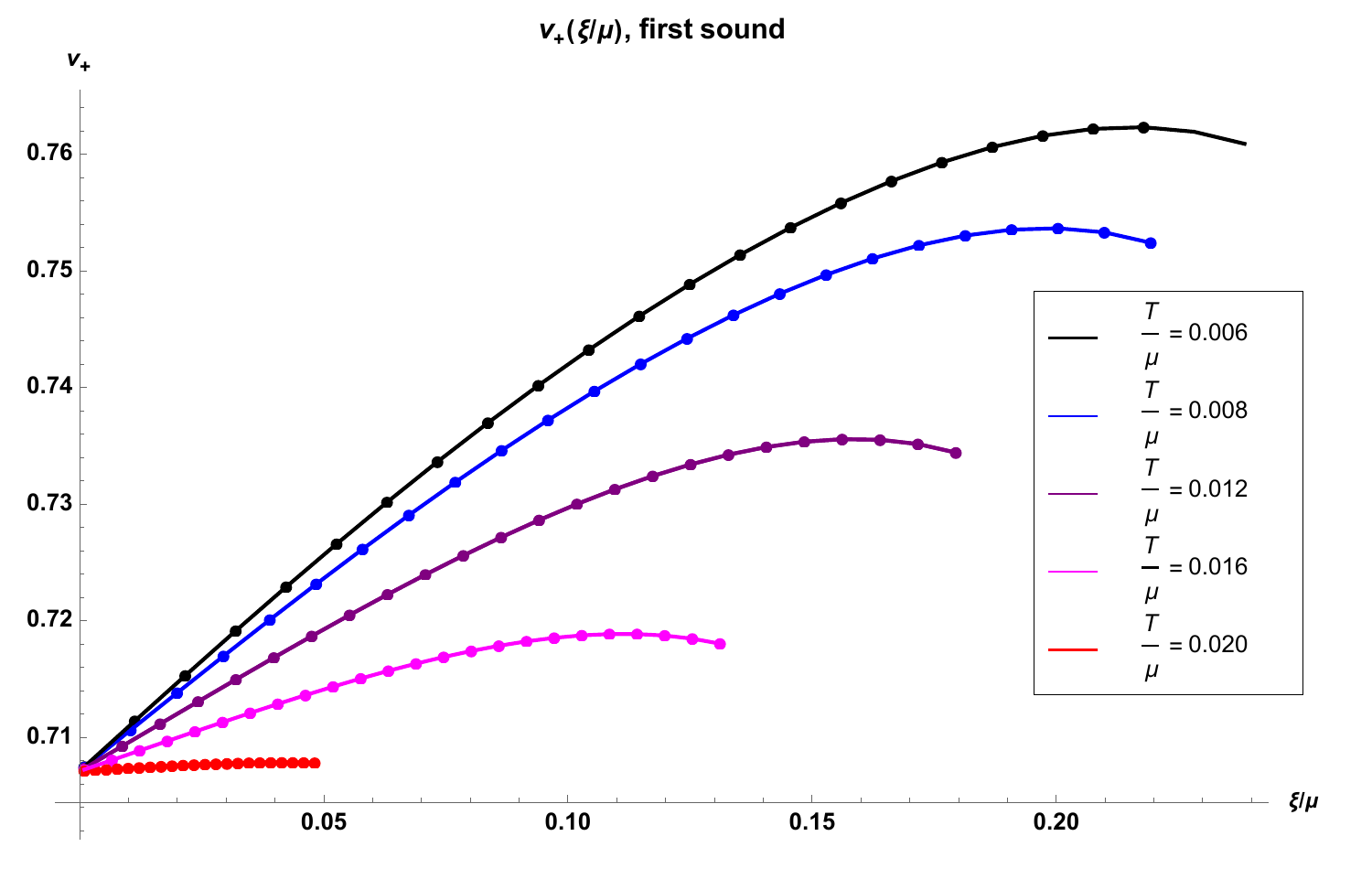}
    \caption{First  sound: the speed of sound of the positive first sound mode as a function of $\xi/\mu$, obtained from the same set of data as Figure \ref{fig:vplusbr}.}
    \label{fig:vplusfirstbr}
\end{figure}

\begin{figure}[h]
    \centering
    \includegraphics[scale=0.59]{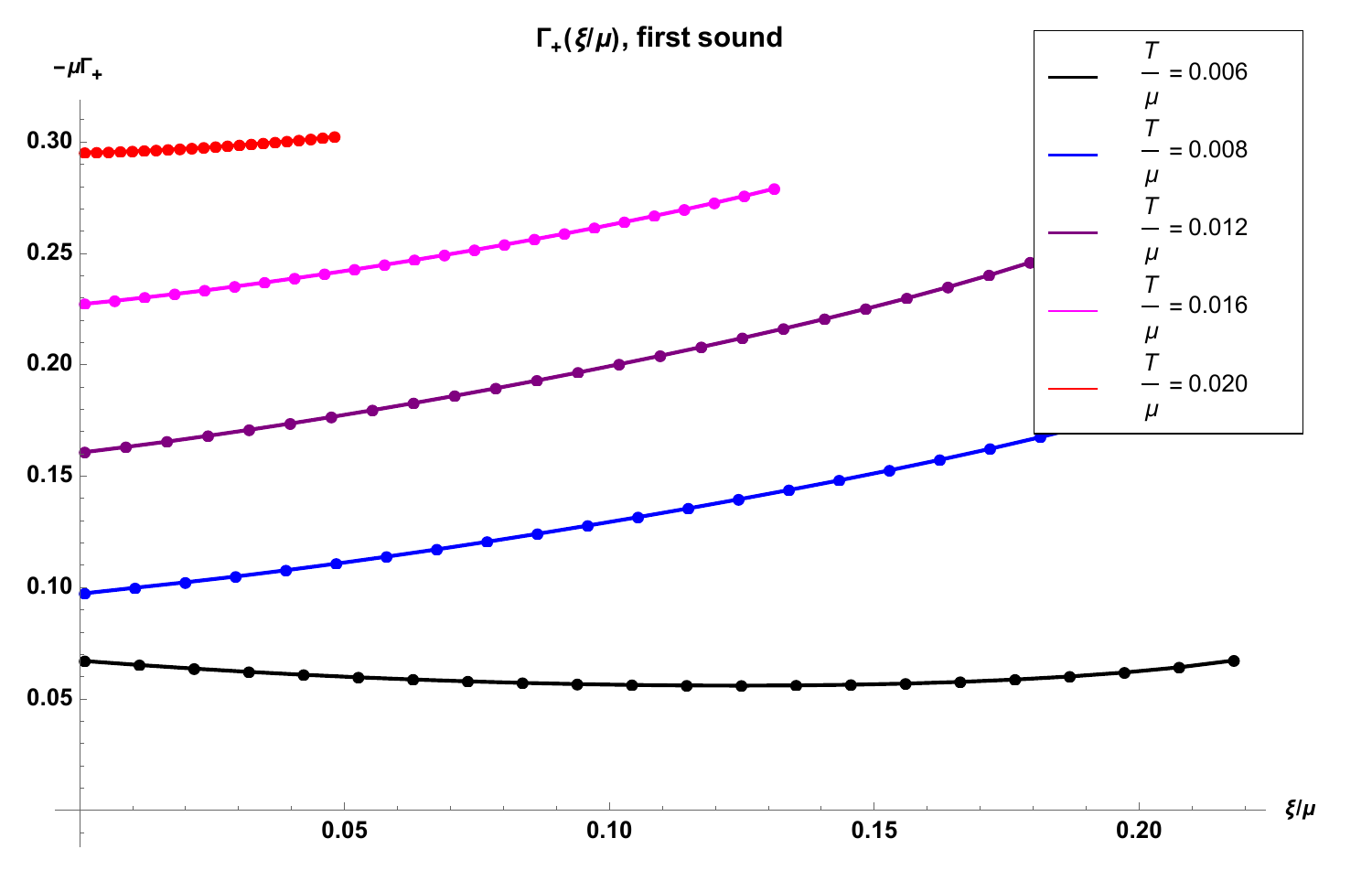}
    \caption{First  sound: the attenuation constant $\Gamma$  of the positive first sound mode as a function of $\xi/\mu$, obtained from the same set of data as Figure \ref{fig:vplusbr}.}
    \label{fig:gammaplusfirstbr}
\end{figure}

\begin{figure}[h]
    \centering
    \includegraphics[scale=0.59]{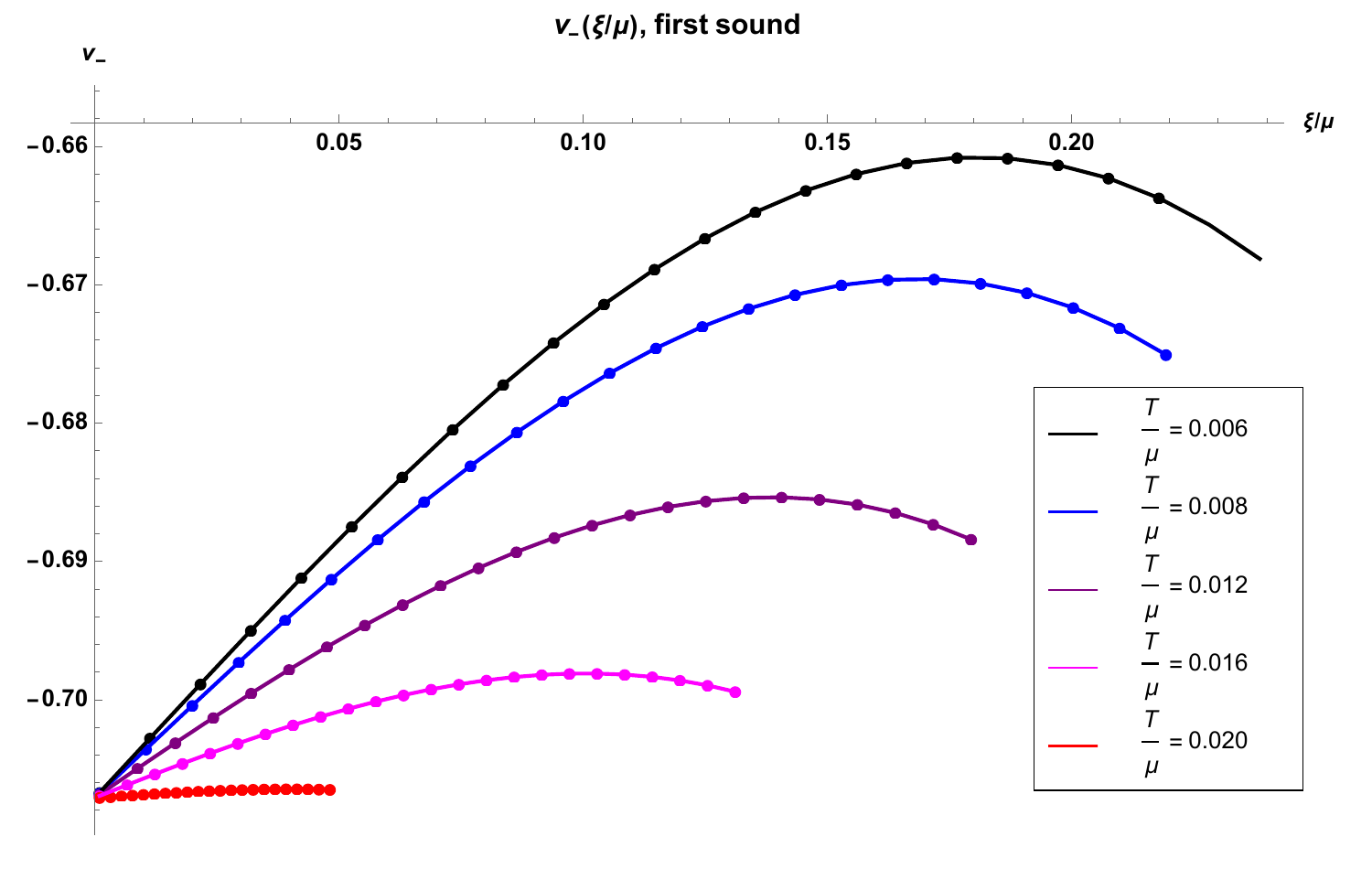}
    \caption{First  sound: the speed of sound of the negative first sound mode as a function of $\xi/\mu$, obtained from the same set of data as Figure \ref{fig:vplusbr}.}
    \label{fig:vminusfirstbr}
\end{figure}

\begin{figure}[h]
    \centering
    \includegraphics[scale=0.59]{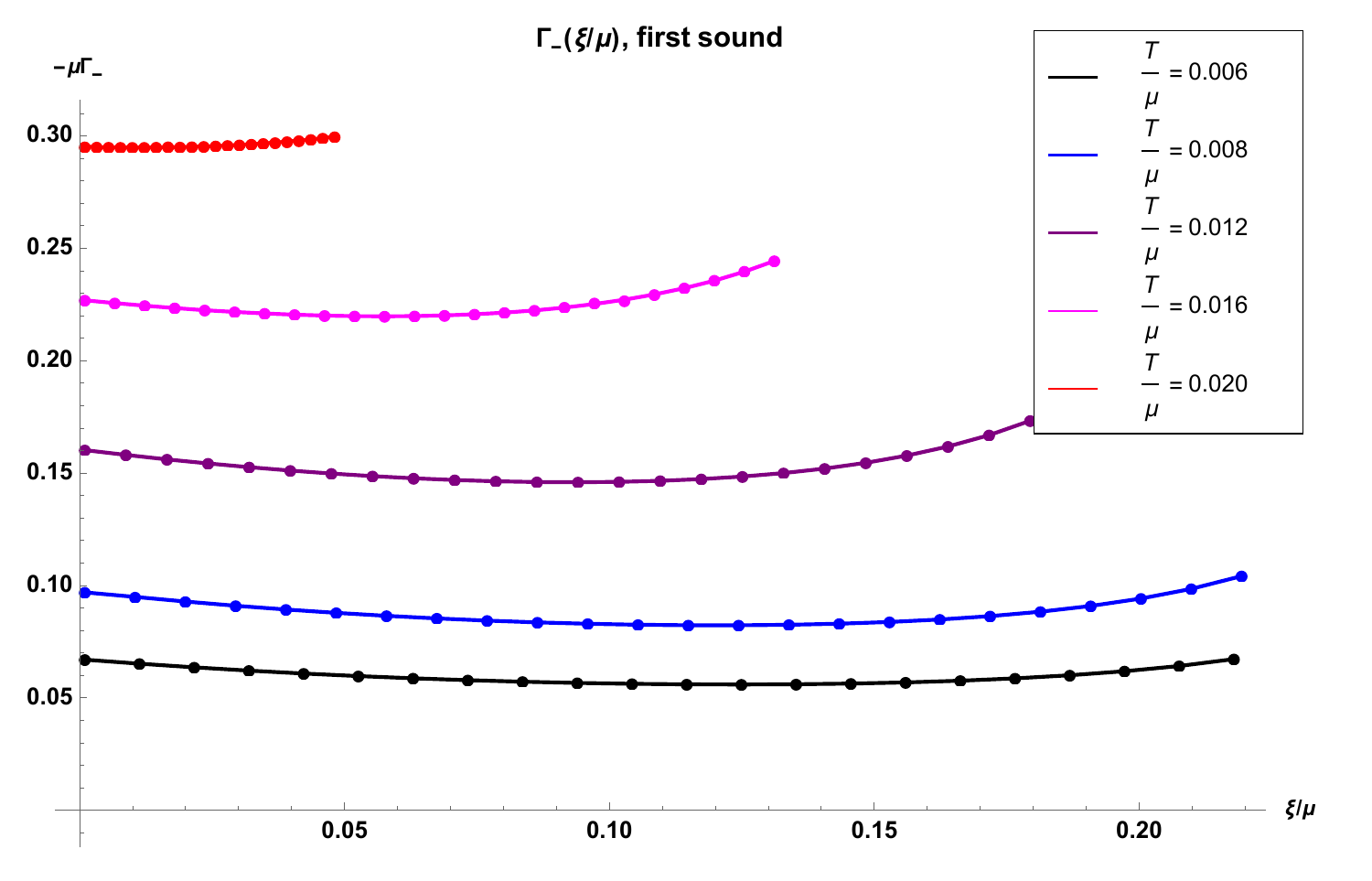}
    \caption{First sound: the attenuation constant $\Gamma$  of the negative first sound mode as a function of $\xi/\mu$, obtained from the same set of data as Figure \ref{fig:vplusbr}.}
    \label{fig:gammaminusfirstbr}
\end{figure}

\begin{figure}[h]
    \centering
    \includegraphics[scale=0.45]{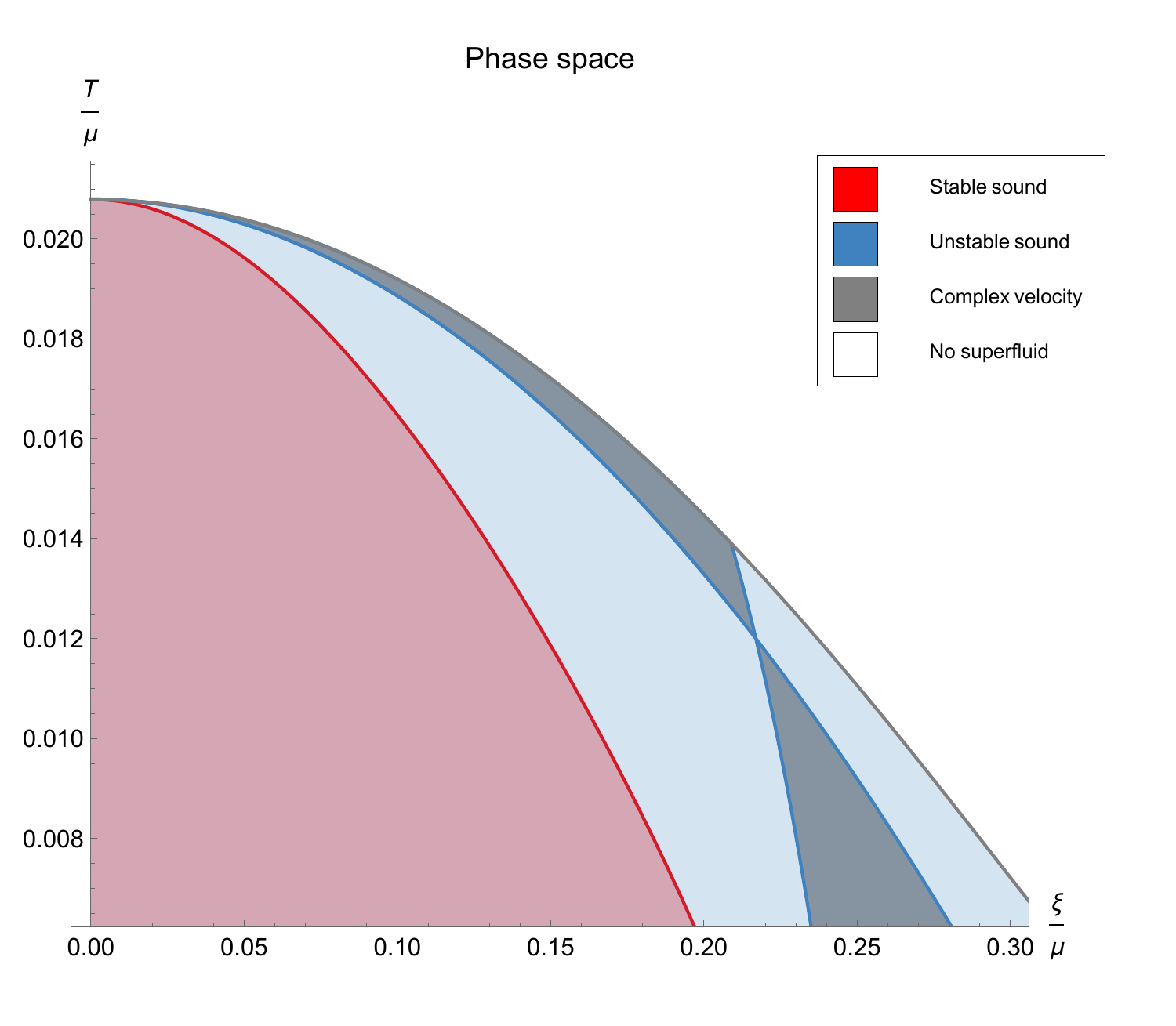}
    \caption{The phase space for $0.33\, T_c \lesssim T \lesssim T_c$. The red area features two stable sound modes; beyond the red line, in the blue area, $v_-$ and $\Gamma_-$ change sign, leading to a dynamical instability.  In the grey region $\text{Im}\,\omega$ acquires a linear contribution in $k$. In the white region the scalar field does not condense and the system reduces to a normal fluid. The curves are obtained from fits of background data, namely the maxima of the curves plotted in Figure \ref{fig:nsvsbr} for the red line and the speeds of sound computed with the matching procedure described in the previous section for the blue lines. The grey line shows, for a given $T/\mu$ ratio, the value of $\xi/\mu$ that gives $\psi(0)=10^{-7}$.}
    \label{fig:phasespacebr}
\end{figure}

\chapter{Conclusions and outlook}
\maketitle

In this thesis we showed that the Landau criterion for superfluid instabilities can be recast as a thermodynamic instability. Most importantly, within hydrodynamics this thermodynamic instability gives rise to a linear dynamical instability. We verified this statement with the help of the AdS-CFT correspondence. 
\bigskip
\\As a first step we gave a general overview of hydrodynamics and, more specifically, superfluid hydrodynamics in the Landau-Tisza framework. We then proceeded by outlining the features of the AdS-CFT correspondence that are relevant to our analysis, most notably the computation of the correlators and the quasinormal modes in a thermal theory. Finally, we were able to find and analyze an instability, arising (for a given temperature) at a critical value of the background superfluid velocity.  We found the same instability both in the so-called probe limit, where the fluctuations of the temperature and the normal fluid velocity are frozen, and in full relativistic superfluid hydrodynamics. Our results were corroborated in both cases by a holographic analysis.  We could indeed observe a very good correspondence between the results yielded by the dual gravity theory and those obtained using hydrodynamics.
\\Furthermore, as we show in the Appendix, our criterion matches the Landau criterion for Galilean superfluids.
\bigskip
\\The instability that we found is at the same time of dynamical and thermodynamic nature: it is signalled by the divergence of $\chi_{\xi \xi}$, the diagonal susceptibility related to the superfluid velocity, which eventually changes sign, thereby becoming negative. Most importantly, we could locate the instability by studying perturbations around thermodynamic equilibrium. 
\\A similar analysis was performed in \cite{schmitt3}. Furthermore, it was already seen in \cite{Amado:2013aea} that the critical velocity is determined by quasinormal mode vanishing in both its real and imaginary part.
\\The same mechanism was found in \cite{lanliu}. By means of a nonlinear simulation in the probe limit, the authors were able to describe the relaxation of the system from a state with a superfluid velocity above the critical velocity down to the stable region. The underlying mechanism was found to be the production of solitons in one spatial dimension, and of vortex-antivortex pairs in 2D.
\bigskip
\bigskip
\\As a final note, our instability bears some resemblance to the Gregory-Laflamme instability of black strings \cite{gregorylaflamme,gregory}.  Gregory and Laflamme showed in 1993 that a 5-dimensional black string can be classically unstable under long wavelength perturbations. The event horizon will ripple, eventually breaking the black string into a sequence of black holes. The interesting feature of this instability is that it can be predicted by means of a thermodynamic argument for black strings with a compactified fifth dimension of size $L$. One can show the ratio of the entropy of the black hole and that of the black string solution (both of mass $M$) behaves as
\be
\frac{S_{\text{black hole}}}{S_\text{black string}} \propto \sqrt{\frac{L}{M}}
\ee
This implies that for large wavelengths one should expect a phase transition away from the black string solution. Indeed, such a behaviour can be detected with perturbation theory around the background solution: there exists a range of wavelengths for which it is possible to find a regular perturbation which is exponentially diverging in the timelike coordinate. These are large wavelengths proper to perturbations around equilibrium - a situation intuitively similar to hydrodynamics. One can thus interpret this behaviour as the onset of a dynamical/thermodynamic instability, in a similar fashion to what we have found to take place in superfluids.
\bigskip
\\Our results could be generalized in a handful of directions. 
\\First, the addition of a transverse component of the wavevector $k_y$ (or, equivalently, the introduction of an angle between the wavevector and the background superfluid velocity) would be an obvious generalization. The inclusion of momentum relaxation in both the hydrodynamic and the holographic analysis (following \cite{Andrade:2013gsa}) would also be a valuable addition to the picture. 
\\One could also consider superfluids on compact spacetimes - for instance a torus or a sphere, which would describe superfluidity in neutron stars. 
\bigskip
\\A deeper understanding of the validity of superfluid hydrodynamics as an effective field theory, and its breakdown, could also shed some more light on the process that we have studied. In particular, it would be valuable to gain a better insight about the role of vortices and their effective description, following \cite{lanliu}. This would involve writing a superfluid hydrodynamic theory which also accounts for free vortices. Such a framework was developed in \cite{blaisedelac2016} and revisited in \cite{Delacretaz:2019brr,armasjain23} using higher-form symmetries.
\bigskip
\\Finally, a comparison with experimental data would obviously be interesting. In fact, our criterion (\ref{thecriterion}) for the superfluid instability is formulated in a quite simple way, only involving macroscopic quantities, that could, in principle, be measured.

\newpage

\chapter*{Appendices}
\addcontentsline{toc}{chapter}{\protect\numberline{}Appendices}

\section*{Appendix A: Hydrodynamic frames}
\addcontentsline{toc}{section}{\protect\numberline{}Appendix A: hydrodynamic frames}
Here we follow Kovtun's approach \cite{Kovtun:2012rj, kovtun19}.
\bigskip
\\Consider normal fluid hydrodynamics.
In equilibrium it is possible to define $T$, $\mu$ and $u^\mu$ across the whole system. Once we break equilibrium this is not possible anymore. We can, however, give valid definitions for the tensors $J^\mu$ and $T^{\mu \nu}$: $\epsilon(T,\mu)$, $p(T,\mu)$ and $\rho(T,\mu)$ will just be parameters entering these definitions. In particular we can redefine them as long as the expressions for the tensors do not change. We do this by adding one-derivative terms.
 \\Let $u^\mu$ denote the fluid velocity with unit norm: $u^\mu u_\mu = -1$. We define the projector onto the subspace of $\mathbb{R}^d$ orthogonal to $u^\mu$ as:
 \be
 \Delta^{\mu \nu} \equiv \eta^{\mu \nu} + u^\mu u^\nu
 \ee
 The most general constitutive relations can be parametrized as:
\begin{equation}
 T^{\mu \nu} = \tilde{\epsilon} u^\mu u^\nu + \tilde{p}\Delta^{\mu \nu} + \big( q^\mu u^\nu + q^\nu u^\mu \big) + \tau^{\mu \nu}
 \end{equation}
 \begin{equation}
 J^\mu = \tilde{\rho} u^\mu + j^\mu
 \end{equation}
where $q_\mu u^\mu = j_\mu u^\mu = 0$ ($q^\mu$ and $j^\mu$ are transverse vectors). Moreover, $\tau^{\mu \nu}$ is taken to be symmetric, transverse and traceless (dissipative contributions to the $T^\mu_\mu$ will appear in $\tilde{\epsilon}$ and $\tilde{p}$). Finally, $q^\mu$, $j^\mu$ and $\tau^{\mu \nu}$ are first order in derivatives.
\bigskip
\\Starting from these relations we obtain:
\be \tilde{\epsilon} \equiv u_\mu u_\nu T^{\mu \nu} \qquad \tilde{p}\equiv \frac{1}{d} \Delta_{\mu \nu}T^{\mu \nu}\qquad \tilde{\rho} \equiv -u_\mu J^\mu \qquad q_\mu \equiv - \Delta_{\mu \rho}u_\nu T^{\rho \nu} \qquad j_\mu \equiv \Delta_{\mu \nu}J^\nu
\ee
These are simply definitions; for instance $\tilde{\epsilon}$ need not exactly coincide with the microscopic definition of the energy density in equilibrium. 
\\With a different choice of ``frame" one shifts $T$, $\mu$ and $u^\mu$ by linear corrections in derivatives ($\delta T(x), \delta \mu(x), \delta u^\mu(x) \sim \mathcal{O}(\partial)$):
\begin{equation}
T(x) \to T'(x) = T(x) + \delta T (x) \quad \mu(x) \to \mu'(x) = \mu(x) + \delta \mu (x) \quad 
u^{\mu}\to u'^{\mu}(x) = u^{\mu}(x) + \delta u^{\mu}(x)
\end{equation}
Requiring $u'^{\mu}u'_{\mu}=-1$ we see that $\delta u_\mu  u^\mu=0$. This implies:
\begin{equation} \label{zerodeltas}
\delta \tilde{\epsilon} = \delta \tilde{p} = \delta \tilde{\rho} =0
\end{equation} 
\begin{equation}
\delta q_\mu = -(\tilde{\epsilon}+ \tilde{p})\delta u_{\mu} \quad \quad \delta j_\mu = - \tilde{\rho}\delta u_\mu \quad \quad \delta \tau_{\mu \nu} = 0
\end{equation} 
For example (under a frame redefinition $\delta T^{\mu \nu} = \delta J^{\mu} = 0$):
\be
\begin{split}
j_\mu \equiv \Delta_{\mu \nu}J^\nu \quad \Longrightarrow \quad \delta j_\mu &= \delta u_\mu u_\nu J^\nu + u_\mu \delta u_\nu J^\nu + \mathcal{O}(\partial^2) \\
&=-\tilde{\rho}\delta u_\mu + \tilde{\rho}\,\cancel{u_\mu \delta u_\nu u^\nu}+\mathcal{O}(\partial^2)
\end{split}
\ee
\be
\begin{split}
q_\mu \equiv -\Delta_{\mu \rho}u_\nu T^{\rho\nu} \quad \Longrightarrow \quad \delta q_\mu &=  -\eta_{\mu \rho}\delta u_\nu T^{\rho \nu} -u_\nu u_\rho \delta u_\mu T^{\rho \nu} -\cancel{u_\nu u_\mu \delta u_\rho T^{\rho \nu}} -\cancel{u_\mu u_\rho \delta u_\nu T^{\rho \nu}} + \mathcal{O}(\partial^2)\\
&=-\tilde{p}\delta u_\mu - (\tilde{\epsilon} + \tilde{p} - \tilde{p}) \delta u_\mu = -(\tilde{\epsilon} + \tilde{p}) \delta u_\mu +\mathcal{O}(\partial^2)\,,
\end{split}
\ee
where the barred terms cancel because $\delta u_\mu u^\mu = 0$.
\\We now see that one can redefine $u_\mu$ such that, for instance, $q^\mu=0$ (Landau frame \cite{landaulif}) or $j^\mu=0$ (Eckart frame \cite{eckart}). In the first case we will choose $\delta u_\mu = q_\mu/(\tilde{\epsilon}+\tilde{p})$; in the latter $\delta u_\mu = j_\mu/\tilde{\rho}$.
\bigskip
\\Finally, we can ask ourselves whether we can pick a frame where we can identify $\tilde{\epsilon}$, $\tilde{p}$ and $\tilde{\rho}$ with their ideal order equivalents $\epsilon$, $p$, $\rho$.In a given frame we can write ($\star= \epsilon, p, \rho$):
\begin{equation}
\tilde{\star} = \star(T,\mu) + f_{\tilde{\star}}(\partial T, \partial \mu, \partial u_\mu)
\end{equation}
where the functions $f_\star$ are first order in derivatives. Let us suppose that we picked $\delta u_\mu$ in order to satisfy some frame condition. We are still free to choose $\delta \mu$ and $\delta T$. From (\ref{zerodeltas}) we know that $\tilde{\epsilon}$, $\tilde{p}$ and $\tilde{\rho}$ do not change. Therefore:
\be
f'_{\star} = f_\star - \pdert{\star}{T}{\mu}\delta T  - \pdert{\star}{\mu}{T}\delta \mu
\ee
We obtain a system of three linear equations for two variables, $\delta T$ and $\delta \mu$. This means that we can redefine the fields so that two of the three functions $f_{\tilde{\epsilon}}$, $f_{\tilde{p}}$ and $f_{\tilde{\rho}}$ are set to 0. It is customary to let $f_{\tilde{\epsilon}}$ and $f_{\tilde{\rho}}$ vanish, effectively setting $\tilde{\epsilon}= \epsilon$ and $\tilde{\rho}= \rho$. 
\bigskip
\\In a general frame one can build three independent scalars at first order in derivatives: $\dnu u^\nu$, $u^\nu \dnu \mu$ and $u^\nu \dnu \mu$. Similarly, there are three transverse vectors $\Delta_{\rho}^{\nu} \dnu \mu$, $\Delta_\rho^{\nu} \dnu T$ and $u^\nu \dnu u_\rho$ and one transverse traceless tensor $\sigma_{\rho \nu}\equiv \Delta_\rho^\alpha \Delta_\nu^\beta(\partial_\alpha u_\beta + \partial_\beta u_\alpha - \frac23 \eta_{\alpha \beta}\partial_\sigma u^\sigma)$. This implies that from $\tilde{\epsilon}$, $\tilde{p}$, $\tilde{\rho}$, $q^\mu$, $j^\mu$ and $\tau^{\mu \nu}$ we get a total of $3+3+3+3+3+1=16$ independent coefficients. As we saw above, we can set $j^\mu$ or $q^\mu$ (or some linear combination thereof) to zero, and we can do the same with the dissipative contributions to two of the three scalars $\tilde{\epsilon}$, $\tilde{p}$ and $\tilde{\rho}$. This procedure eliminates $3+3+3=9$ coefficients, leaving us with a total of seven. Furthermore, the equations of motion $\dmu J^\mu=0$ and $\dmu T^{\mu \nu}=0$ can be decomposed in two relations $\dmu J^\mu =u_\mu \dmu T^{\mu \nu}=0$ between the three scalars $\dnu u^\nu$, $u^\nu \dnu \mu$ and $u^\nu \dnu \mu$, as well as one relation $\dmu T^{\mu \nu}=0$ between the three vectors  $\Delta_{\rho}^{\nu} \dnu \mu$, $\Delta_\rho^{\nu} \dnu T$ and $u^\nu \dnu u_\rho$. This reduces the total count of independent coefficients to $7-3=4$. Finally, imposing positivity of entropy production one can eliminate one further coefficient, resulting in a total amount three independent coefficients on-shell \cite{kovtun19}.
\bigskip
\\A similar reasoning can be applied to superfluid hydrodynamics.

\section*{Appendix B: recovering Landau's criterion}
\addcontentsline{toc}{section}{\protect\numberline{}Appendix B: recovering Landau's criterion}
An interesting point to make is that our criterion correctly reproduces the Landau criterion for Galilean superfluids (for example $^4$He with roton excitations). To do so, we consider Galilean superfluids in the superfluid rest frame. As required by the Galilean boost Ward identity \cite{geracie}, in the Galilean limit we impose that the momentum density be equal to the current: $g^i\equiv T^{ti}=j^i$ (setting the electron charge and the particle mass $e=m=1$). 
\bigskip
\\Going to the superfluid rest frame involves boosting from the lab frame to a frame moving with velocity $\bar{\mathbf{v}}_\text{s}$, parametrized by coordinates $t'=t$, ${\bf x'}={\bf x}-\bar{\bf v}_s t$ with $\partial_t' = \partial_t - \bar{v}_s^i\partial_i', \,\, \partial_i'=\partial_i$. Note that in this rest frame the chemical potential is given by $\mu_0 = \mu + \bar{\mathbf{v}}_\text{s}\cdot \mathbf{v}_\text{n}-\bar{\mathbf{v}}_\text{s}^2/2$. 
\bigskip
\\Given a wavevector $\mathbf{q}$, we write the pressure in the superfluid rest frame (in the ensemble with fixed $\mathbf{v}_\text{s}$) as:
\be
\tilde{p} = -T\int \frac{\text{d}^d q}{(2\pi)^d}\ln (1-e^{-(\epsilon_q-{\bf q}\cdot{\bf w})/T}).
\ee
where $\mathbf{w}\equiv \mathbf{v}_\text{n}-\bar{\mathbf{v}}_\text{s}$ is the relative velocity between the normal fluid and the superfluid and $\epsilon_q(\mu_0, T)$ is the dispersion relation of the quasiparticle excitations, which the normal fluid is made of \cite{landau1980course9,Schmitt:2014eka}.
It is clear from the form of the pressure that as:
\begin{align}
\label{galileanlandau}
\mathbf{w} \to \mathbf{w}_c \equiv \text{min}_q \frac{\epsilon_q(\mu_0,T)}{q}
\end{align}
the susceptibilities will diverge:
\begin{align}
\lim_{w\to w_c}\tilde{\chi}_{ij} \propto \int \frac{\text{d}^d q}{(2\pi)^d}\frac{\mathcal{F}_{ij}(q,\mu_0,T)}{(\epsilon_q - {\bf q}\cdot{\bf w})^2}.
\end{align}
Notice that smoothness of $\epsilon_q(T,\mu_0)$ implies that $\mathcal{F}$ is also smooth near $\mathbf{w}_c$.
\bigskip
\\Furthermore, with a similar analysis to what we did in (\ref{soundinst},\ref{gammainst}) one can show that at $\chi_{hh}\equiv \tilde{\chi}_{v_\text{s}v_\text{s}}^{-1} \to 0$ one sound mode vanishes and becomes unstable, thus validating the consistency of our criterion, as we showed in \cite{shortpaper}.
\\In \cite{andreev1,andreev2}, the critical velocity for which $\chi_{hh}=\tilde{\chi}_{v_sv_s}=0$ was related to the Landau critical velocity in a different thermodynamic ensemble, but only found equality at zero temperature, assuming a phonon-roton quasiparticle dispersion relation. Here we see that we do not need to assume any specific dispersion relation.
\bigskip
\\For more details on Galilean superfluid hydrodynamics we refer the reader to \cite{clark, putterman,Banerjee:2016qxf, landau}.

\bigskip
\section*{Appendix C: pseudospectral methods}
\addcontentsline{toc}{section}{\protect\numberline{}Appendix C: Pseudospectral methods}
Pseudospectral methods come in handy because they allow to numerically approximate the solution of differential equations with good precision, while keeping a smaller number of grid points \cite{krikun}. 
\bigskip
\\Let us consider a general differential equation:
\begin{equation} \label{diffeqnum}
D[f(x)]= 0
\end{equation}
where $D$ is a differential operator. Numerical methods are necessary to obtain an approximation of its solution in a given interval when the analytic solution is not available.
\bigskip
\\Finite element methods divide the interval into a grid and approximate the value of the function on the $n+1$-th grid point by means of a quadrature formula relying on the values of the previously computed points.
\\On the other hand, pseudospectral methods compute the values of $f(x)$ at once on all grid points. To do so one picks an orthonormal basis of functions $f_n(x)$ on the chosen interval and formally writes the solution as:
\begin{equation}
f(x) = \sum_{n=0}^{\infty} \alpha_n f_n(x)
\end{equation}
Determining the coefficients $\alpha_n$ is equivalent to finding the solution of the differential equation. From a practical point of view, one selects a grid of $N+1$ points in the chosen interval. This allows to determine the coefficients of the first $N+1$ basis functions:
\begin{equation}
f(x) \approx f_N(x) = \sum_{n=0}^{N} \alpha_n f_n(x)
\end{equation}
The differential equation is exactly solved on the grid points. It is intuitively evident that a larger number of grid points produces a more accurate approximation of $f(x)$ (although in practice higher precision may be required with a larger grid).
\\Note that most of the times the grid points are not equidistant. Instead, their choice depends on the choice of basis functions. The most commonly used basis functions $f_n(x)$ are the \textit{Chebyshev polynomials} $T_n(x)$. They are defined by the relation \cite{boyd}:
\begin{equation}
T_n(\cos \theta) = \cos(n\theta)   \qquad n = 0, 1, \cdots
\end{equation}
So that on the interval $-1\leq x\leq1$ one can write:
\begin{equation} \label{chebpoints}
T_n(x) = \cos \big(n \arccos(x)\big)      
\end{equation}
One can show \cite{boyd} that the optimal choice for the grid points, i.e. the one that minimizes the difference $f(x) - f_N(x)$, is given by the roots of the Chebyshev polynomials of degree $N+1$:
\begin{equation}
x_i = - \cos \left( \frac{(2i+1)\pi}{2(N+1)}\right) \qquad \qquad i = 0, \cdots, N
\end{equation}
One can see that the grid points get denser near the endpoints, the spacing there being proportional to $1/N^2$.
\\On a general interval $a<x<b$ one defines the map $y(x):  \quad [a,b]\, \to \,[-1,1]$
\begin{equation}
y(x) = \frac{2}{b-a} \, x + \frac{a+b}{a-b}
\end{equation}
and sets:
\begin{equation}
y_i = y(x_i)
\end{equation}
The bulk equations of motion presented in this thesis were solved on the $[0,1]$ interval, whose two endpoints correspond respectively to the UV AdS boundary and the IR black hole horizon.
\bigskip
\\The differential equation (\ref{diffeqnum}) evaluated at the $N+1$ grid points yields $N+1$ equations for the coefficients $c_n$. In practice, however, one does not usually solve for the coefficients $\alpha_n$. Instead, via decomposition in Chebyshev polynomials, one can show that the values of $f_N'(x)$ on the grid can be obtained from those of $f_N(x)$ via a linear map:
\begin{equation}
f_N'(x_i) = ({D_N})_{ij}\, f_N(x_j)
\end{equation}
with the matrix $({D_N})_{ij}$ defined as \cite{trefethen}: 
\begin{align}
\begin{split}
(D_N)_{00} = \frac{2N^2+1}{6} \qquad  (D_N)_{NN} = -\frac{2N^2+1}{6} \\
(D_N)_{ii}  = -\frac{x_j}{2(1-x_i^2)} \qquad \text{for } 1\leq i\leq N-1 \\
(D_N)_{ji}  = \frac{c_i}{c_j}\frac{(-1)^{i+j}}{x_i -x_j} \qquad \text{for } i \neq j 
\end{split}
\end{align}
with 
\begin{equation}
c_i = \begin{cases}
 1 \quad \text{for } 1\leq i \leq N-1 \\
 2 \quad \text{otherwise} 
  \end{cases}
\end{equation}
Similar matrices can be computed for higher derivatives: one simply has ${D_k}_N=(D_N)^k$, where ${D_k}_N$ is the matrix corresponding to the $k^{\text{th}}$ derivative.
Furthermore, ${D_0}_N = \text{Id}_{(N+1)\times(N+1)}$.
\\For a linear differential operator the original equation (\ref{diffeqnum}) is thus reduced to a linear problem:
\begin{equation}
\left(\sum_{k=0}^{k_{\text{max}}} \beta_k \, \partial_k \right) f(x) =0
\end{equation}
One finds the $N+1$ values of $f(x)$ on the grid points by solving for 
\begin{equation}
\tilde{D}_{ij}\,f(x_j) \equiv \left(\sum_{k=0}^{k_\text{max}} \beta_k\, ({D_k}_N)_{ij}\right) f(x_j) =0
\end{equation}
\\The boundary conditions are implemented by substituting the lines of $\tilde{D}$ corresponding to the grid points at which the conditions are imposed with new lines corresponding to the given boundary problems. For instance: a Dirichlet boundary condition $f(a) = c$ is implemented  by setting $(\tilde{D})_{0j}=\delta_{0j}$ and by solving the linear system of equations:
\begin{equation}
\tilde{D}_{ij} \, f(x_j)= \lambda_i
\end{equation}
where $\lambda_i = c\, \delta_{i0}$. Similarly, a Neumann boundary condition $f'(a) = c$  is implemented  by setting $(\tilde{D})_{0j}=D_{1j}$. 
\bigskip
\\A final note about coupled equations: in the case of $n$ differential equations for $n$ functions $f_\ell(x)$, $1\leq \ell\leq n$, one proceeds in a similar fashion. One casts the numerical values of each function on the grid into a $n(N+1)$-dimensional array
\begin{equation}
f \equiv (f_1(x_0), \cdots, f_1(x_N), \cdots, f_n(x_0), \cdots, f_n(x_N))^\text{T}
\end{equation}
The derivative matrices will be $(N+1)\times n(N+1)$ matrices. For instance matrix giving the $k$-th derivative of $f_\ell(x)$ is:
\begin{align}
\begin{split}
D_{k,\ell} \equiv (0, \cdots, \,&D_{k}, \cdots, 0) \\
&\downarrow \\
&\hspace*{-0.85cm}\ell\text{-th position}
\end{split}
\end{align}
The final matrix $\tilde{D}$ encoding the system of differential equations will be a $n(N+1) \times n(N+1)$ matrix formed by $n$ building blocks of dimensions $(N+1)\times n(N+1)$ corresponding to the individual equations. 

\subsection*{Computing the quasinormal modes with pseudospectral methods}
Solving for the quasinormal modes (QNMs) of the systems means finding the poles of the Green's functions. Equivalently, casting the equations of motion as
\begin{equation} \label{eqwithDtilde}
\tilde{D}(\omega, k) \, f = \lambda
\end{equation}
one looks for the values $\omega(k)$ that send $\det \tilde{D}$ to zero.
\bigskip
\\Following \cite{jansen}, let us first assume that $\tilde{D}$ is linear in $\omega$:
\begin{equation}
\tilde{D} = \tilde{D}_0 + \omega \tilde{D}_1
\end{equation}
Imposing $\det \tilde{D} = 0$ is equivalent to solving the generalized eigenvalue problem $\tilde{D}_0 f = - \omega \tilde{D}_1 f$. This operation is built in in most scientific programming languages. In Mathematica\tiny{{\textregistered}} \normalsize it is implemented in the command \texttt{Eigenvalues}.
\bigskip
\\In the more general case where $\tilde{D}$ is non-linear in $\omega$ 
\begin{equation}
\tilde{D} = \sum_{i=0}^{i_\text{max}} \omega^i \tilde{D}_i
\end{equation}
We can then define the vector 
\begin{equation}
\mathcal{F} \equiv (f, \omega f, \cdots, \omega^{i_\text{max}-1}f)
\end{equation}
and recast (\ref{eqwithDtilde}) as (finite values of the boundary conditions do not affect the computation of the modes, so for simplicity we set $\lambda=0$):
\begin{equation}
\mathcal{D}\mathcal{F} \equiv (\mathcal{D}_0 + \omega \, \mathcal{D}_1) \mathcal{F}=0
\end{equation}
with 
\begin{align}
\mathcal{D}_0 \equiv \begin{pmatrix}
\tilde{D}_0 & \tilde{D}_1& \tilde{D}_2 & \ldots &\tilde{D}_{i_\text{max}-1} \\
0 & \text{Id} & 0 &\ldots & 0 \\
0 & 0 & \text{Id} & \ldots &0 \\
\vdots & \vdots  & \vdots & \ddots & \vdots \\
0 & 0 & 0& 0&\text{Id}
\end{pmatrix}  \quad \, \,
\mathcal{D}_1 \equiv \begin{pmatrix}
0& 0&0 & \ldots & 0&\tilde{D}_{i_\text{max}} \\
-\text{Id} & 0 & 0 & \ldots &0 &0 \\
0&-\text{Id} & 0 & \ldots &0&0 \\
\vdots & \vdots &\vdots& \ddots &\vdots &\vdots \\
0 & 0 & 0& \ldots&-\text{Id} &0
\end{pmatrix} \, \text{,}
\end{align}
thus obtaining a generalized eigenvalue problem as above.

\addcontentsline{toc}{chapter}{\protect\numberline{}Bibliography}
\bibliographystyle{JHEP}
\bibliography{local}

\includepdf{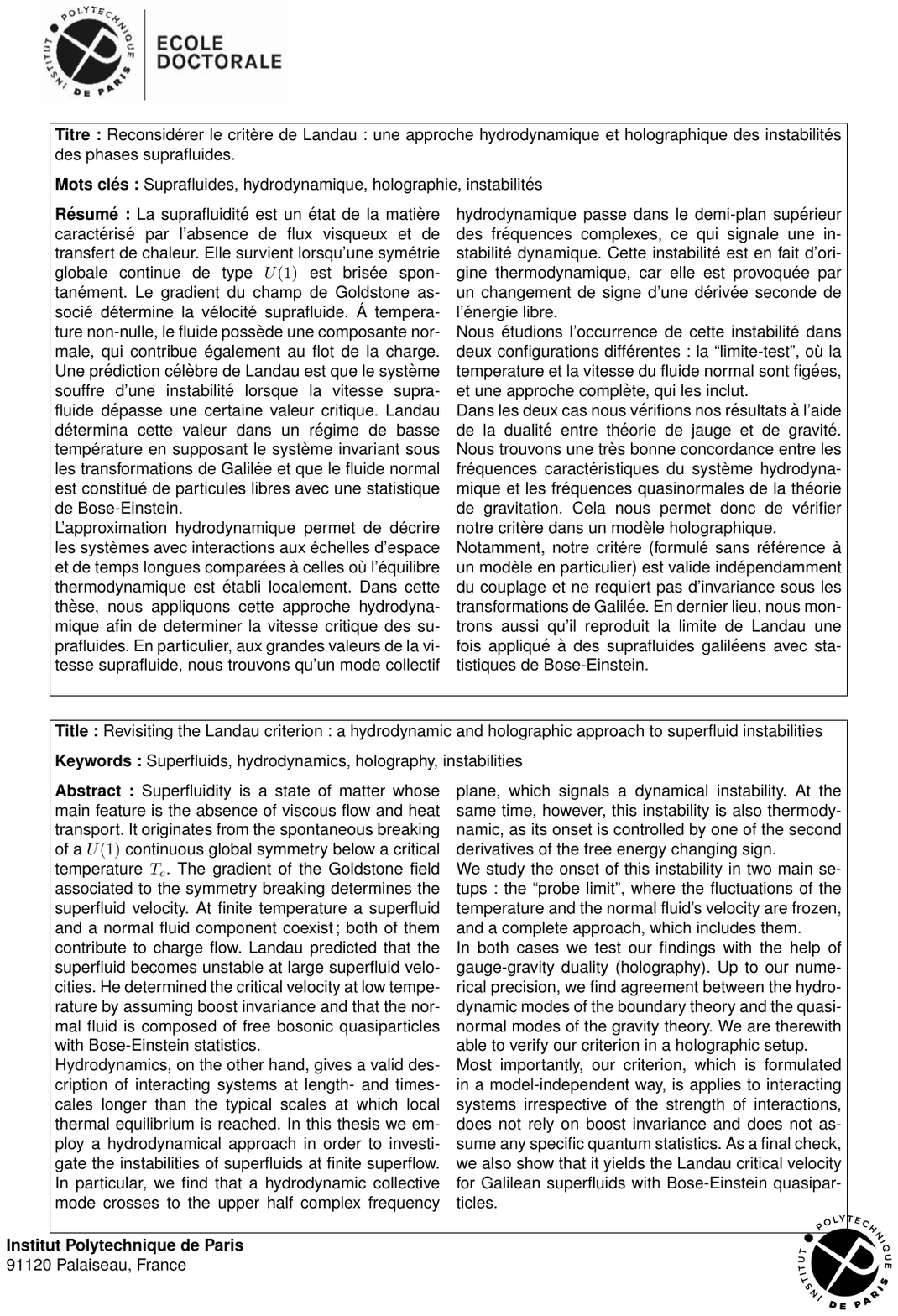}

\end{document}